\documentclass[useAMS,usenatbib,usegraphicx]{mn2e}

\title[Constraints on dark matter annihilation]{Multi-messenger constraints on dark matter annihilation into electron-positron pairs}

\author[M.~Wechakama and Y.~Ascasibar]
{
M.~Wechakama$^{1,2\star}$ and Y.~Ascasibar$^{3\dagger}$ \\
$^1$Leibniz-Institut f\"{u}r Astrophysik Potsdam, An der Sternwarte 16, Potsdam
14482, Germany\\
$^2$Department of Physics, Faculty of Science, Kasetsart University, Chatuchak, Bangkok, 10900, Thailand \\
$^3$Departamento de F\'{i}sica Te\'{o}rica, Universidad Aut\'{o}noma de Madrid,
Madrid 28049, Spain
}

\date{\bf MNRAS in Press}

\newcommand{\be}{\begin{equation}}
\newcommand{\ee}{\end{equation}}
\newcommand{\bea}{\begin{eqnarray}}
\newcommand{\eea}{\end{eqnarray}}

\newcommand{\dd}{{\rm d}}
\newcommand{\deriv}[2]{\frac{\dd#1}{\dd#2}}

\usepackage{mathrsfs}

\usepackage{color}
\newcommand{\Referee}[1]{{\color{black}#1}}
\begin{document}

\maketitle

\begin{abstract}
We investigate the production of electrons and positrons in the Milky Way within the context of dark matter annihilation.
Upper limits on the relevant cross-section are obtained by combining observational data at different wavelengths (from Haslam, WMAP, and Fermi all-sky intensity maps) with recent measurements of the electron and positron spectra in the solar neighbourhood by PAMELA, Fermi, and HESS.
We consider synchrotron emission in the radio and microwave bands, as well as inverse Compton scattering and final-state radiation at gamma-ray energies.
According to our results, the dark matter annihilation cross-section into electron-positron pairs should not be higher than the canonical value for a thermal relic if the mass of the dark matter candidate is smaller than a few~GeV.
In addition, we also derive a stringent upper limit on the inner logarithmic slope $\alpha$ of the density profile of the Milky Way dark matter halo ($\alpha<1$ if $m_{\rm dm}<5$~GeV, $\alpha<1.3$ if $m_{\rm dm}<100$~GeV and $\alpha<1.5$ if $m_{\rm dm}<2$~TeV) assuming that $\langle \sigma v \rangle_{e^\pm}=3 \times 10^{-26}$~cm$^3$~s$^{-1}$.
A logarithmic slope steeper than $\alpha \sim 1.5$ is hardly compatible with a thermal relic lighter than $\sim 1$~TeV, regardless of the dominant annihilation channel.
\end{abstract}

\begin{keywords}
dark matter -- astroparticle physics -- radiation mechanisms: non-thermal -- Galaxy: structure
\end{keywords}

\footnotetext[1]{E-mail: mwechakama@gmail.com (MW)}
\footnotetext[2]{E-mail: yago.ascasibar@uam.es (YA)}

%--------------------------------------------------------------------------
 \section{Introduction}
%--------------------------------------------------------------------------

Dark matter can be indirectly detected through the signatures of standard model
particles produced by its annihilation or decay \citep[see e.g.][]{Bertone+05,
Bertone2010}.
A great deal of work has focused on the emission of gamma rays from the Galactic
centre \citep[e.g.][among many others]{Berezinsky+94, Bergstrom+98, BaltzEdsjo99, GondoloSilk99, Morselli+02, Ullio+02, Stoehr+03, Peirani+04, Prada+04, Cesarini+04, Bergstrom+05b, Bergstrom+05, Profumo05, Aharonian+06, ZaharijasHooper06, Boyarsky+08, Pospelov+08, Springel+08, BellJacques09, CirelliPanci09, Fornasa+09, BernalPalomares-Ruiz_10, Abazajian+10, Cirelli+10, PapucciStrumia10, Abramowski+11, Hooper+11, HooperLinden11, HooperLinden11b, Ackermann+12Con, AbazajianHarding12}, the Milky Way
satellites \citep[e.g.][]{Baltz+00, Tyler02, BaltzWai04, Hooper+04, Bergstrom+06, SanchezConde+07, Strigari+07, Strigari+08, Wood+08, Martinez+09, Abdo+10, Acciari+10, Essig+09, Essig+10, Ackermann+11, HESS+11}, and galaxy
clusters \citep[e.g.][]{Colafrancesco+06, Jeltema+09, Ackermann+10b, SanchezConde+11, Pinzke+11}.
Prospects for indirect dark matter detection in the microwave background have also been considered by several authors \citep[e.g.][]{Blasi+03, Colafrancesco04, PadmanabhanFinkbeiner05, Mapelli+06, Zhang+06, Zhang+07, Cholis+09a, Galli+09, Slatyer+09, Kanzaki+10, Lavalle10, Hutsi+11, Galli+11, McQuinn+11, Delahaye+12}, as well as
X-ray \citep[e.g.][]{Abazajian+01, Boyarsky+07, Boyarsky+08b, Zavala+11}, radio \citep[e.g.][]{ColafrancescoMele01, Aloisio+04, Bergstrom+09, Borriello+09, Ishiwata+09, Fornengo+12} and multi-wavelength signatures \citep[e.g.][]{Regis+08, RegisUllio08, Bertone+09, Pato+09, Profumo+10, Crocker+10}.

The recent results from indirect detection experiments in the solar neighbourhood have also suggested the possibility that such a signature has been seen.
In particular, the PAMELA experiment has pointed a significant excess of electrons and positrons above the expected smooth astrophysical background \citep{Adriani+09_positrons}.
If these results, confirmed by Fermi \citep{Ackermann+12} and AMS-02 \citep{Aguilar+13}, are interpreted in terms of dark matter annihilation, then an abundant population of high-energy $e^\pm$ is being created everywhere in the Galactic dark matter halo, with the associated final-state radiation (FSR), as well as synchrotron emission in the Galactic magnetic field and inverse Compton scattering (ICS) of the photons of the interstellar radiation field (ISRF).

Although the currently most favoured explanation for the origin of Galactic positrons, traced by the positron annihilation emission line at 511 keV \citep[see][for a recent review]{Prantzos+11} is low-mass X-ray binaries \citep{Weidenspointner+08}, and the local positron excess at high energies is most likely due to the contribution of nearby pulsars \citep[see e.g.][]{Profumo12}, several works have considered the possibility that dark matter annihilation makes a sizeable contribution to the positron budget of the Milky Way \citep[e.g.][]{Boehm+04, BoehmAscasibar04, Beacom+05, PicciottoPospelov05, Ascasibar+06, BeacomYuksel06, Sizun+06, FinkbeinerWeiner07, PospelovRitz07, Barger+09, Bergstrom+09b, ChenTakahashi09, Cholis+09, Cirelli+09, Donato+09, Grasso+09, Malyshev+09, MertschSarkar09, RegisUllio09, Yin+09, Chen+10, Meade+10, Cline+11, Vincent+12}.

The present work focuses on the astrophysical signatures of dark matter annihilation into electron-positron pairs, neglecting other processes, such as dark matter decay, or other annihilation products, such as protons and antiprotons \citep[whose contribution is severely constrained by recent observational data; see e.g.][]{Adriani+09_protons}.
We try to impose robust, yet stringent \Referee{constraints} on the relevant cross-section by comparing the
predictions of an analytic model of particle propagation with a multi-wavelength
set of observational data obtained from the literature.
More precisely, we compare the expected emission from synchrotron radiation, 
ICS and FSR within the
Milky Way with 18 maps of the sky at different frequencies: the Haslam radio map
at 408 MHz, the 7-year data from the Wilkinson Microwave Anisotropy Probe (WMAP) in
its 5 bands (23 GHz, 33 GHz, 41 GHz, 61 GHz, and 94 GHz), and gamma-ray maps
from the Fermi Large Area Telescope (LAT) binned in 12 different channels (from
0.3 to 300 GeV).
A straightforward statistical criterion is used in order to mask the most
obvious astrophysical signals (i.e. emission from the Galactic disc and point
sources), and observational upper
limits are derived from the remaining spherically-symmetric component.

In addition to the photon data, we also consider the recent measurements of the local electron and positron spectra performed by PAMELA \citep{Adriani+09_positrons, Adriani+10, Adriani+11, Adriani+13}, Fermi \citep{Ackermann+10, Ackermann+12}, and HESS \citep{Aharonian+08}.
As will be shown below, considering the positron spectrum separately (rather than the combined electron+positron spectrum) yields a significant improvement on the maximum value allowed for the positron injection rate or, equivalently, the dark matter annihilation cross-section.

Rather than focusing on a particular dark matter candidate, we adopt a model-independent
approach \citep[see e.g.][]{WechakamaAscasibar11}, in which all the injected particles are
created with the same initial energy $E_0$, of the order of the mass of the
dark matter particle.
Since this mass is usually much larger than the rest mass of the electron, electrons and positrons will be relativistic at the moment of their creation.
However, they can efficiently lose their energy through different processes, such as ICS, synchrotron radiation, Coulomb collisions, bremsstrahlung, and ionization.
 Throughout this paper, we will often use the Lorentz factor
$\gamma$ to express the energy $E = \gamma m_{\rm e}c^2$ of the annihilation products, where
$m_{\rm e}$ denotes the rest mass of electron, and $c$ is the speed of light.
We will first discuss the results obtained for a `canonical' model of the Milky Way and then explore the effects of varying the intensity of magnetic field, the diffusion coefficient, the ISRF, and the inner logarithmic slope of dark matter density profile.

The remainder of this paper is structured as follows: Section~\ref{secModel}
describes the procedure followed to estimate the electron-positron spectrum,
the surface brightness profiles, and the parameters of the Milky Way model.
Our analysis of the observational data is fully described in
Section~\ref{secObservation} (tables with precise numeric values are provided as
an appendix), and Section~\ref{secConstraints} is devoted to the constraints on the dark matter annihilation cross-section.
The effect of the different astrophysical parameters is discussed in Section \ref{secAstroParams}, while Section~\ref{secAlpha} focuses on the constraints that one can impose on the slope of the dark matter density profile by assuming that dark matter particles are produced as thermal relics in the primordial Universe.
Particular annihilation channels are discussed in Section~\ref{secAnnihilationChannels}, and our main conclusions are succinctly summarized in Section~\ref{secConclusions}.

%--------------------------------------------------------------------------
 \section{Model Predictions}
 \label{secModel}
%--------------------------------------------------------------------------

%--------------------------------------------------------------------------
 \subsection{Electron-positron propagation}
 \label{subsecPropagation}
%--------------------------------------------------------------------------

As in our previous work \citep{WechakamaAscasibar11}, the propagation of
electrons and positrons through the interstellar medium (ISM) is
determined by the diffusion-loss equation
\bea
\nonumber
\frac{ \partial }{ \partial t }
\frac{ {\rm d}n }{\rm d\gamma }(\textit{\textbf{x}},\gamma)
&=&
\nabla
\left[
K(\textit{\textbf{x}},\gamma) \nabla\frac{ {\rm d}n }{\rm d\gamma
}(\textit{\textbf{x}},\gamma)
\right]
\\
& + & \nonumber
\frac{ \partial }{ \partial\gamma }
\left[
b(\textit{\textbf{x}},\gamma) \frac{ {\rm d}n }{\rm d\gamma
}(\textit{\textbf{x}},\gamma)
\right]
\\
& + & Q(\textit{\textbf{x}},\gamma).
\eea
We assume a diffusion coefficient of the form
\be
K(\gamma) = K_0 \gamma^\delta
\label{eqDiffusion}
\ee
independent of Galactic location.
The values of $K_0$ and $\delta$ corresponding to the three models discussed by \citet{Donato+04} and \citet{Delahaye+08} are provided in Table~\ref{tabDiffusion} below.
The energy loss rate
\be
b(\textit{\textbf{x}},\gamma) \equiv -\frac{ {\rm d}\gamma }{ {\rm d}t
}(\textit{\textbf{x}},
\gamma) = \sum_i b_i(\textit{\textbf{x}}, \gamma)
\ee
is a sum over the
relevant physical processes, and the source term $Q(\textit{\textbf{x}},\gamma)$
represents the instantaneous electron-positron injection rate.

Given enough time \citep[of the order of 100~Myr; c.f. Figure 2 in][]{WechakamaAscasibar11}, the electron-positron population will approach
a steady-state distribution, $\frac{\partial}{\partial
t}\frac{{\rm d}n}{\rm d\gamma}(\textit{\textbf{x}},\gamma)=0$. Assuming
that $b(\textit{\textbf{x}}, \gamma)$ varies smoothly in space, the
particle spectrum fulfills the relation
\be
\frac{ \partial
y(\textit{\textbf{x}},\gamma) }{ \partial\gamma } + \frac{ K(\gamma) }{
b(\gamma) } \nabla^2 y(\textit{\textbf{x}}, \gamma) = -Q(\textit{\textbf{x}},
\gamma), \ee where \be y(\textit{\textbf{x}},\gamma) \equiv b(\gamma)
\frac{ {\rm d}n }{\rm d\gamma }(\textit{\textbf{x}},\gamma).
\ee

Imposing
$\frac{{\rm d}n}{\rm d\gamma}(\textit{\textbf{x}},\gamma)=0$ at infinity,
one obtains the Green's function
\be
G(\textit{\textbf{x}}, \gamma,
\textit{\textbf{x}}_{\rm s}, \gamma_{\rm s}) = \frac { \exp\left(-\frac
{\mid\textit{\textbf{x}}-\textit{\textbf{x}}_s\mid ^2} {2\Delta\lambda^2}
\right) }
{ \left(2\pi\Delta\lambda^2\right)^{3/2} } \
\Theta(\gamma-\gamma_{\rm s})
\ee
and either the image charges
method or an expansion over the eigenfunctions of the linear
differential operator may be used to derive the Green's function
for other boundary conditions \citep[see e.g.][]{BaltzEdsjo99,Delahaye+09}.
The electron-positron spectrum is thus given by
\be
\frac{{\rm d}n}{\rm d\gamma}(\textbf{x},\gamma)
\!=
\!\frac{ 1 }{ b(\textbf{x},\gamma) }
\!\int^\infty_\gamma\!\!\!\!\!\!\!\rm d\gamma_{\rm s}
\!\int_0^\infty\!\!\!\!\!\!\!\rm d^3\!\!\textbf{x}_s
\frac
{
 \exp
 \left(
   \!-\frac{\mid\textbf{x}-\textbf{x}_s\mid^2}{2\Delta\lambda^2}
  \right)
}
{ \left(2\pi\Delta\lambda^2\right)^{3/2} }
\textit{Q}(\textbf{x}_{\rm s}, \gamma_{\rm s}),
\ee
where the quantity
\be
 \Delta\lambda^2 = \lambda^2(\gamma)-\lambda^2(\gamma_{\rm s})
\ee
is related to the characteristic diffusion length of the electrons and
positrons, $\gamma_{\rm s}$ denotes their initial energy, and the variable $\lambda$ is defined as
\be
\lambda^2(\gamma) = \int_\gamma^\infty \frac{ 2 K(\gamma) }{ b(\gamma) }\rm
d\gamma.
\ee

Considering the dark matter halo as a spherically-symmetric source, the spatial
integral can be reduced to one dimension, and the
electron-positron spectrum is finally given by the expression
\bea
\nonumber
\frac{ {\rm d}n }{\rm d\gamma }(r,\gamma) \!\!\!\!&=&\!\!\!\!
\frac{ 1 }{ b(\gamma) } \frac{ \exp\left( - \frac{ r^2 }{ 2\Delta\lambda^2 }
\right)}
     { \left( 2\pi r^2 \Delta\lambda^2 \right)^{1/2} }
\\
&\times&\!\!\!\!\!\! \nonumber
\Big\{
\, \int^\infty_\gamma\!\!\!\! {\rm d} \gamma_{\rm s}
\int_0^\infty\!\!\!\! {\rm d} r_{\rm s}
\ r_{\rm s}
\ \exp\left( -\frac{ r_{\rm s}^2 }{ 2 \Delta\lambda^2 } \right)
\\
&\times&\!\!\!\!\!\!
\left[
\exp\left( \frac{ r r_{\rm s} }{ \Delta\lambda^2 } \right)
-
\exp\left( - \frac{ r r_{\rm s} }{ \Delta\lambda^2 } \right)
\right]
Q(r_{\rm s}, \gamma_s)\Big\}.
\label{eqSpectrum}
\eea

%--------------------------------------------------------------------------
 \subsection{Loss rates}
 \label{secLoss}
%--------------------------------------------------------------------------

Electrons and positrons can lose their energy by several physical processes as
they move through the ISM. We consider ICS of cosmic microwave background (CMB),
starlight and infrared photons, synchrotron radiation,
Coulomb collisions, bremsstrahlung, and ionization of neutral hydrogen atoms.

The energy loss rates depend on the energy of the particle.
High-energy electrons and positrons mainly lose energy by ICS
\citep[e.g.][]{Sarazin99}.
We compute the total power radiated by a single electron using the formalism described in Section~\ref{secBrightness}, based on the Klein-Nishina cross-section.
In the non-relativistic regime, the loss function can be approximated as
\be
b_{\rm ICS}(\gamma) =
\frac{ 4 }{ 3 } \frac{ \sigma_{\rm T} }{ m_{\rm e} c } \gamma^2 U_{\rm rad},
\label{eqICloss}
\ee
where $\sigma_{\rm T}$ is the Thomson cross-section.
The combined radiation energy density of the CMB, starlight (SL), and infrared (IR) light from thermal dust emission \citep[see e.g.][]{PorterStrong05, PorterMoskalenko08} is represented by three grey bodies,
\be
 U_{\rm rad}
=
 \frac{ 4\, \sigma_{\rm SB} }{ c }
 \left(\
   T_{\rm CMB}^4 + \mathscr{N}_{\rm SL} T_{\rm SL}^4 + \mathscr{N}_{\rm IR} T_{\rm IR}^4\
 \right)
\label{eqUrad}
\ee
where $T_i$ and $\mathscr{N}_i$ represent the effective temperature and the normalization of each component, respectively, and $\sigma_{\rm SB}$ is the Stefan-Boltzmann constant.
The cosmic microwave background is modeled as a perfect black body with temperature $T_{\rm{CMB}} = 2.726$~K \citep{Fixsen09}, and we follow \citet{CirelliPanci09} for the two other components (see Table~\ref{tabParameters} below).
Expression~(\ref{eqICloss}) provides a good approximation for low values of the Lorentz factor~$\gamma$, but it severely overestimates it for $\gamma m_{\rm e} c^2 \ge$~TeV, where relativistic effects become important.

Synchrotron radiation is another important loss mechanism at high energies.
The expression for the loss rate is similar to that of non-relativistic ICS, substituting the
radiation energy density in equation~(\ref{eqICloss}) by the magnetic energy density, $U_{\rm B}=B^2/(8\pi)$, where $B$ is the intensity of the
magnetic field:
\be
b_{\rm syn}(\gamma) =
\frac{ 4 }{ 3 } \frac{ \sigma_{\rm T} }{ m_{\rm e} c } \gamma^2 U_{\rm B}.
\label{eqsynloss}
\ee
% Both equations (\ref {eqICloss}) and (\ref {eqsynloss}) assume $\gamma\gg 1$.

For lower-energy electrons and positrons, Coulomb interactions with the thermal
plasma must be taken into account.
The loss rate is approximately \citep{Rephaeli79}
\be
b_{\rm Coul}(\gamma) \approx
1.2 \times 10^{-12} n_{\rm e}
\left[ 1 + \frac{ \ln( \gamma / n_{\rm e} ) }{ 75 } \right]\ \ \rm{s^{-1}},
\ee
where $n_{\rm e}$ is the number density of thermal electrons.

Collisions with thermal ions and electrons also produce radiation through
bremsstrahlung.
The loss rate due to bremsstrahlung can be approximated as
\citep{BlumenthalGould70}
\be
b_{\rm brem}(\gamma) \approx 1.51 \times 10^{-16} n_{\rm e} \gamma \left[
\ln(\gamma) + 0.36 \right]\ \ \rm{s^{-1}}.
\ee

Additional energy losses come from the ionization of hydrogen atoms. The loss
rate is given in \citet{Longair81},
\bea
\nonumber b_{\rm
ion}(\gamma) \!\!\!\!&=&\!\!\!\! \frac{ q_{\rm e}^4 n_{\rm H} }
     { 8\pi \epsilon_0^2 m_{\rm e}^2 c^3 \sqrt{ 1 - \frac{1}{\gamma^2} } }
\times
\Big[~
\ln \frac{ \gamma ( \gamma^2 - 1 ) }{ 2 \left( \frac{ I }{ m_{\rm e} c^2 }
\right)^2 }\\
& & \!\!\!\! - \left( \frac{ 2 }{ \gamma } - \frac{ 1 }{ \gamma^2 } \right) \ln2
+ \frac{ 1 }{ \gamma^2 } + \frac{ 1 }{ 8 } \left( 1 - \frac{ 1
}{ \gamma } \right)^2 ~\Big],
\eea
where $n_{\rm H}$ is the number density of
hydrogen atoms, $q_{\rm e}$ is the electron charge, $\epsilon_0$
is the permittivity of free space, and $I$ is the ionization energy of the
hydrogen atom. The number density of thermal electrons and neutral
atoms can be expressed in terms of the total ISM gas density $\rho_{\rm g}$ and
the ionization fraction $X_{\rm ion}$ as
\be
n_e = \frac{
\rho_{\rm g} }{ m_{\rm p} } X_{\rm ion} \ee 
and 
\be n_{\rm H} = \frac{ \rho_{\rm
g} }{ m_{\rm p} } ( 1 - X_{\rm ion}),
\ee
respectively.

%--------------------------------------------------------------------------
 \subsection{Source term}
%--------------------------------------------------------------------------

Since the electrons and positrons in our model originate from the annihilation
of dark matter particles, the instantaneous production rate at any given point can be expressed as
\be
Q(r,\gamma)= \eta\ n_{\rm dm}(r)\ n_{\rm dm^*}(r)
\ \langle \sigma v \rangle_{e^\pm}
\ \deriv{N_{e^\pm}}{\gamma}(\gamma),
\ee
where $n_{\rm dm}$ and $n_{\rm dm^*}$ denote the number densities of dark matter
particles and anti-particles, respectively, $\langle\sigma v\rangle_{e^\pm}$ is
the thermal average of the annihilation cross-section times the dark matter
relative velocity, and $\deriv{N_{e^\pm}}{\gamma}$ is the injection spectrum of
electrons and positrons in the final state.
For self-conjugate dark matter particles,
$ n_{\rm dm} = n_{\rm dm^*} = \frac{ \rho_{\rm dm} }{ m_{\rm dm} } $ and $\eta=1/2$ in order to avoid double counting; else, $ n_{\rm dm} = n_{\rm dm^*} = \frac{1}{2} \frac{ \rho_{\rm dm} }{ m_{\rm dm} } $ and $\eta=1$.

We consider self-conjugate dark matter particles throughout this work and assume that each annihilation event injects one electron and one positron with roughly the same energy $\gamma_0 \sim m_{\rm dm} / m_{\rm e}$,
\be
\deriv{N_{e^\pm}}{\gamma}(\gamma) = 2\ \delta(\gamma-\gamma_0),
\label{eqInjectionSpectrum}
\ee 
where $\delta(\gamma-\gamma_0)$ denotes a Dirac delta function.
Although this is a rather coarse approximation, it has the advantage of being 
model-independent.
For self-conjugate dark matter particles, we obtain
\be
Q(r,\gamma) = \, \left[ \frac{ \rho_{\rm dm}(r) }{ m_{\rm dm} } \right]^2 \langle \sigma v \rangle_{e^\pm}\ \delta(\gamma-\gamma_0).
\label{eqQ0} 
\ee 

We consider a spherically-symmetric halo, described by a density profile of the form 
\be 
\rho_{\rm {dm}}(r) = \frac { \rho_{\rm s} } {
\left( \frac { r }{ r_{\rm s} } \right)^\alpha
\left( 1 + \frac { r }{ r_{\rm s} } \right)^{3-\alpha}
},
\label{eqDMProfile1}
\ee
where $r_{\rm s}$ and $\rho_{\rm s}$
denote a characteristic density and radius of the halo,
respectively, and $\alpha$ is the inner logarithmic slope of
the density profile. Local inhomogeneities that would boost the
expected signal, such as small-scale clumpiness or the presence of
subhaloes, are not taken into account.
The shape of the dark matter density profile in the inner regions is far from
being a settled question.
N-body simulations suggest that, at least in the
absence of baryons, the profile should be quite steep near the centre ($\alpha
\sim 1$), in apparent contradiction with observations.
Traditionally, it has been argued that the presence of gas and stars makes the
profile even steeper due to the effects of adiabatic contraction
\citep{Blumenthal+86}, although some recent claims have also been made in the
opposite direction \citep[e.g.][]{El-Zant+01, Mashchenko+06, Oh+_10}.
Given the current uncertainties, we have left the inner slope of the density
profile as a free parameter of the model.

%--------------------------------------------------------------------------
 \subsection{Surface brightness profile}
 \label{secBrightness}
%--------------------------------------------------------------------------

Once the electron-positron spectrum is computed, the emission coefficient\footnote{Energy radiated per unit volume per unit frequency per unit time per unit solid angle.} for
photons of frequency $\nu$ is given by the integral
\be
j_{\nu}(r,\nu) = \frac{1}{4\pi}
\int_1^\infty \frac{{\rm{d}}n}{\rm{d}\gamma}(r,\gamma)
\ l(\gamma, \nu)\ \rm{d}\gamma
\ee
of the electron-positron spectrum $\frac{{\rm{d}}n}{\rm{d}\gamma}(r,\gamma)$ times the specific luminosity $l(\gamma, \nu)$ emitted at frequency $\nu$ by a single electron or positron with Lorentz factor $\gamma$.
The intensity from any given direction in the sky is simply the integral along
the line of sight of the emission coefficient. Since we assume a spherically-symmetric source and boundary conditions, it will only depend on the angular separation $\theta$ with respect to the
Galactic centre,
\be
I_\nu (\theta, \nu) = \int_0^\infty j_{\nu}(r,\nu)\ {\rm{d}}s,
\label{eqIntensity}
\ee   
where $s$ represents the distance along the line of sight, and the radial distance $r$ to the centre of the Milky Way at any point along the ray is
\be
r = \sqrt{x^2 + y^2},
\ee
with $x = s \sin\theta$, $ y = s \cos \theta - R_\odot $, and $R_\odot = 8.5$~kpc (the distance of the Sun from the Galactic centre).

The contribution of synchrotron radiation, which dominates at low photon energies, can be estimated as \citep[see e.g.][]{Sarazin99}
\be
l_{\rm{syn}}(\gamma, \nu) = \frac{\sqrt{3}\,q_{\rm e}^3B}{m_{\rm e} c^2} \
R[\chi(\gamma)],
\ee
where $m_{\rm e}$ and $q_{\rm e}$ denote the electron mass and charge, respectively, $B$ is the intensity of the
magnetic field, and the function $R(\chi)$ is defined as \citep[e.g.][]{Ghisellini+88} 
\be
% R(\chi) \equiv 2\chi^2
% \left[
%  K_{4/3}(\chi) K_{1/3}(\chi)
%  - \frac{3}{5} \chi
%  \left\{ K^2_{4/3}(\chi) - K^2_{1/3}(\chi) \right\}
% \right]\!\!\!.
R(\chi) \equiv 2\chi^2
\left[
 K_{\frac{4}{3}}(\chi) K_{\frac{1}{3}}(\chi)
 - \frac{3}{5} \chi
 \left\{ K^2_{\frac{4}{3}}(\chi) - K^2_{\frac{1}{3}}(\chi) \right\}
\right].
\ee
In this expression, $K$ refers to the modified Bessel function, and the normalized frequency
\be
\chi \equiv \frac{\nu}{3\gamma^2\nu_c}
\ee
is expressed in terms of the cyclotron frequency
\be
\nu_c \equiv \frac{q_eB}{2\pi m_{\rm e}c}.
\ee

At high photon energies (i.e. gamma rays), we consider the contributions of inverse Compton scattering and final-state radiation.
For ICS \citep[see e.g.][]{BlumenthalGould70}
\be
l_{\rm{ICS}}(\gamma, \nu)
=
\frac{ 3\sigma_{\rm T}c\, h\nu}{ 4\gamma^2 }
\int_0^\infty \frac{ n(\nu_0) }{ \nu_0 }\ \mathscr{F}(\Gamma,q)\ \dd\nu_0,
\ee
where $n(\nu_0)$ is the photon number density of the interstellar radiation field
being scattered, which we represented as the sum of three grey bodies
\bea
\nonumber n(\nu_0)
\!\!\!\!&=&\!\!\!\!
\frac{ 8\pi \nu_0^2 }{ c^3 }
\Big[~ \frac{ 1 }{ \exp( h\nu_0/kT_{\rm CMB} ) - 1 }\\
& & \!\!\!\!
+\ \frac{ \mathscr{N}_{\rm SL} }{ \exp( h\nu_0/kT_{\rm SL} ) - 1 } 
+  \frac{ \mathscr{N}_{\rm IR} }{ \exp( h\nu_0/kT_{\rm IR} ) - 1 } ~\Big]
\label{eqISRF}
\eea
and
\bea
\nonumber \mathscr{F}(\Gamma,q)
\!\!\!\!&\equiv&\!\!\!\!
\left[\ 2q \ln q + (1+2q)(1-q) + \frac{(\Gamma q)^2\, (1-q)}{2\, (1+\Gamma q)}\ \right]\\
&& \!\!\!\!
\times\ \Theta(q-\frac{1}{4\gamma^2})\ \Theta(1-q)
\eea
with $\Gamma \equiv \frac{4\gamma h\nu_0}{m_{\rm e} c^2}$, $\Gamma q \equiv \frac{h\nu}{\gamma m_e c^2 - h\nu}$, $\sigma_{\rm{T}}$ the Thomson cross-section, $h$ the Planck constant, $k$ the Boltzmann constant, and the product of Heaviside functions ensures that only kinemattically-allowed collisions $\frac{1}{4\gamma^2} \le q \le 1$ are taken into account.

For FSR, the emission coefficient for photons of frequency $\nu$ is given by 
\be
j_{\nu}(r,\nu) = \frac{ h \nu }{ 4\pi }
\left[ \frac{ \rho_{\rm dm}(r) }{ m_{\rm dm} } \right]^2 
\deriv{ \langle \sigma v \rangle_{\rm FSR} }{ \nu }.
\ee
with each annihilation event yielding a photon spectrum given by
\be
\deriv{ \langle \sigma v \rangle_{\rm FSR} }{ \nu }
\!=\!
\langle \sigma v \rangle_{e^\pm}
\, \frac{ \alpha }{ \pi }
\, \frac{ \kappa^2 - 2\kappa + 2 }{ \nu }
\ln\! \left[\! \left( \frac{ 2 m_{\rm dm} }{ m_{\rm e} }\! \right)^{\!\!2} (1-\kappa)\! \right]
\ee
where $\alpha$ is the fine-structure constant and $\kappa = h \nu/m_{\rm{dm}} c^2$ \citep[see e.g.][]{Peskin}.

%--------------------------------------------------------------------------
 \subsection{Astrophysical parameters}
 \label{secPara}
%--------------------------------------------------------------------------

The emission coefficient associated to final-state radiation is fully specified by the initial energy and injection rate of the electron-positron pairs, related to the nature of the dark matter particle (mass and cross-section) and the parameters describing the density profile of the Galactic halo.
In contrast, the photon intensity from the synchrotron and ICS emission also depends on the astrophysical parameters that determine the propagation and energy losses of the relativistic particles.
We will first define a canonical model based on observations of the Milky Way and then investigate the effect of each individual component by varying the values of the adopted parameters.
In all cases, we calculate the electron-positron spectrum as described in
expression~(\ref{eqSpectrum}), and then estimate the photon intensity according
to expression~(\ref{eqIntensity}).

Our canonical model assumes a dark matter density profile with $\alpha=1$
\citep{Navarro+97}, $r_{\rm s}=17$~kpc and $\rho_{\rm s}c^2=0.35$~GeV~cm$^{-3}$,
consistent with dynamical models of the Milky Way
\citep[e.g.][]{DehnenBinney98, Klypin+02}. The virial mass of the
Galaxy is thus $10^{12}$~M$_\odot$, and the local dark matter density is
$\rho_{\rm dm}(r_\odot)\,c^2 = 0.3$~GeV~cm$^{-3}$. The ISM is mainly
composed of neutral hydrogen atoms ($X_{\rm ion} = 0$) with number density
$\rho_{\rm g}/m_{\rm p} \sim 1$~cm$^{-3}$
\citep{DehnenBinney98,Ferriere01,Robin+03}, and it is permeated by a
tangled magnetic field whose intensity is $B \sim 6~\mu$G
throughout the Galaxy \citep{Ferriere01,Beck01,AscasibarDiaz10}.

%__________________________________
\begin{table}
\begin{center}
\begin{tabular}{lcc}
Model    		&{$K_0$~[kpc$^2$~s$^{-1}$]} 	& $\delta$\\ \hline
{M1}             	&{$7.42\times10^{-17}$} 	&  {$0.46$} \\
{MED (canonical)} 	&{$1.76\times10^{-18}$} 	&  {$0.70$} \\
{M2}             	&{$2.92\times10^{-18}$} 	&  { $0.55$} \\
\end{tabular}
\end{center}
\caption{Three different models of the diffusion coefficient, following the parameterization $K(\gamma) = K_0 \gamma^\delta$. The model MED has been proposed by \citet{Donato+04}, and models M1 and M2 are adoped from \citet{Delahaye+08}.}
\label{tabDiffusion}
\end{table}
%__________________________________

%__________________________________
\begin{table}
\begin{center}
\begin{tabular}{lcc}
% Model    & $\mathscr{N}_{\rm SL}$ & $\mathscr{N}_{\rm IR}$ & $\mathscr{N}_{\rm CMB}$ \\
%           & $T_{\rm SL} = 3481$~K & $T_{\rm IR} = 40.6$~K & $T_{\rm CMB} = 2.726$~K \\\hline
% ISRF (I)  & $1.7 \times 10^{-11}$ & $7.0 \times 10^{-5}$ & 1 \\ 
% Canonical & $2.7 \times 10^{-12}$ & $7.0 \times 10^{-5}$ & 1 \\ 
% ISRF (II) & $8.9 \times 10^{-13}$ & $1.3 \times 10^{-5}$ & 1 \\
Model    & $\mathscr{N}_{\rm SL}$ & $\mathscr{N}_{\rm IR}$ \\
          & $T_{\rm SL} = 3481$~K & $T_{\rm IR} = 40.6$~K \\\hline
{ISRF (I)}  & {$2.7 \times 10^{-12}$} & {$7.0 \times 10^{-5}$} \\ 
{Canonical} & {$1.7 \times 10^{-11}$} & {$7.0 \times 10^{-5}$} \\
ISRF (II) & $8.9 \times 10^{-13}$ & $1.3 \times 10^{-5}$ \\
\end{tabular}
\end{center}
\caption{Normalization of the grey-body models describing the interstellar radiation field, adopted from \citet{CirelliPanci09}.
In our canonical model, we use the values appropriate for the Galactic centre in order to compute the ICS and synchrotron emission. For the electron-positron spectrum at the Solar neighbourhood, we use ISRF(I).
}
\label{tabParameters}
\end{table}
%__________________________________

Apart from the canonical model, we consider the effect that the magnetic field, the diffusion coefficient, and the interstellar radiation field have on the synchrotron and ICS emission.
The intensity of the magnetic field $B$ is varied from 1 to 100 $\mu$G.
For the diffusion coefficient (see equation~\ref{eqDiffusion}), we consider the three models discussed by \citet{Donato+04} and \citet{Delahaye+08}, summarized in Table~\ref{tabDiffusion}.
We will also use three different models of the ISRF \citep[adopted from][]{CirelliPanci09} where the photon intensity is represented by three grey-body components (see equations~\ref{eqUrad} and~\ref{eqISRF}).
The normalizations and effective temperatures of the light emitted by the Galactic stars and dust are quoted in Table~\ref{tabParameters}.

%__________________________________
\begin{table}
\begin{center}
\begin{tabular}{cccr}
      & $\alpha$ & $\rho_{\rm s} c^2$~[GeV~cm$^{-3}$] & $r_{\rm s}$~[kpc]\\ \hline
      &0.00 & 2.346 & 8.64 \\
      &0.20 & 1.737 & 9.56 \\
      &0.50 & 1.042 & 11.41\\
      &0.70 & 0.702 & 13.08\\
      &1.00 & 0.349 & 16.67\\
      &1.20 & 0.197 & 20.33\\
      &1.25 & 0.169 & 21.49\\ 
      &1.50 & 0.066 & 29.81\\
      &1.70 & 0.025 & 42.57\\
      &1.90 & 0.006 & 70.30\\
{Einasto}& {0.17}& {0.060}& {20.00}\\
\end{tabular}
\end{center}
\caption{Characteristic density and radius of the dark matter density
profile~(\ref{eqDMProfile1}) as a function of its asymptotic logarithmic inner
slope $\alpha$.}
\label{tabRhosRs}
\end{table}
%__________________________________

Most importantly, we also investigate the effect of the inner slope of dark matter density profile on the production rate of electron-positron pairs.
We vary the inner logarithmic slope $\alpha$ from 0.5 to 1.5. When varying
$\alpha$ we also modify the
characteristic density and radius in expression~(\ref{eqDMProfile1}) so that the
dark matter density at the solar radius is equal to 0.3~GeV~cm$^{-3}$ and the
virial mass of the Galaxy is $10^{12}$~M$_\odot$. The appropriate values of
$\rho_{\rm s}$ and $r_{\rm s}$ are quoted in Table~\ref{tabRhosRs} for several
values of the inner logarithmic slope $\alpha$.
In addition, we also consider the so-called Einasto profile
\be
\rho(r) = \rho_{\rm s}\, \exp\left\{\frac{-2}{\alpha}\left[\left(\frac{r}{r_{\rm s}}\right)^\alpha-1\right]\right\},
\label{eqEinasto}
\ee
where $\alpha=0.17$.

%--------------------------------------------------------------------------
 \section{Observational data}
 \label{secObservation}
%--------------------------------------------------------------------------

In order to constrain the production of relativistic electrons and positrons in
the Milky Way, we consider observations of the whole sky at very different
wavelengths.
More precisely, the Haslam map in the radio band, the 5 WMAP channels at
microwave wavelengths, and 12 energy bins of the Fermi LAT observations in the
gamma-ray regime.
The Haslam and WMAP maps are dominated by synchrotron emission, whereas Fermi
traces ICS and FSR.

The Haslam 408 MHz radio continuum all-sky map \citep{Haslam+81,
Haslam+82} combines data from four different surveys.
The data were obtained from the archives of the
NCSA ADIL in equatorial 1950 coordinates, and they were subsequently processed
further
in the Fourier domain to mitigate baseline striping and strong point sources.
For the WMAP data, we take the full-resolution coadded temperature maps for each
of the 5 frequency bands (23, 33, 41, 61, and 94 GHz) corresponding to the
7-year observations \citep{Jarosik+11}.
The Fermi gamma-ray maps were computed by \citet{Dobler+10} from all ``Class 3''
(diffuse) photon events in the first-year data release.
We use the 12 logarithmically-spaced frequency bands, from 0.3
to 300 GeV, of the smoothed maps without point source
subtraction.

Since we are interested in a spherically-symmetric component, we may follow a
simple, conservative procedure in order to mask the emission from the Galactic
disk and individual point sources without relying on any particular foreground
model.
For each frequency, we compute the average intensity $I(\theta)$ in 180 bins as
a function of the angular separation $\theta$ from the Galactic centre.
We also estimate the standard deviation $\sigma(\theta)$ within each bin, as
well as the average
standard deviation
\be
\sigma_{\rm ave} = \frac{ \sum_{i=1}^{n}\sigma(\theta_i) }{ n },
\ee
where $n=180$ is the total number of the bins.
We then start an iterative procedure, where all pixels more than $3\,\sigma_{\rm{ave}}$ away from $I(\theta)$ are discarded until convergence is achieved.

This method seems to correctly identify and remove the most obvious structures
in all but the two highest-energy Fermi bands, where the photon statistics is so
poor that it is extremely difficult to distinguish diffuse emission from
individual point sources.
For these two bands, we opted to use the original average intensity
$I_0(\theta)$ without applying any mask.
Raw intensity maps, masked residual maps, i.e. $I-I(\theta)$, and the average
intensity $I(\theta)$ for each wavelength are shown in
Figures~\ref{figHaslam&WMAPMap}, \ref{figFermiMap1}, and \ref{figFermiMap2} of
Appendix~\ref{secTables}.
Numeric values of $I_0(\theta)$, $I(\theta)$, and $\sigma(\theta)$ are quoted in
Tables~\ref{tabHaslam&WMAPIntensity}, \ref{tabWMAP&FermiIntensity},
\ref{tabFermiIntensity1}, and
\ref{tabFermiIntensity2}.

Besides these observational data, we also consider the energy spectra of cosmic-ray electrons and positrons in the solar neighbourhood; in particular, we use the combined electron+positron spectrum measured by the Fermi \citep{Ackermann+10} and HESS \citep{Aharonian+08} collaborations, as well as the positron-only spectrum determined from Fermi \citep{Ackermann+12} and PAMELA \citep{Adriani+13} data.
The positron fraction has also been recently measured by the AMS-02 collaboration \citep{Aguilar+13}, and it is foreseen that electron, positron, and combined spectra will be available in the near future.
% For PAMELA, we combine the electron-only spectrum obtained by \citet{Adriani+11} with the positron fraction discussed in \citet{Adriani+10} in order to derive the positron spectrum.

%--------------------------------------------------------------------------
 \section{Constraints on the dark matter cross-section}
 \label{secConstraints}
%--------------------------------------------------------------------------

%__________________________________
\begin{figure}
\centering \includegraphics[width=8cm]{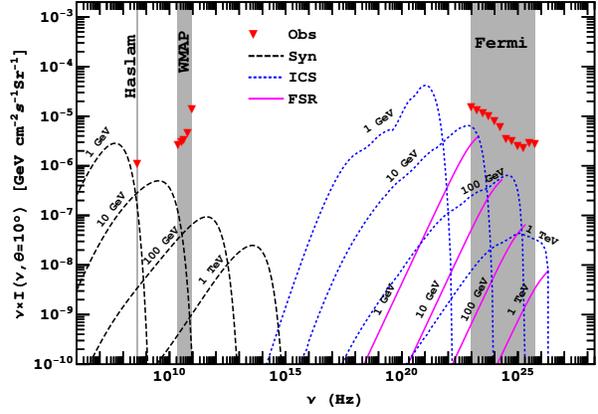}
\caption
{
Theoretical photon spectra of synchrotron radiation (dashed black lines), ICS (dotted blue lines) and FSR (solid magenta lines) for our canonical model with $\langle \sigma v \rangle_{e^\pm}=3 \times 10^{-26}\ \rm{cm^3\ s^{-1}}$ and different injection energies, evaluated at $10^\circ$ from the Galactic centre.
Grey bands illustrate the frequency ranges of Haslam, WMAP and Fermi.
The observational data at $\theta=10^\circ$ are plotted as red triangles.
}
\label{figSpectra}  
\end{figure}
%__________________________________

%__________________________________
\begin{figure*}

\includegraphics[width=.32\textwidth]{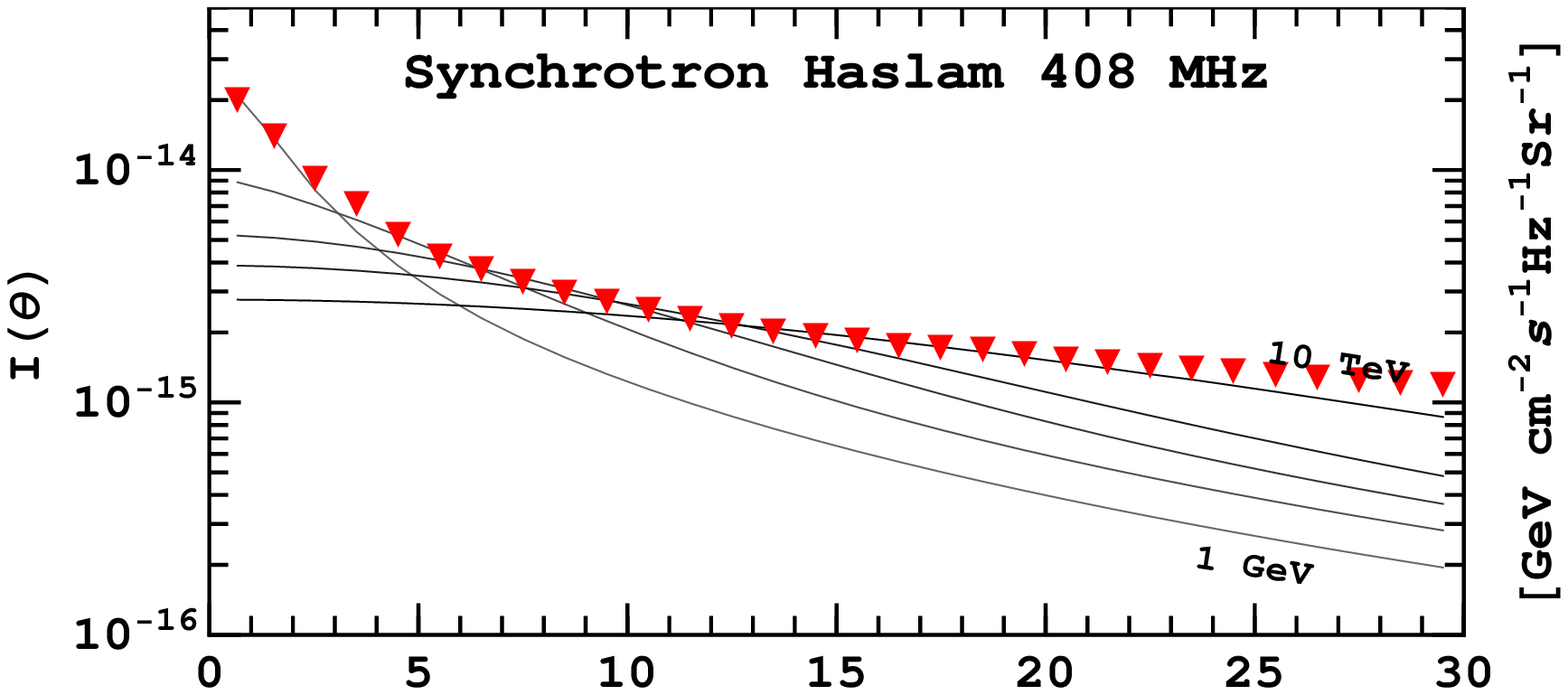}\hfill
\includegraphics[width=.32\textwidth]{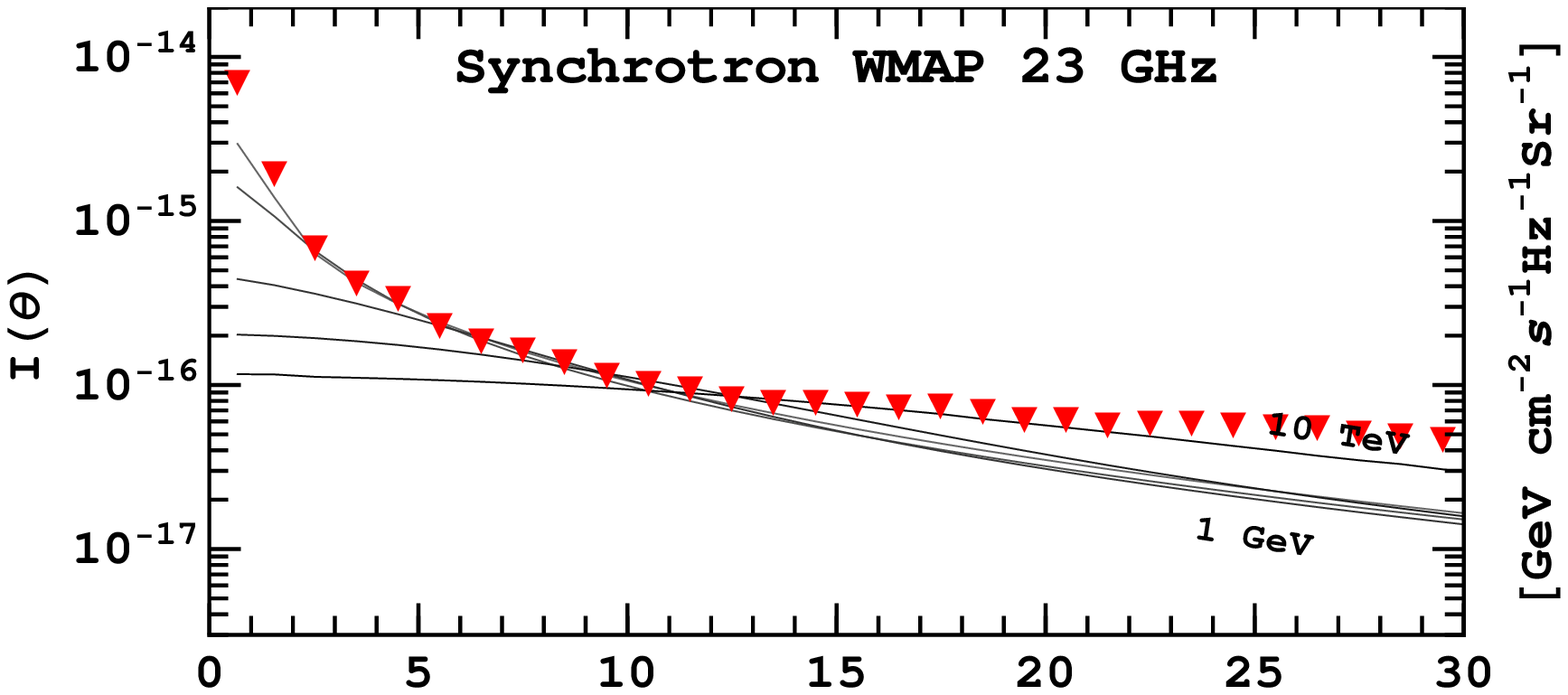}\hfill
\includegraphics[width=.32\textwidth]{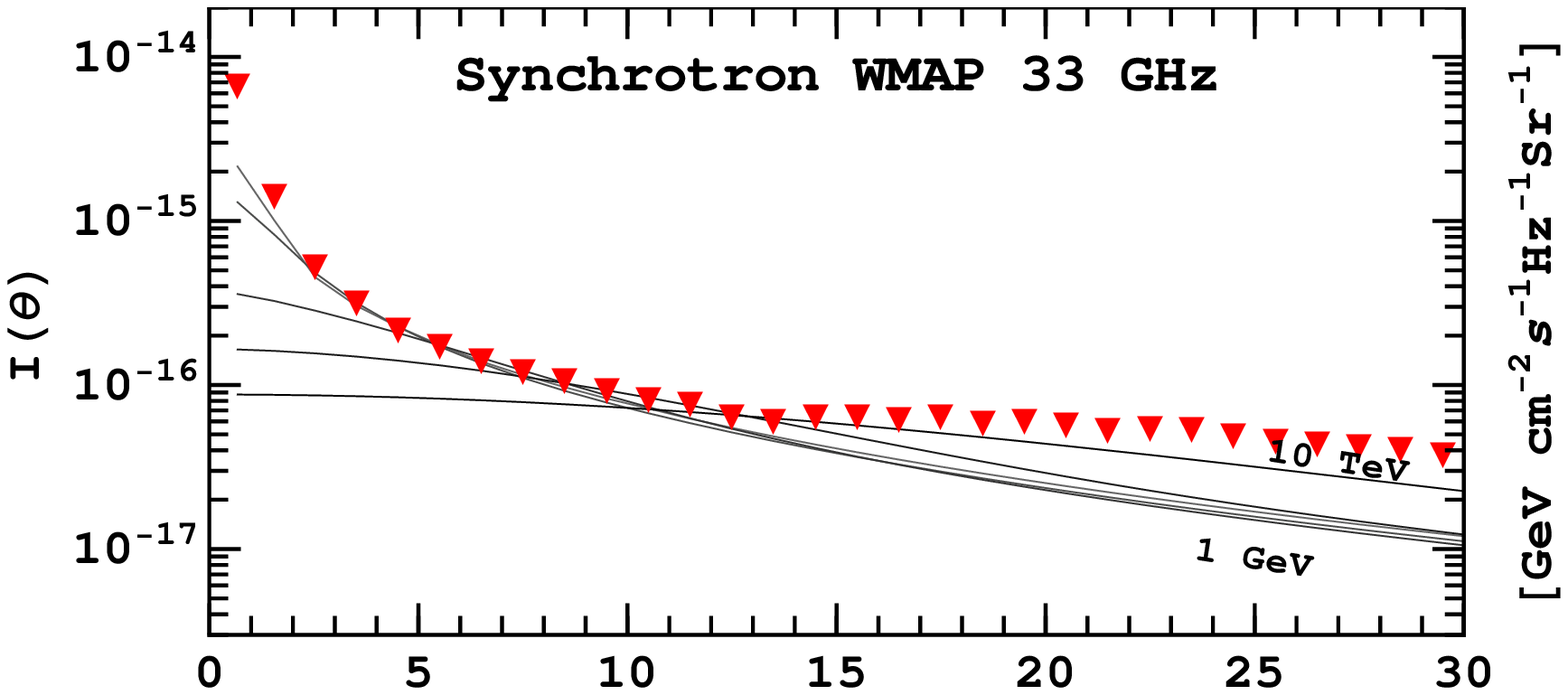}\hfill

\includegraphics[width=.32\textwidth]{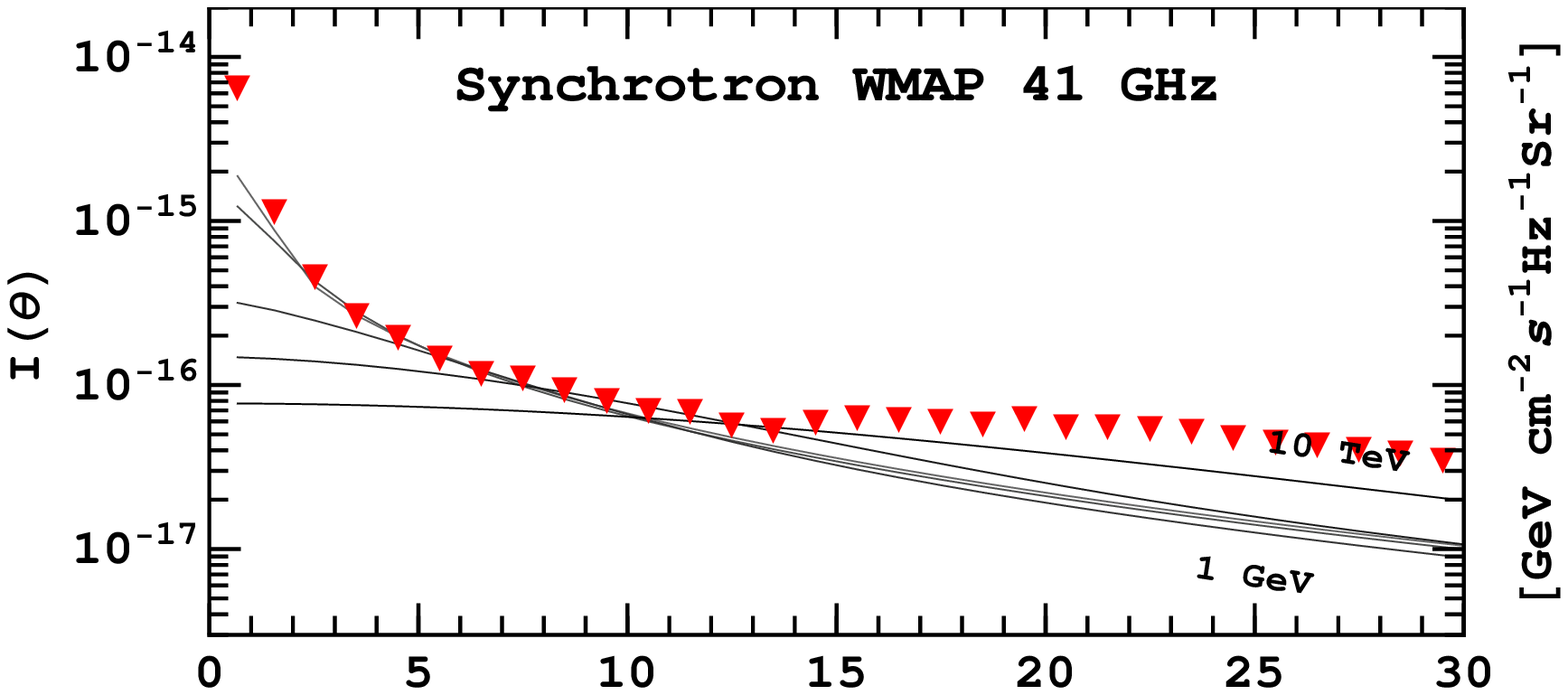}\hfill
\includegraphics[width=.32\textwidth]{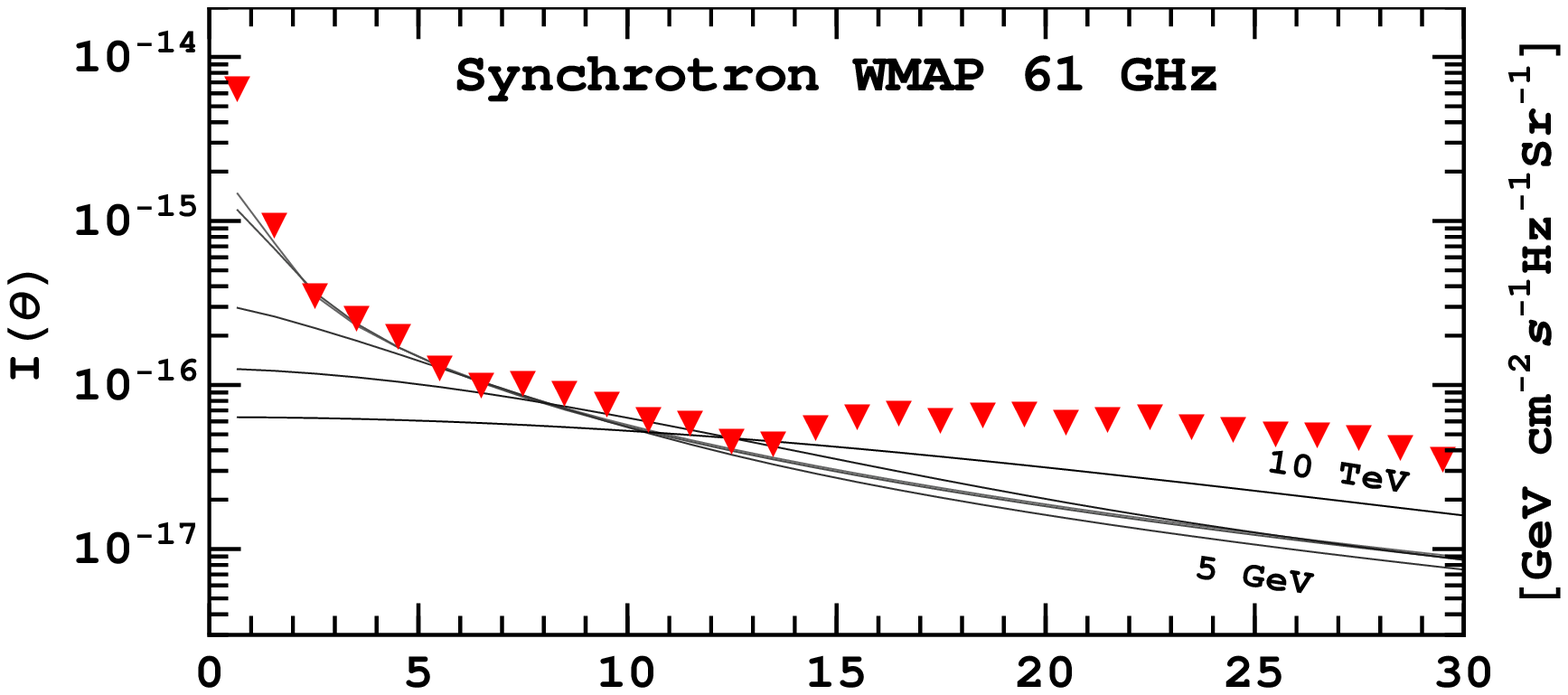}\hfill
\includegraphics[width=.32\textwidth]{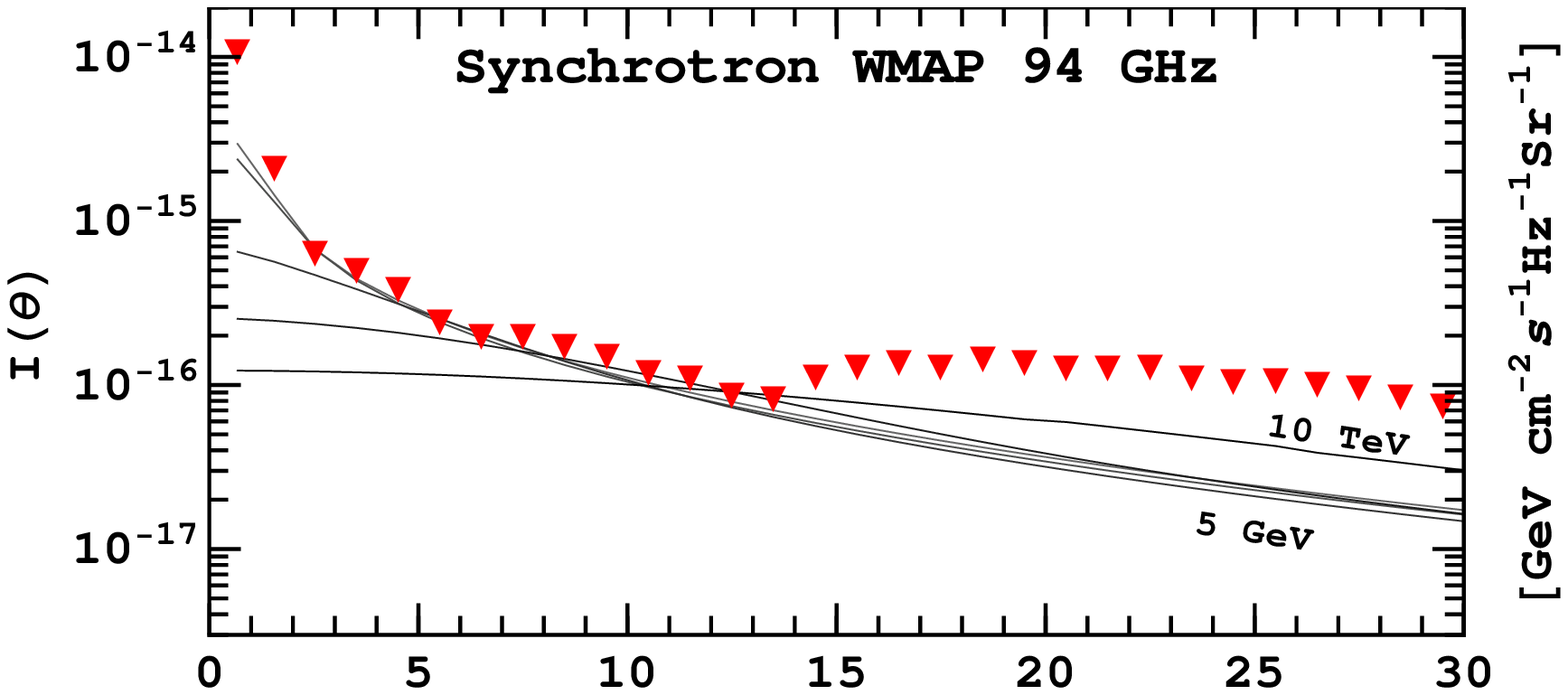}

\includegraphics[width=.32\textwidth]{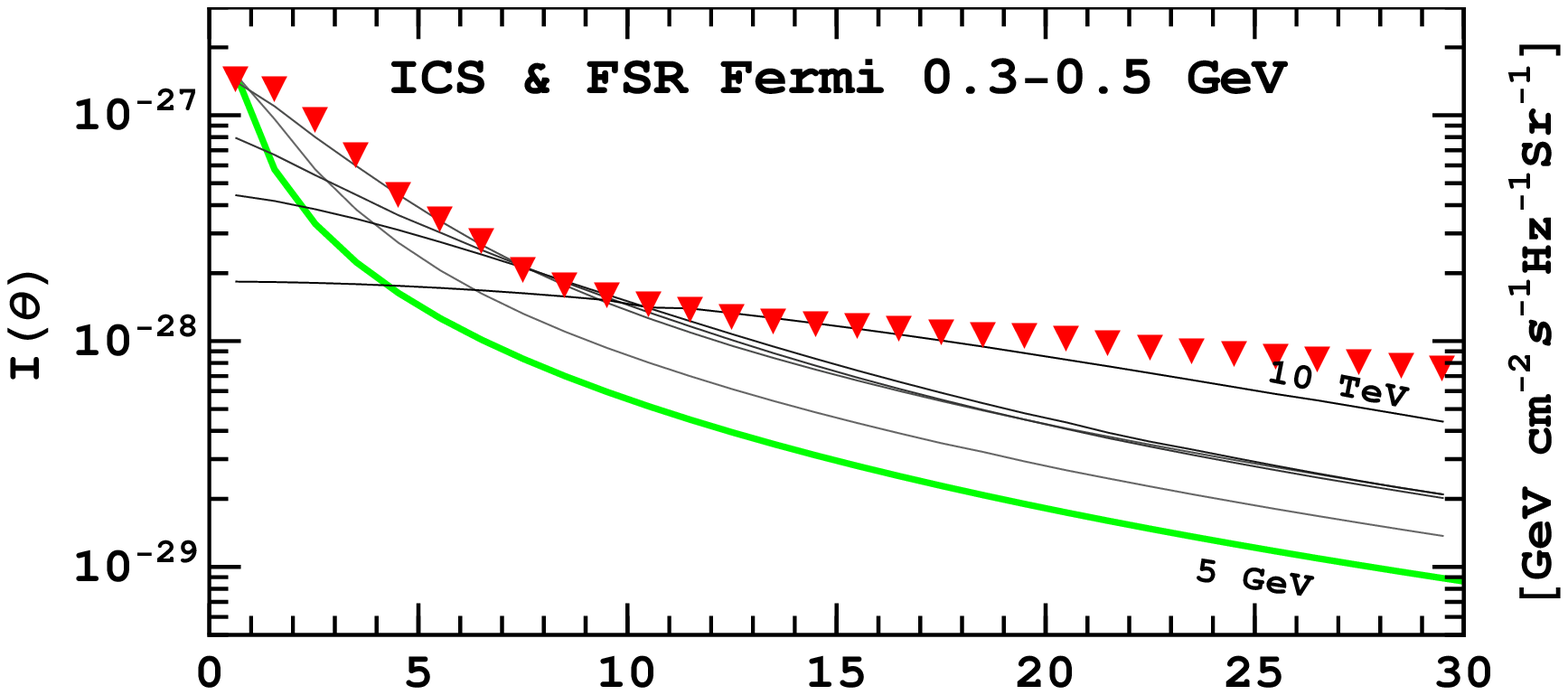}\hfill
\includegraphics[width=.32\textwidth]{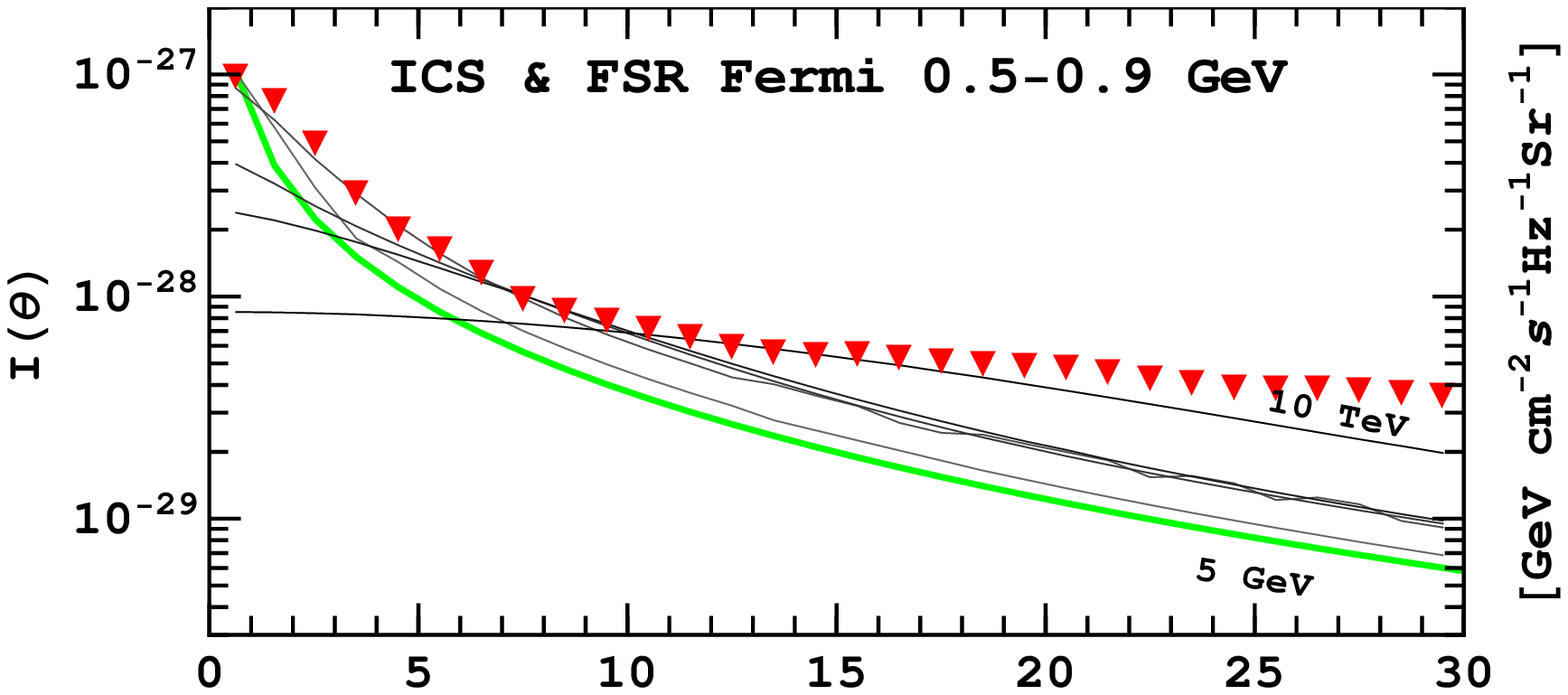}\hfill
\includegraphics[width=.32\textwidth]{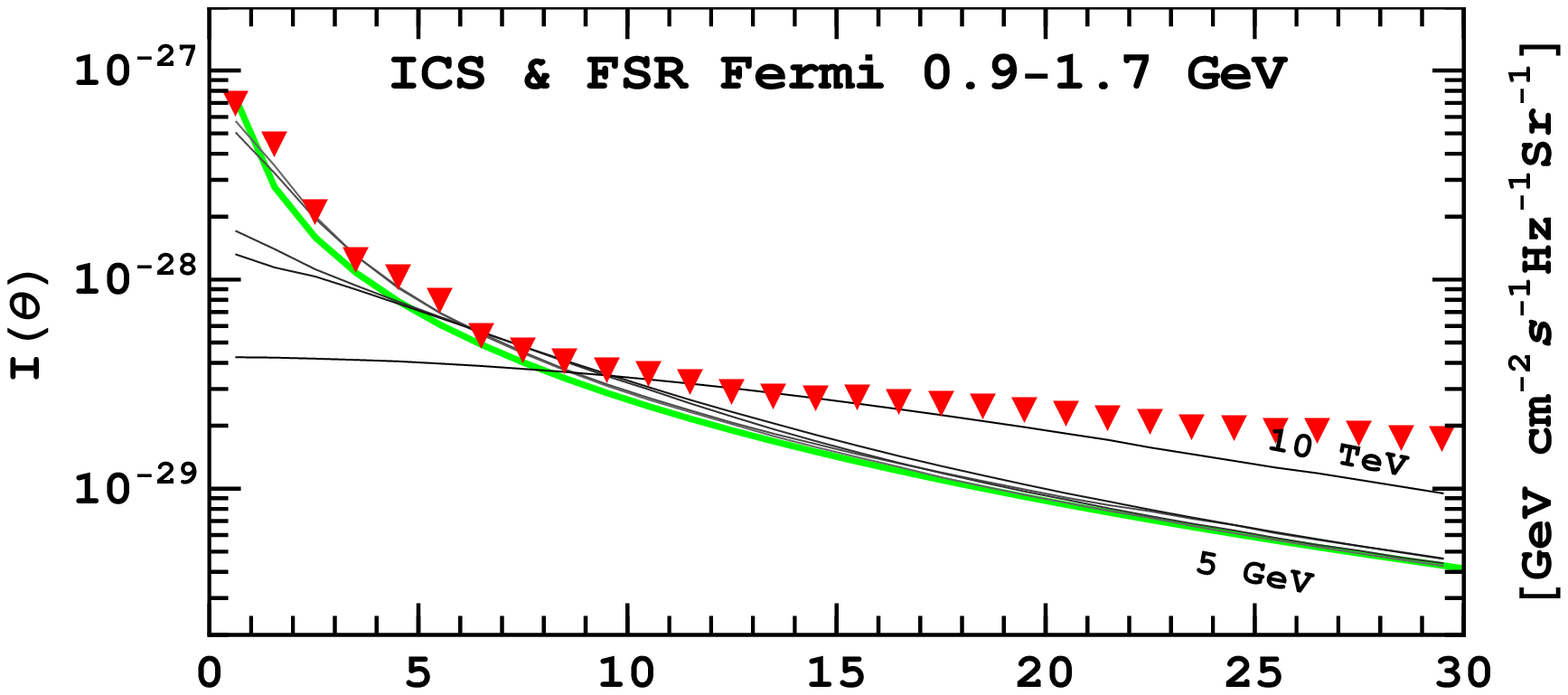}\hfill

\includegraphics[width=.32\textwidth]{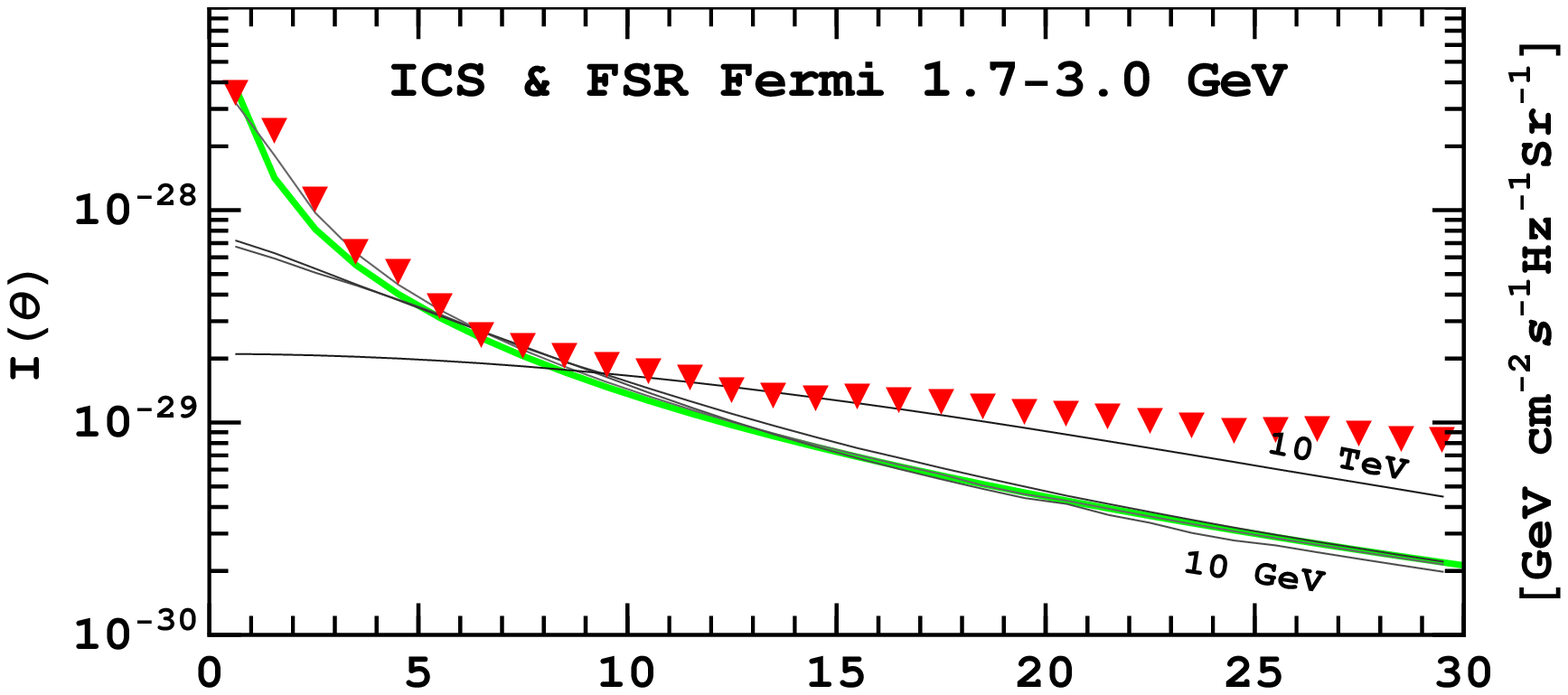}\hfill
\includegraphics[width=.32\textwidth]{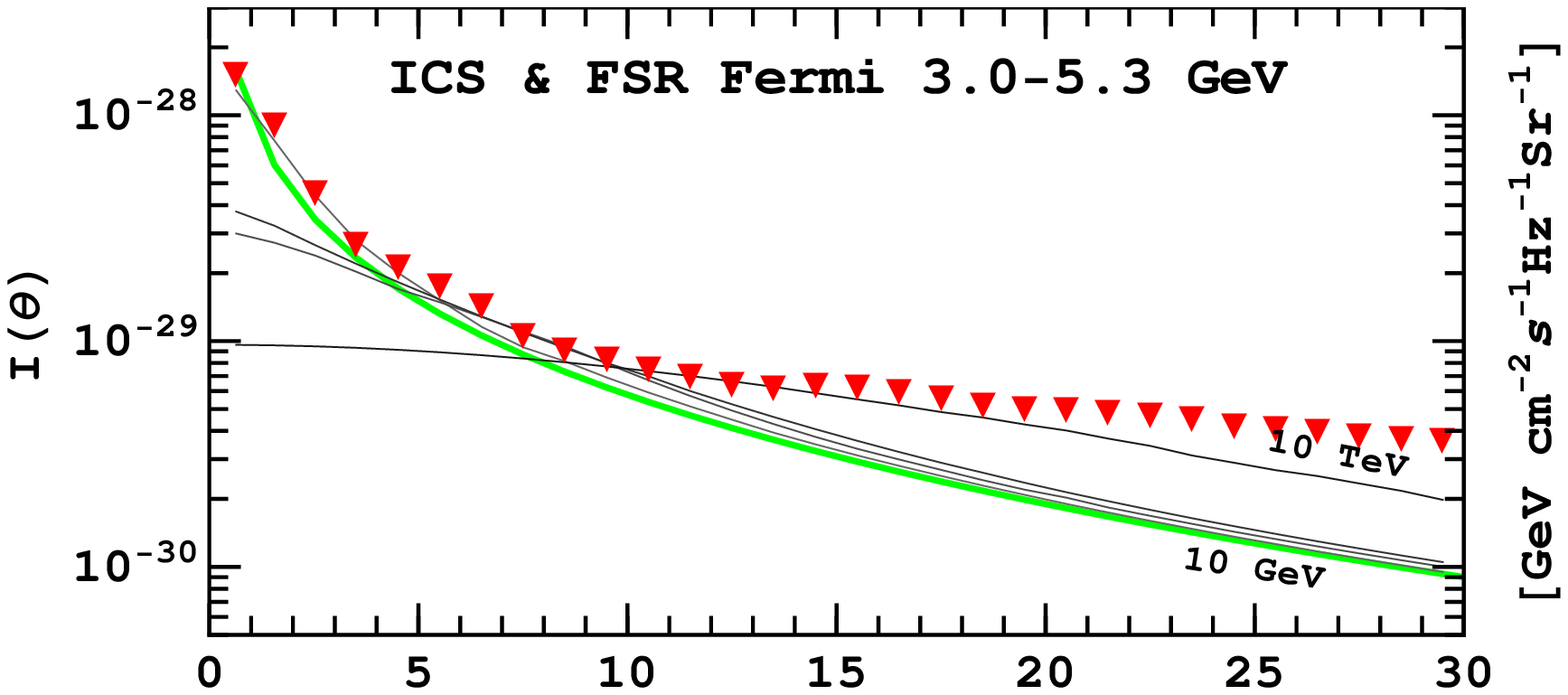}\hfill
\includegraphics[width=.32\textwidth]{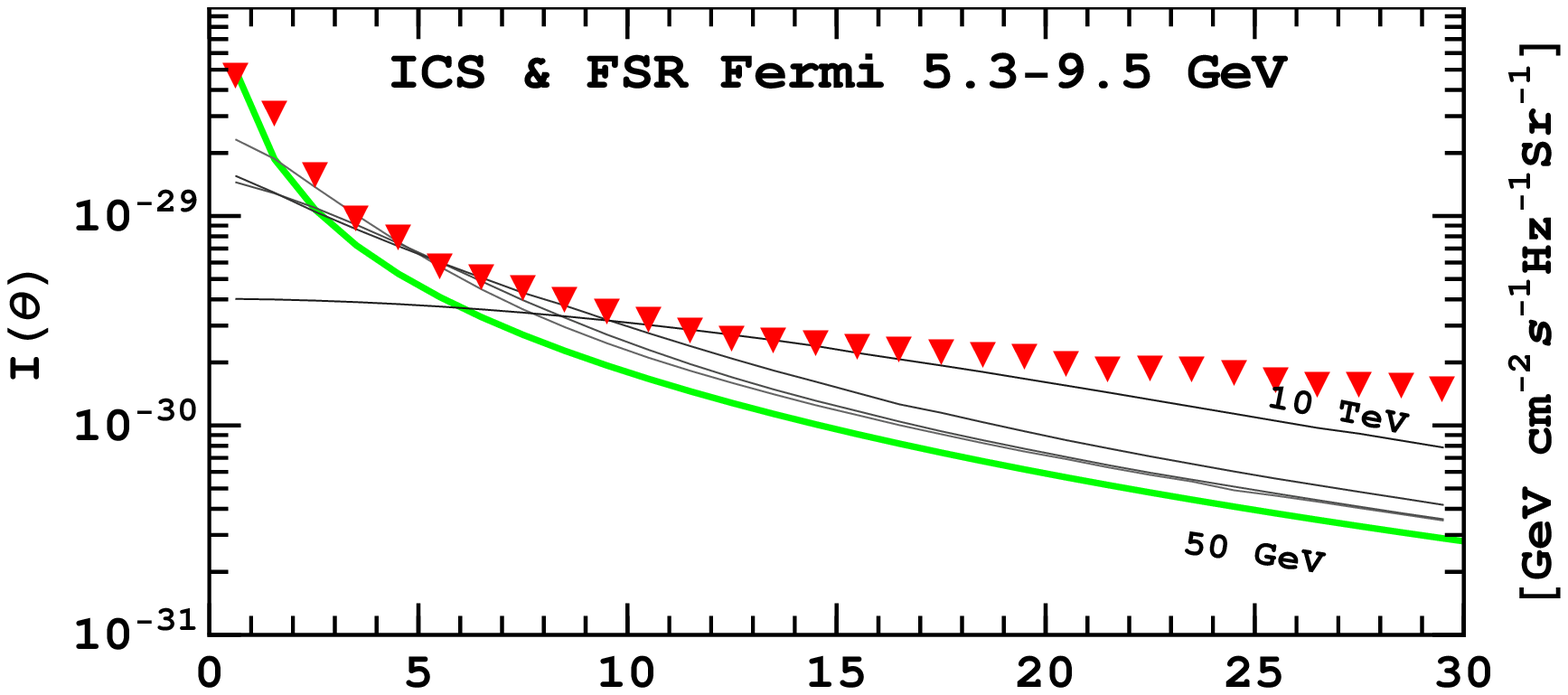}\hfill

\includegraphics[width=.32\textwidth]{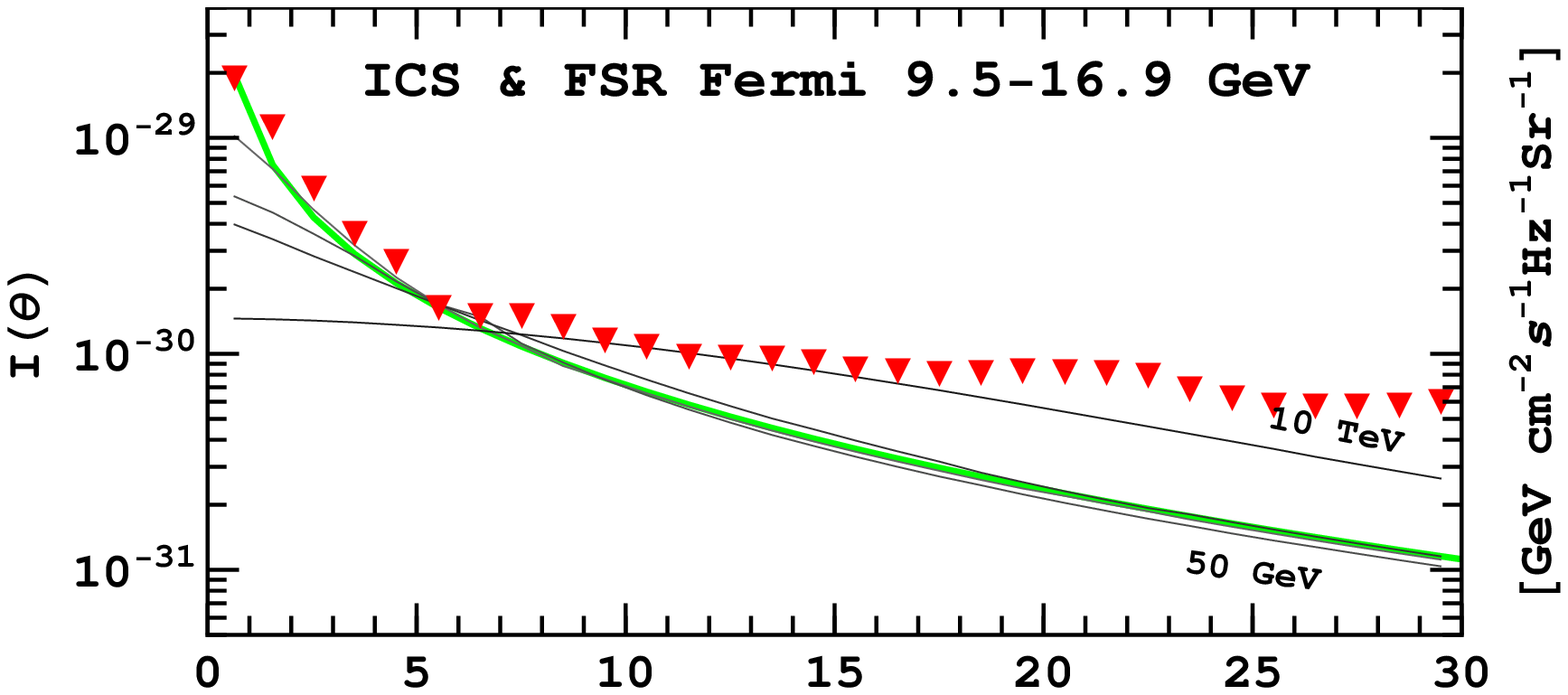}\hfill
\includegraphics[width=.32\textwidth]{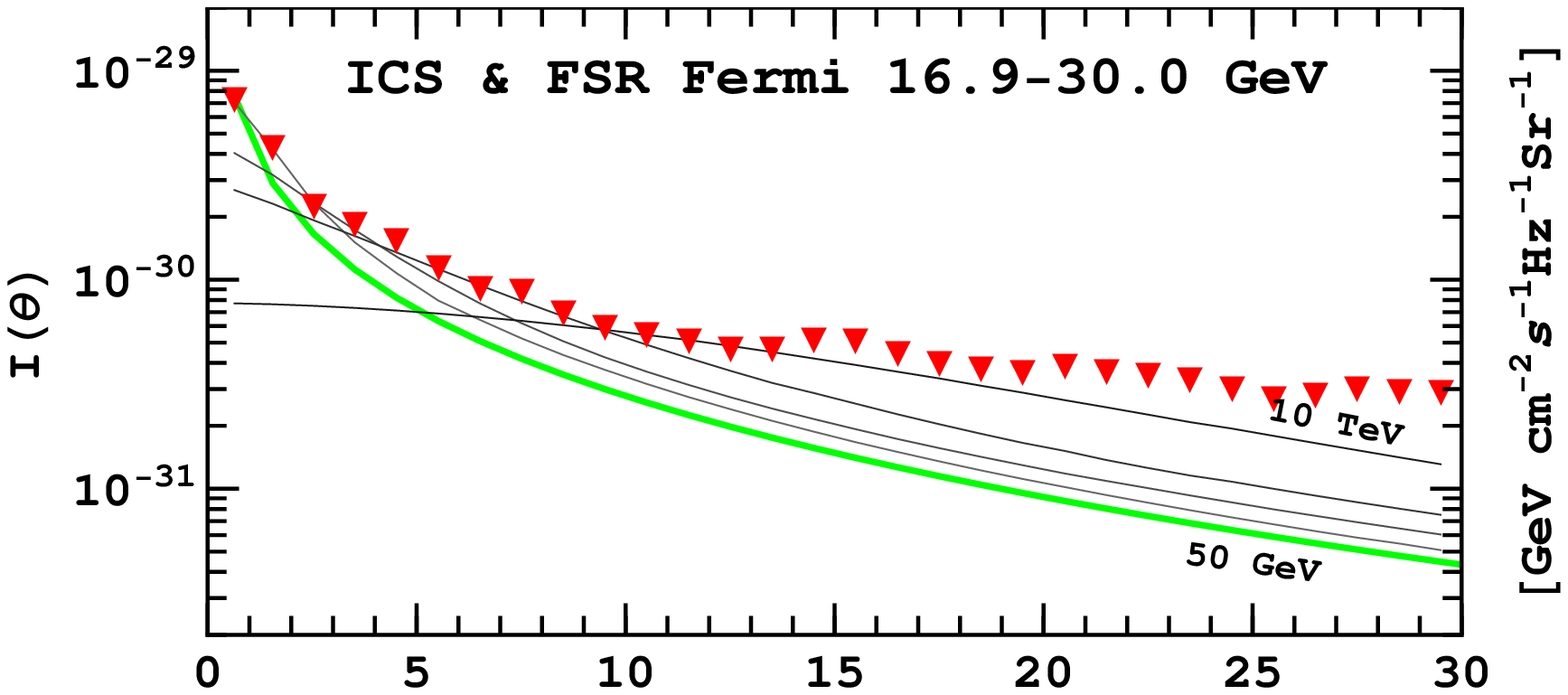}\hfill
\includegraphics[width=.32\textwidth]{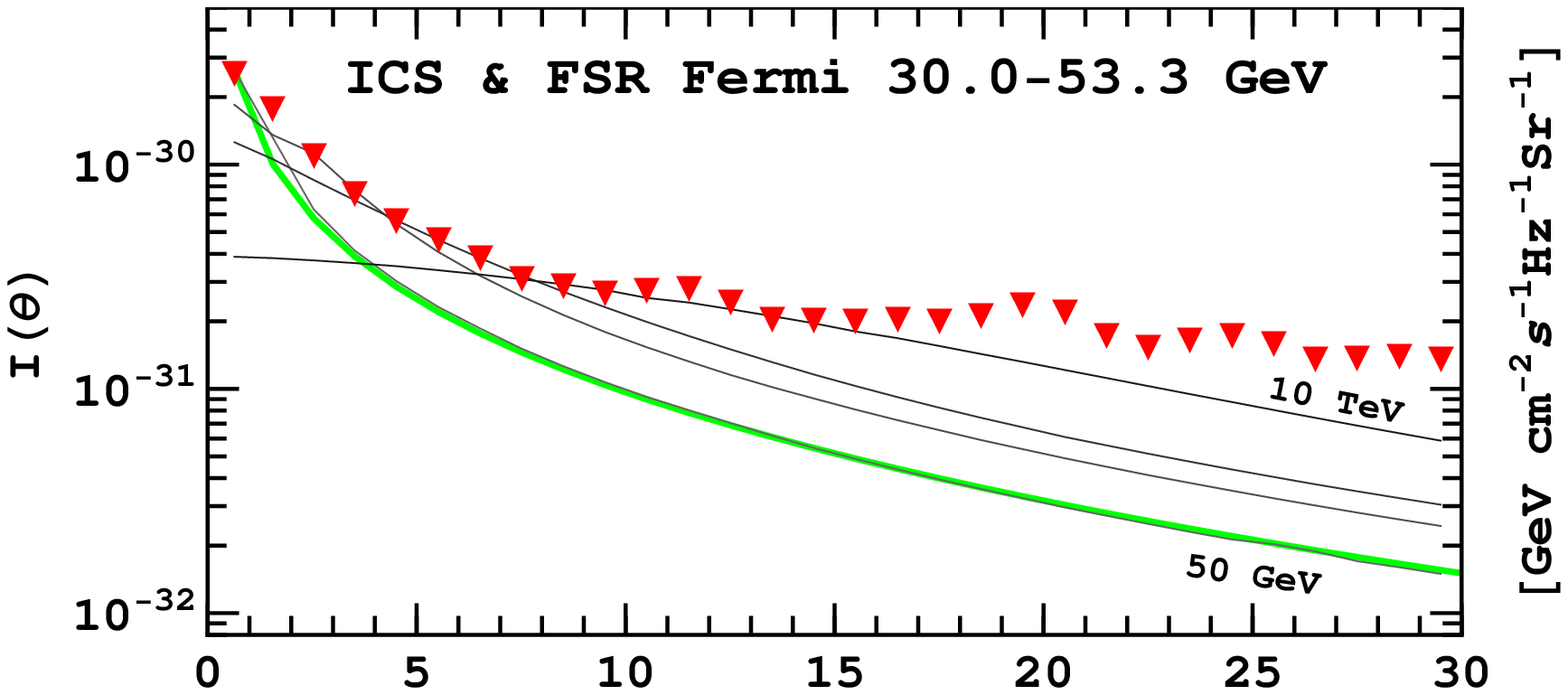}\hfill

\includegraphics[width=.32\textwidth]{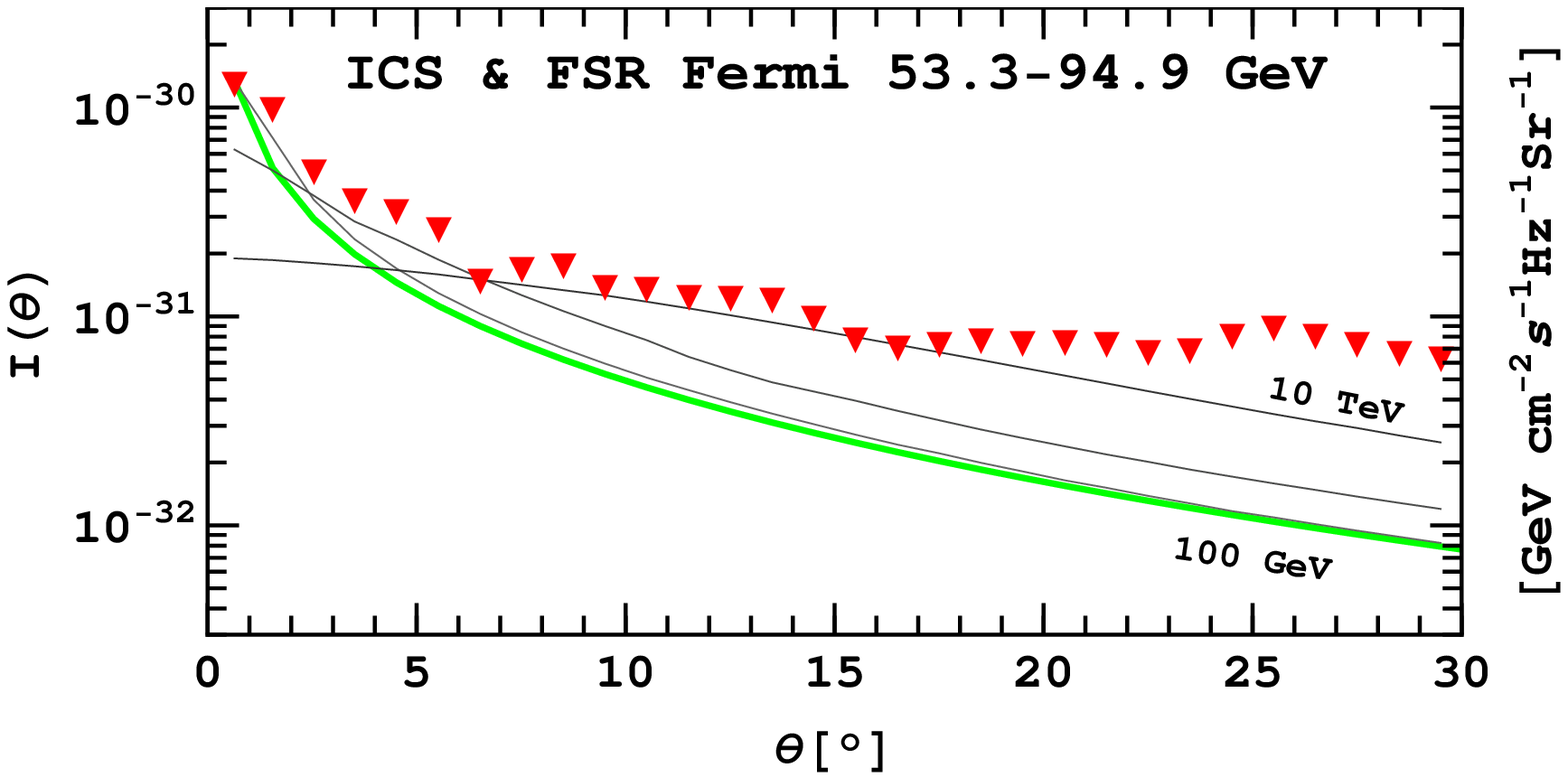}\hfill
\includegraphics[width=.32\textwidth]{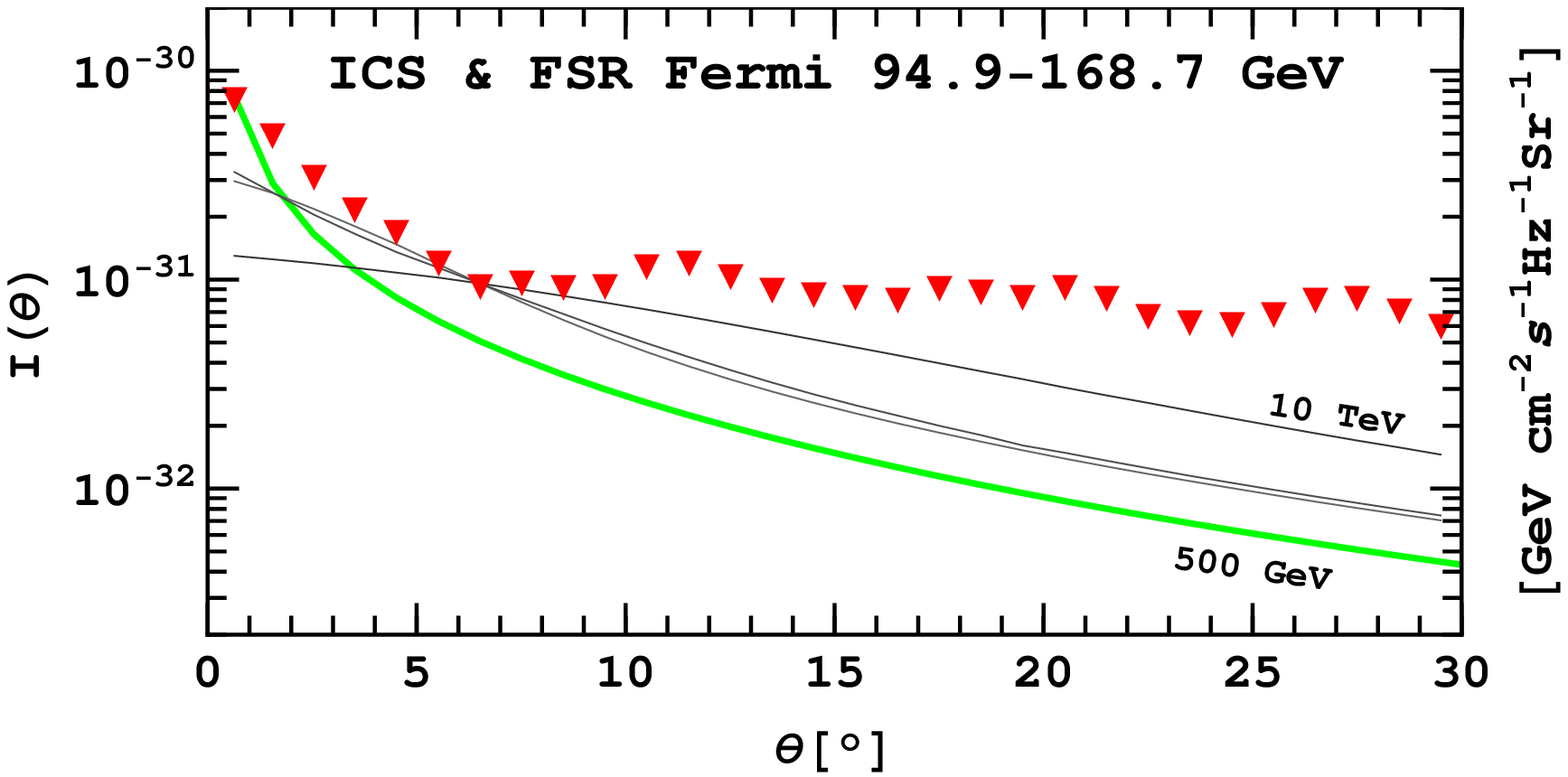}\hfill
\includegraphics[width=.32\textwidth]{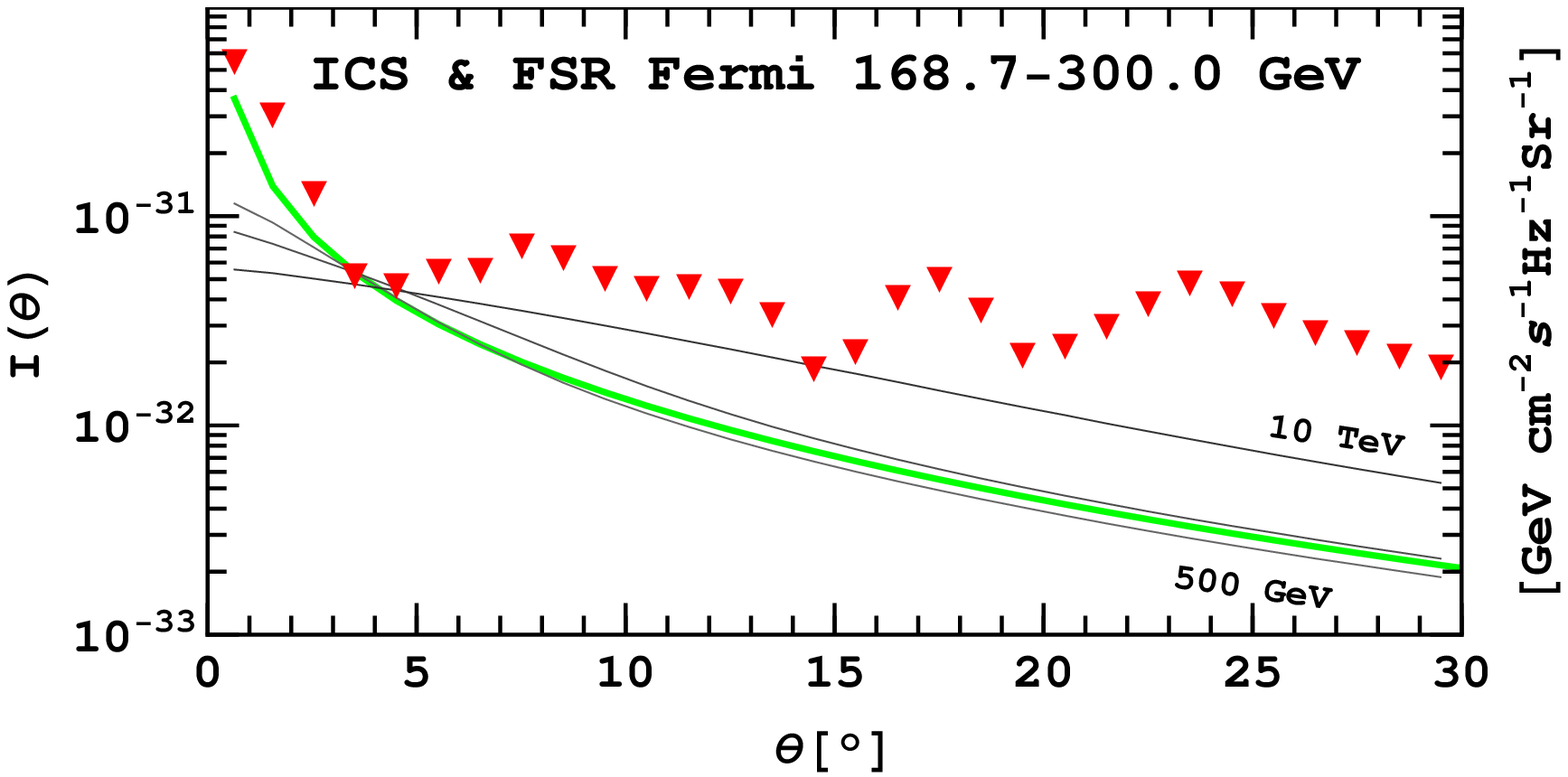}
\caption
{
Surface brightness profiles of synchrotron, ICS, and FSR as a function of the angular separation $\theta$ from the Galactic centre.
Red triangles correspond to the mean observational intensity after discarding the contribution of the Galactic disk and prominent point sources as discussed in Section~\ref{secObservation}.
Theoretical profiles are normalized to the maximum value of the annihilation cross-section (see Figure~\ref{figCanonical}) allowed by these data.
The angular separation that provides the tightest constraint -- i.e. the tangent point between models and observations -- is depicted in Figure~\ref{figConstraintTheta}.
For synchrotron and ICS emission, the intensities obtained for the canonical Milky Way model are expressed in grey to black lines, where a darker colour represents a higher value of the injection energy $E_0$.
The normalized intensity of FSR, shown as a green solid line, does not depend on $E_0$.
}
\label{figIntensity}
\end{figure*}
%__________________________________

Once the emission from the Galactic disc and the most prominent point sources is excluded, the remaining spherically-averaged component can be used to place upper limits on the cross-section for dark matter annihilation into electron-positron pairs.

First of all, model intensities are computed according to the scheme described in Section~\ref{secModel}.
We consider the injection energy (i.e. the mass of the dark matter particle) as a free parameter and investigate values of the initial Lorentz factor $\gamma_0$ between $2\times 10^{3}$ and $2\times 10^{7}$, corresponding to injection energies $E_0 = \gamma_0 m_{\rm e}c^2$ from 1~GeV to 10~TeV.
As an example, Figure~\ref{figSpectra} displays the results of our canonical Milky Way model for the synchrotron, inverse Compton, and final-state radiation contributions to the photon intensity at 10$^\circ$ from the Galactic centre, assuming a dark matter annihilation cross-section of $\langle \sigma v \rangle_{e^\pm} = 3 \times 10^{-26}\ \rm{cm^3\ s^{-1}}$.

%__________________________________
\begin{figure}
\centering \includegraphics[width=8cm]{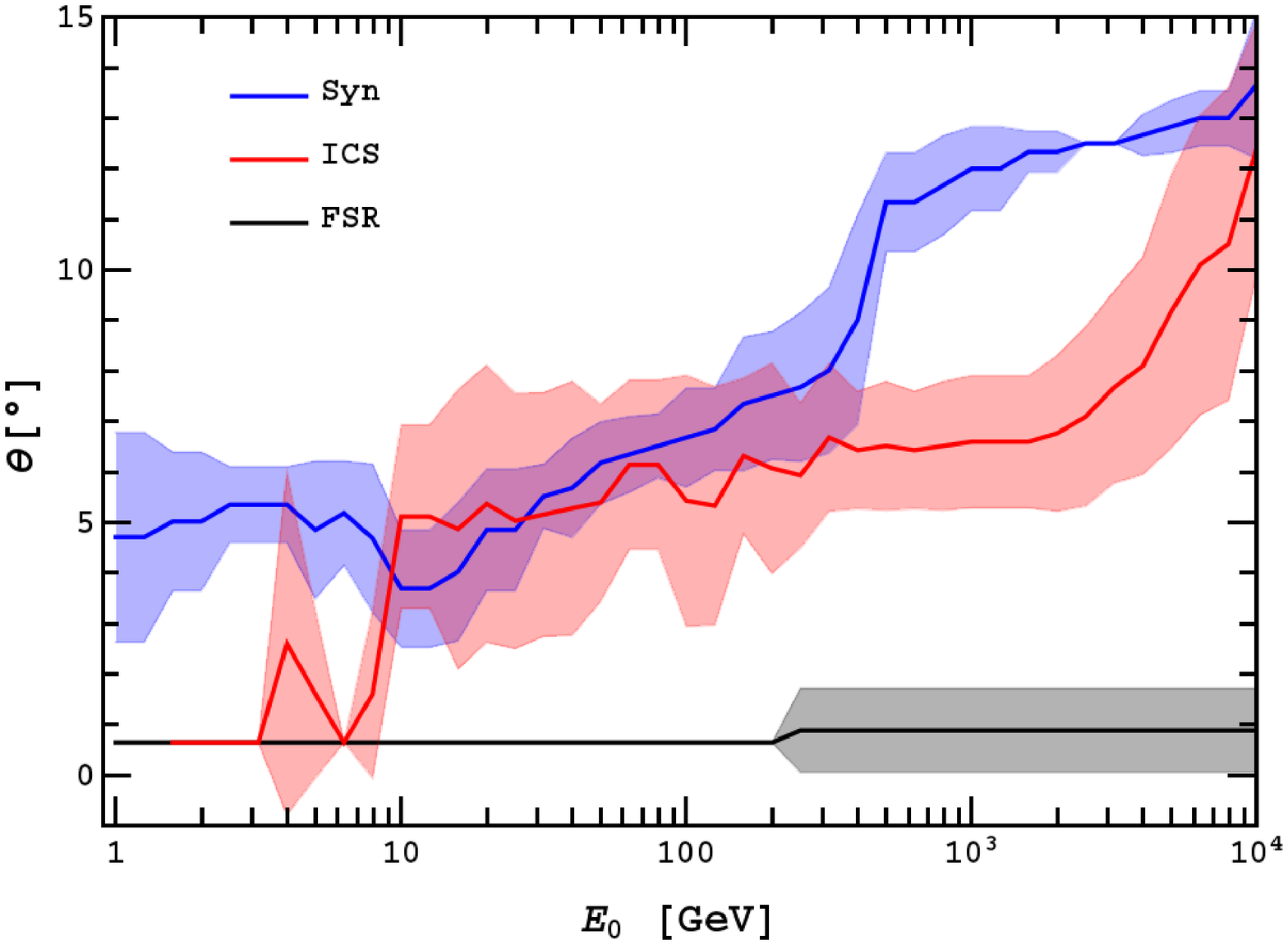}
\caption
{
Angular separation $\theta$ that provides the upper limits for synchrotron, ICS and FSR emission.
The optimal value is computed independently for each of the observed wavelengths.
Solid lines and shadowed regions show the average $<\theta> = 1/N_\lambda \sum_i \theta(\lambda_i)$ and standard deviation $<\theta^2>-<\theta>^2$ across different channels, respectively.
}
\label{figConstraintTheta}  
\end{figure}
%__________________________________

One can readily see that the Haslam radio map will be most sensitive to synchrotron emission by particles with an initial energy between 1 and 10~GeV, whereas WMAP data will cover the range $E_0\sim 10-100$~GeV.
On the other hand, the gamma rays observed by the Fermi LAT will constrain the maximum ICS and FSR emission allowed.
The final-state radiation is sharply peaked at the injection energy, and it traces values of $E_0$ between 1~GeV and 1~TeV.
The inverse Compton spectrum is broader; it features three distinct emission peaks, due to the scattering of CMB, starlight, and infrared photons, and it is best suited to probe injection energies above $\sim 10$~GeV.

Since the value of the annihilation cross-section only sets the normalization of the spectra, and it does not alter its shape, it is relatively easy to set an upper limit by imposing that the model intensities do not exceed the observed values (red triangles in Figure~\ref{figSpectra}) at \emph{any} angular separation $\theta$.
Not surprisingly, the tightest constraint will always be provided by a small value of $\theta$, i.e. close to the Galactic centre.
The dark matter density, and thus the injection rate, are higher there than anywhere else in the Galaxy.
However, the observed intensity also reaches a maximum at $\theta=0$, and particles may diffuse from their injection point, effectively smoothing the density cusp.
The predicted surface brightness profiles of synchrotron, ICS, and FSR emission, normalized according to such prescription, are plotted in Figure~\ref{figIntensity} together with the observational data, and the angle that sets the maximum normalization that would be compatible with the observations (i.e. the upper limit of $\langle \sigma v \rangle_{e^\pm}$) is plotted in Figure~\ref{figConstraintTheta}.

Final-state radiation is produced at the very moment of pair creation, and thus it directly traces the positron injection profile, which is, in turn, proportional to the square of the dark matter density.
Therefore, the intensity of the FSR emission does not depend on the injection energy of the particles or any astrophysical parameter other than the inner logarithmic slope $\alpha$ of the dark matter density profile.
For this reason, the normalized surface brightness profiles of FSR depicted in Figure~\ref{figIntensity} do \emph{not} depend on $E_0$.
In our canonical model (where $\alpha=1$), and even more so if $\alpha>1$, the tightest constraints on the final-state radiation come from the very centre of the Galaxy ($\theta<1^\circ$) in almost all cases, yielding a null standard deviation in Figure~\ref{figConstraintTheta} for most values of $E_0$.

For synchrotron and ICS emission, particle diffusion makes the intensity profile shallower, especially at high injection energies.
In general, one can say that photons of a given frequency trace electrons and positrons within a certain energy range.
If that range is close to $E_0$, these particles would have just been injected, and therefore the effects of particle propagation should be small, whereas, away from $E_0$, these electrons and positrons would have traveled a significant distance from the point of injection, and the surface brightness profile will become considerably shallower.

This trend is indeed evident in Figure~\ref{figIntensity}: surface brightness profiles become progressively shallower as one moves from $E_0=1$~GeV to 10~TeV, and the effect is more pronounced for those channels that trace low-energy particles, i.e. Haslam, WMAP, and the lowest-energy Fermi bands.
In the most extreme cases, diffusion keeps the electron-positron spectrum (and the
ensuing intensity) roughly constant within the innermost $10-20^\circ$.
For synchrotron emission, the tightest constraints on the annihilation cross-section come from $\theta \sim 5-13^\circ$, whereas for inverse Compton scattering the optimal angle increases from $1$ to $12^\circ$ (see Figure~\ref{figConstraintTheta}.

In addition to the photons arriving from the centre of the Milky Way, the dark matter annihilation cross-section $\langle \sigma v \rangle_{e^\pm}$ is also strongly constrained by the observed abundance of relativistic electrons and positrons in the solar neighbourhood.
In particular, we consider the recent measurements of the \emph{positron} spectrum by the Fermi collaboration \citep{Ackermann+12} and the PAMELA experiment \citep{Adriani+13}.
% The latter data have not been published as such in the literature, but they can be trivially derived from the quoted positron fraction \citep{Adriani+10} and electron spectrum \citep{Adriani+11}.
Since the positron fraction is of the order of 10 percent or less at the energies below $\sim 10$~GeV, the constraints from the positron-only spectrum will be much tighter than those derived from the combined electron+positron data.
For the sake of comparison, we also show these for PAMELA \citep{Adriani+11}, Fermi \citep{Ackermann+10}, and HESS \citep{Aharonian+08}.
Note that, in the latter case, the measurements are able to probe higher ($\sim$~TeV) energies, but it is not possible to discriminate between the electron and positron signatures.

Our constraints are derived by imposing that the predicted amount of electrons and/or positrons does not exceed the observed values for \emph{any} Lorentz factor $\gamma$.
Given the energy dependence of the observed spectrum, $\left[ \deriv{n}{E}
\right]_{\rm obs} \sim E^{-3}$, and the energy losses, $b(E)\sim E^2$, the most
restrictive constraint comes from the spectrum near the injection energy, where
propagation can be safely neglected and  $\left[ \deriv{n}{E} \right]_{\rm
model} \approx \frac{Q_0}{b} \propto E^{-2}$.
The maximum production rate allowed by the data can then be expressed as
\be
Q_0(r_\odot) < b(\gamma_0)\, \left[ \deriv{n}{E} \right]_{\rm
obs}\!\!\!\!\!\!\!\!(\gamma_0),
\ee 
and one arrives to the condition
\be
\langle \sigma v \rangle_{e^\pm}(\gamma_0)
<
\left[ \frac{ m_{\rm dm} }{ \rho_{\rm dm}(r_\odot) } \right]^2
b(\gamma_0)\, \left[ \deriv{n}{E} \right]_{\rm
obs}\!\!\!\!\!\!\!\!(\gamma_0)
\label{eqProductionRate3}
\ee
in order not to overproduce the observed signal.

%__________________________________
\begin{figure}
\centering \includegraphics[width=8cm]{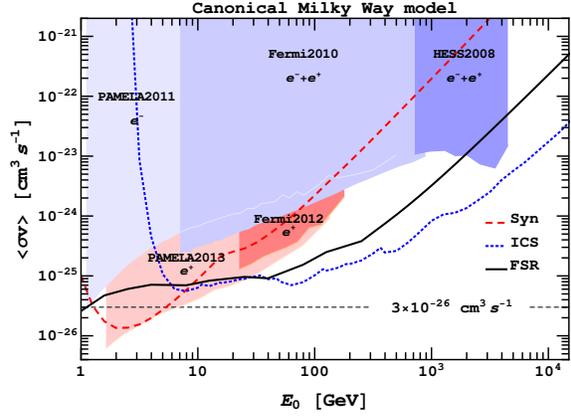}
\caption
{
Upper limits on the dark matter annihilation cross-section derived by comparing the predicted synchrotron (red dashed line), ICS (blue dotted line) and FSR (black solid line) emission with multi-wavelength observational data.
The areas shaded in blue show the constraints obtained from the measurements of the combined electron+positron spectrum at the solar neighbourhood by PAMELA, Fermi and HESS.
The upper limits obtained from the positron spectrum are shown by the red areas.
The horizontal dotted line indicates the value $\langle \sigma v \rangle_{e^\pm} = 3\times 10^{-26}$~cm~s$^{-1}$.
} 
\label{figCanonical}
\end{figure}
%__________________________________

The results are plotted in Figure~\ref{figCanonical}, together with the upper limits on the dark matter annihilation cross-section derived from the comparison of the predicted synchrotron, ICS and FSR emission, assuming our canonical Milky Way model for particle propagation, with multi-wavelength observations by Haslam, WMAP, and Fermi.
As can be readily seen in the figure, the tightest constraints are provided by final-state radiation and inverse Compton scattering for injection energies above $20-30$~GeV, whereas the positron spectrum in the solar neighbourhood and synchrotron emission limit the production cross-section at lower energies.

Similar (or stronger) constraints can also be obtained from the analysis of the CMB \citep[e.g.][]{Galli+09,Galli+11,Slatyer+09} and the gamma-ray emission from the Galactic centre at $\sim$~TeV energies.
In particular, a stringent upper limit in this mass range has been derived by comparing HESS measurements from suitably defined ``source'' and ``background'' regions \citep{Abramowski+11,AbazajianHarding12}.

According to Figure~\ref{figCanonical}, the typical value for thermal relics, $\langle \sigma v \rangle_{e^\pm} = 3\times 10^{-26}$~cm~s$^{-1}$, is ruled out for particle masses lighter than a few GeV.
Both particle physics processes and astrophysical boost factors have previously been advocated to increase the current annihilation rate in the Milky Way by more than a factor of 10 with respect to the early Universe.
Such models would be excluded for any dark matter candidate below the $\sim$~TeV regime annihilating primarily into electron-positron pairs.
Since our analysis involves a very conservative treatment of the astrophysical signal, merely excluding the emission from the disk and prominent point sources, it is expected that a deeper understanding of the astrophysical sources of electrons and positrons would make possible to probe the interesting region of the parameter space below $\langle \sigma v \rangle_{e^\pm} = 3\times 10^{-26}$~cm~s$^{-1}$.

%--------------------------------------------------------------------------
 \section{Effect of the astrophysical parameters}
 \label{secAstroParams}
%--------------------------------------------------------------------------

All the constraints represented in Figure~\ref{figCanonical} are based on the `canonical' Milky Way model discussed in Section~\ref{secPara}.
The final-state radiation from the Galactic centre and the local positron spectrum directly trace the instantaneous injection rate, and therefore they do not depend on the propagation parameters.
However, the surface brightness profiles of synchrotron and ICS emission are sensitive to the precise values adopted for the intensity of the magnetic field, the diffusion coefficient, and the interstellar radiation field.
The inner logarithmic slope of dark matter density profile has a very strong impact on the injection rate close to the centre, and thus it affects all the tracers considered in the present work except the positron spectrum in the solar neighbourhood.

Here we investigate the effect of the various astrophysical parameters of our propagation model on the upper limits obtained for the dark matter annihilation cross-section.
As we did for the canonical model, we consider different initial energies $E_0$ from 1~GeV to 10~TeV and compare the predicted emission with the full observational data set, but now we vary each of the astrophysical parameters in turn in order to assess their influence on the results.

%__________________________________
\begin{figure}
\centering \includegraphics[width=8cm]{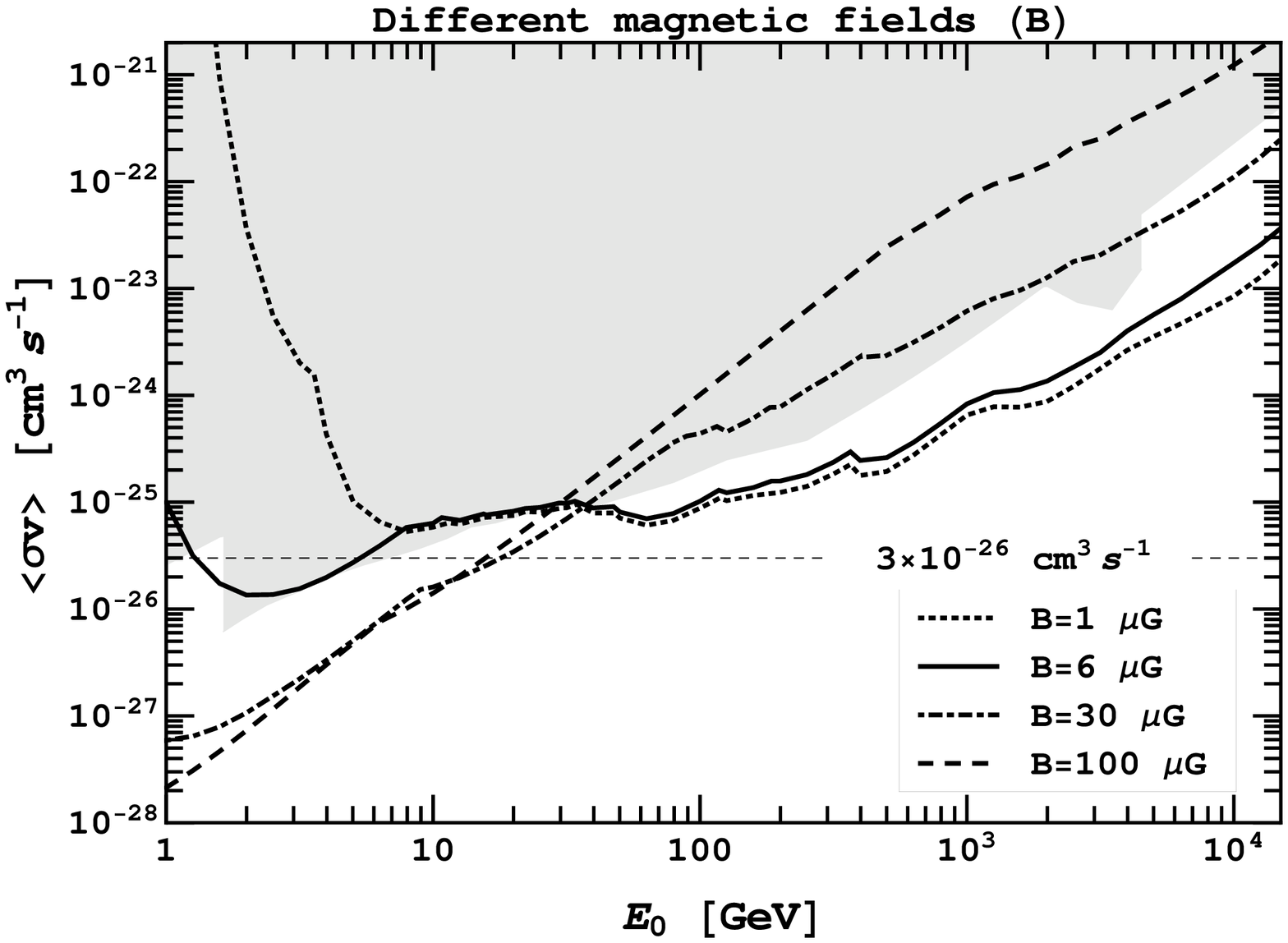}
\centering \includegraphics[width=8cm]{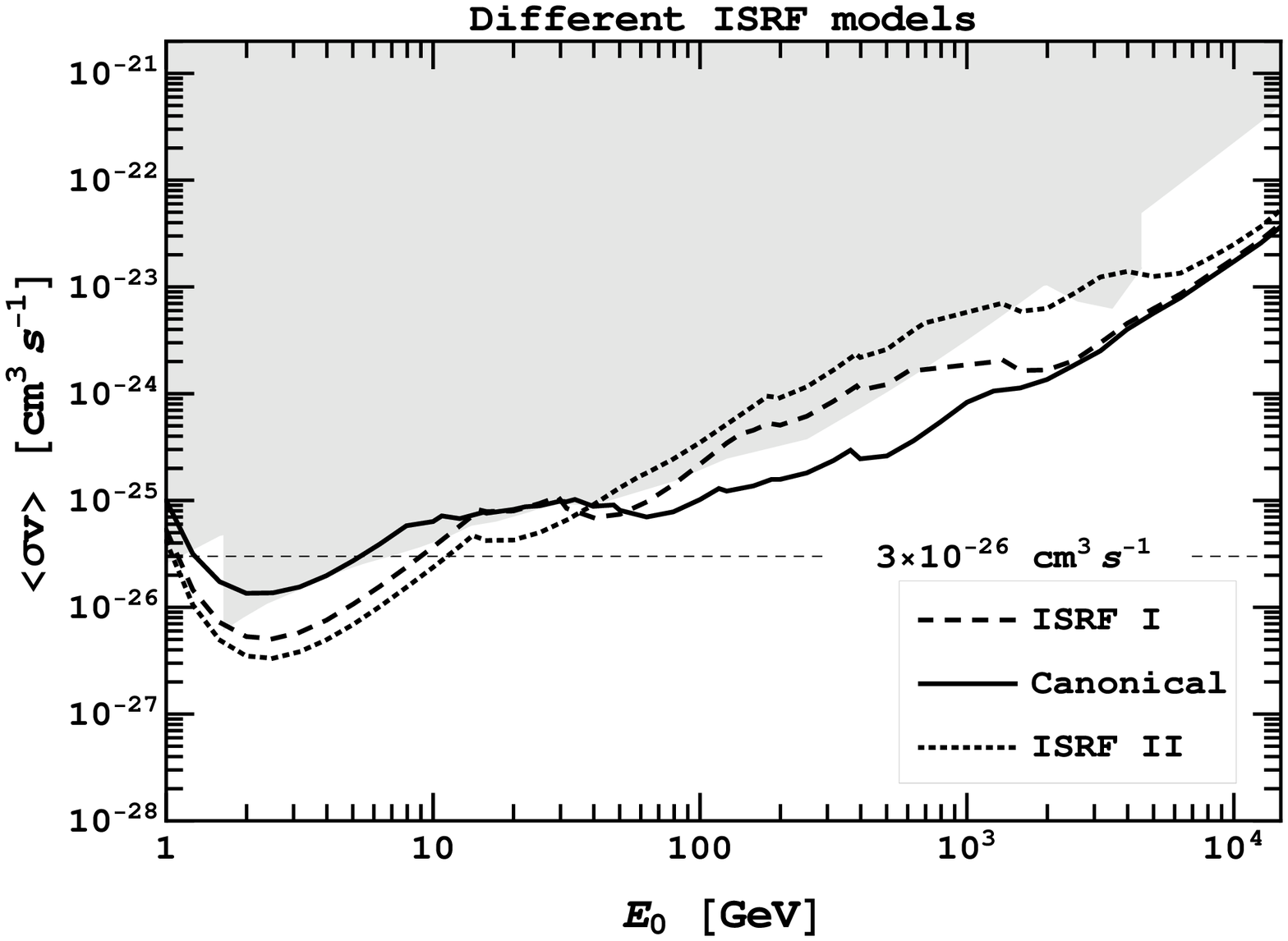}
\centering \includegraphics[width=8cm]{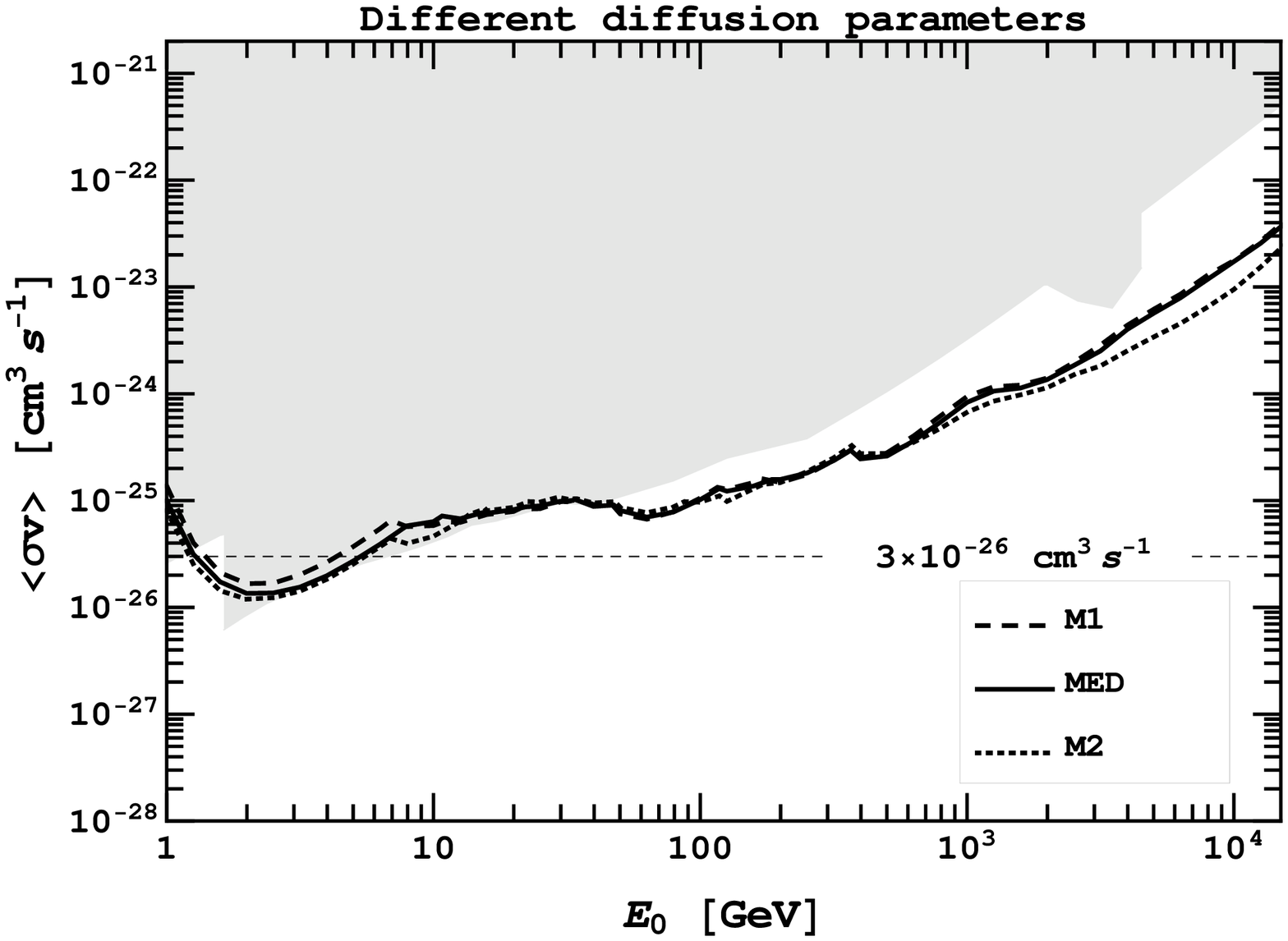}
\caption
{
Upper limits on the dark matter annihilation cross-section from synchrotron and ICS, for different values of the magnetic field (top), interstellar radiation field (middle), and diffusion coefficient (bottom).
Constraints from FSR and the local positron spectrum are indicated by the shaded area.
} 
\label{figDependenceAstro}
\end{figure}
%__________________________________

Let us start with the intensity of the magnetic field $B$.
This parameter plays an important role in the energy losses, and it sets the total amount of energy that is radiated away as synchrotron emission.
The top panel on Figure~\ref{figDependenceAstro} shows the upper limits derived by combining the constraints obtained from synchrotron and inverse Compton scattering.
\Referee{The results obtained for $B = 1$, 6} (our canonical model), 30 and $100~\mu$G are plotted as dotted, solid, dash-dotted, and dashed lines, respectively.
All the other constraints (FSR and local positron spectrum) are independent of $B$, and are shown by the shaded area.

Synchrotron constraints are most important at the lowest injection energies ($E_0\sim 1-30$~GeV), while the upper limits at higher initial energies (from $\sim 30$~GeV to 10~TeV) are due to ICS in the gamma-ray regime.
The intensity of the magnetic field affects both processes in an opposite way: for low values of the magnetic field, all energy is lost by inverse Compton scattering, and synchrotron emission is almost irrelevant; as one increases the value of $B$, synchrotron constraints become more important at the expense of ICS emission.
In the most extreme case ($B=100~\mu$G), gamma-ray constraints are negligible, and the upper limits derived from synchrotron radiation are well approximated by a pure power law.
For large values of the magnetic field, the synchrotron constraints are more stringent than the upper limits derived from the positron spectrum.

In the middle panel of Figure~\ref{figDependenceAstro}, we investigate the upper limits of synchrotron and ICS for different models of the interstellar radiation field.
As mentioned in Section~\ref{secPara}, we adopted the parameterization proposed by \citet{CirelliPanci09} in terms of three black-body components.
The temperatures and normalizations of each component are summarized in Table~\ref{tabParameters}.
The effect of the ISRF is similar to that of the magnetic field, but in the opposite direction: a higher photon density results in a larger amount of energy being lost by inverse Compton scattering rather than synchrotron emission.
Nevertheless, for reasonable values of the model parameters, the upper limits on $\langle\sigma v\rangle$ do not vary by more than a factor of three.

As shown on the bottom panel of Figure~\ref{figDependenceAstro}, the effect of the diffusion coefficient is even smaller.
The upper limits are slightly more stringent when the electrons and positrons are allowed to travel a shorter distance from the place where they were injected, but the difference between the three propagation models is barely noticeable.
Thus, we conclude that our results are not severely affected by the astrophysical uncertainties associated to particle propagation.
An additional source of uncertainty would be related to our choice of spherical boundary conditions.
Although we have not investigated this issue in detail, comparison with other studies based on cylindrical boundary conditions \citep[e.g.][]{Fornengo+12, Mambrini+12, Ackermann+12Con} suggest that the effect of this choice on the annihilation cross-section is relatively minor (see Appendix~\ref{secCylindrical}).

%__________________________________
\begin{figure}
\centering \includegraphics[width=8cm]{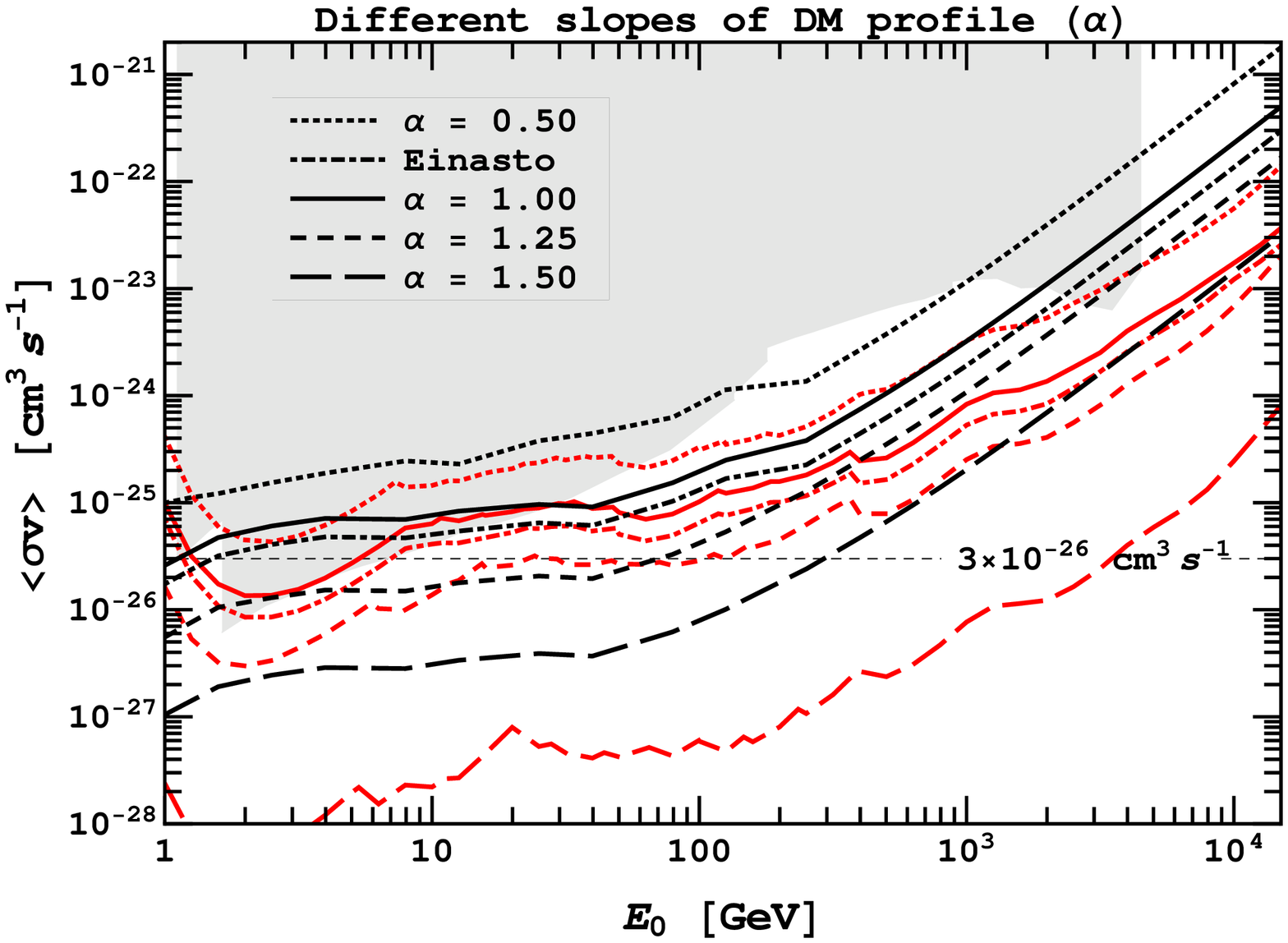}
\centering \includegraphics[width=8cm]{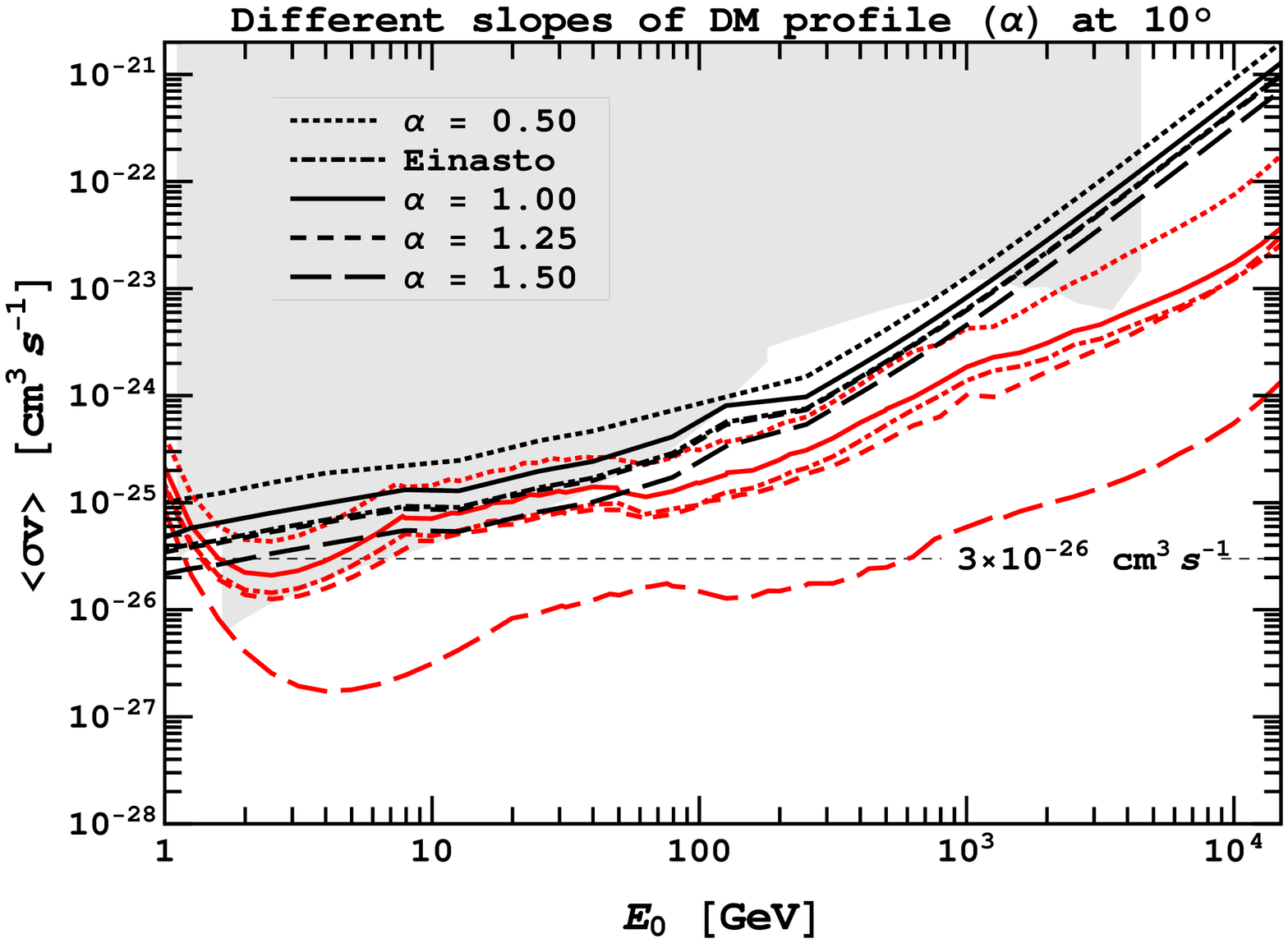}
% \hfill \includegraphics[width=8cm]{figs/UpperLimitAlpha10Deg.eps}
\caption
{
Upper limits on the dark matter annihilation cross-section for different values of the inner logarithmic slope $\alpha$ of the dark matter density profile.
On the top panel, the constraints are derived from the angle that provides the most stringent limit (see Figure~\ref{figConstraintTheta} for the canonical case $\alpha=1$), whereas all the constraints on the bottom panel are obtained from the observed emission at $\theta=10^\circ$ from the Galactic centre.
In both cases, black and red lines represent the limits associated to FSR and Synchrotron+ICS emission, respectively.
The constraints from the local positron spectrum (independent on $\alpha$) are shown by the shadowed areas.
} 
\label{figDependenceAlpha}
\end{figure}
%__________________________________

In contrast, the exact value of the inner slope $\alpha$ of the dark matter density profile plays a very important role in setting the actual constraints on $\langle\sigma v\rangle$.
We have investigated several values in the interval $0<\alpha<2$ (the appropriate values of $\rho_s$ and $r_s$ are quoted in Table~\ref{tabRhosRs}) as well as the Einasto profile given by equation~(\ref{eqEinasto}).
We report in Figure~\ref{figDependenceAlpha} the upper limits obtained from the comparison of the predicted final-state radiation, synchrotron, and inverse Compton scattering emission for $\alpha=0.50$, 1.00, 1.25, and 1.50 with our multi-wavelength observational data set.
Results for FSR and the combination of synchrotron and ICS emission are plotted separately. Constraints from the local positron spectrum are independent of $\alpha$ and are shown as a solid area.

The top panel of the figure shows the upper limits obtained by the same procedure applied to the canonical model, i.e. choosing the angular separation $\theta$ that provides the tightest constraint.
Not surprisingly, larger values of $\alpha$ result in lower values of $\theta$.
The constraints from FSR and Synchrotron+ICS emission come from innermost $1^\circ$ for $\alpha>$0.5 and $\alpha>$1.25, respectively.

Since the particle production rate near the centre of the Milky Way increases dramatically with the value of the inner slope of the density profile, this is, by far, the most relevant astrophysical parameter.
For $\alpha > 1.25$, a cross-section larger than $3 \times 10^{-26}$~cm$^3$~s$^{-1}$ is ruled out for any dark matter candidate lighter than $\sim 100$~GeV.
On the contrary, if the dark matter density profile of the Milky Way was shallow, with a logarithmic slope significantly below $\alpha = 1$, the positron spectrum in the solar neighbourhood would provide the most stringent limits on dark matter annihilation, and therefore the constraints would not depend at all on the actual value of the logarithmic slope.

One may remove the dependency of the results on the precise shape of the dark matter density profile by fixing $\theta = 10^\circ$ when comparing model predictions with observational data.
As shown in the bottom panel of Figure~\ref{figDependenceAlpha}, we find, in agreement with previous work \citep[e.g.][]{SerpicoZaharias08,Ackermann+12Con}, that the uncertainty associated to the precise value of $\alpha$ reduces to about a factor of 2 when the comparison is restricted to the photon intensity at $\theta = 10^\circ$.
While this is therefore a good choice when the goal is to provide a conservative upper limit on the dark matter annihilation cross-section, we would like to stress that any prior knowledge of the dark matter density profile may lead to much stronger constraints if the inner slope was steeper than $\alpha=1$, as evidenced in the upper panel.

Finally, let us note that the local dark matter density is subject to relatively large uncertainties \citep[c.f.][]{DehnenBinney98, Klypin+02, Salucci+10, Iocco+11}, which translate trivially to the upper limits on the cross-section.
In addition, departures from spherical symmetry (including the presence of substructures) will also have a significant effect on the derived constraints \citep[see e.g.][]{Diemand+05, Lavalle+08}.

%--------------------------------------------------------------------------
 \section{Constraints on the inner slope of the density profile}
 \label{secAlpha}
%--------------------------------------------------------------------------

As pointed out in \citet{Ascasibar+06}, the photons from the central region of the Galaxy contain information on \emph{both} the dark matter annihilation cross-section and the shape of the density profile.
By assuming a given value of the cross-section, one can constrain the value of $\alpha$ from the total intensity and the morphology of the observed surface brightness.

In this work, we will focus only on the total intensity in order to derive a robust upper limit.
More detailed constraints could be obtained from the shape of the surface brightness profiles at different wavelengths once the astrophysical contribution is adequately subtracted.
We set the dark matter annihilation cross-section into electron-positron pairs to the value expected for a thermally-produced relic, $\langle \sigma v \rangle_{e^\pm}=3 \times 10^{-26}$~cm$^3$~s$^{-1}$, and compute the value of $\alpha$ for which the predicted emission rises above the observed level.

%__________________________________
\begin{figure}
\centering \includegraphics[width=8cm]{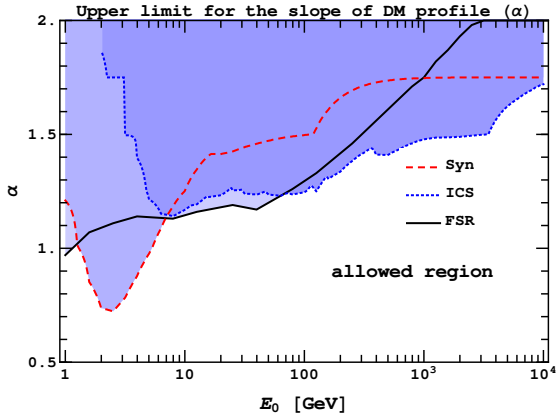}
\caption
{
Upper limits on the inner logarithmic slope of dark matter density profile $\alpha$, obtained by imposing that FSR, ICS and synchrotron emission do not overproduce the observed signal (according to the observational data) for a thermal dark matter relic (i.e. $\langle \sigma v \rangle_{e^\pm}=3 \times 10^{-26}$~cm$^3$~s$^{-1}$).
} 
\label{figUpperLimitAlpha}
\end{figure}
%__________________________________

The corresponding upper limits are plotted in Figure~\ref{figUpperLimitAlpha} as a function of the initial energy $E_0$ associated to the mass of dark matter candidate.
Our results show that, for a thermal relic with $m_{\rm dm} < 100$~GeV, the dark matter density profile of the Milky Way \emph{must} be shallower than $\alpha\sim 1.3$ in order not to overproduce the observed signal.
It is worth noting that, since final-state radiation only depends on the injection rate, this constraint on the inner logarithmic slope $\alpha$ is independent on the other astrophysical parameters.
Synchrotron and inverse Compton scattering yield stronger limits than FSR at low and high injection energies, respectively, although of course these results depend much more on the details of the adopted propagation model (most notably, the intensity of the magnetic field).
For our canonical set-up, synchrotron radiation imposes extremely tight constraints for a limited range of dark matter masses, around a few GeV (observational data at lower frequencies would probably make possible to extend these constraints toward lower masses).
In particular, the standard case $\alpha=1$ would be excluded for $E_0<5$~GeV.
At high energies, ICS emission rules out slopes steeper than $\alpha=1.5$ for dark matter masses below $\sim 2$~TeV.
The regime $\alpha>1.8$ seems to be excluded in any case.

The fact that we are considering the total radio and gamma-ray emission, without taking into account the contribution of astrophysical origin, implies that these are conservative upper limits, and therefore we can conclude that, if dark matter particles annihilate primarily into electrons and positrons (or, more generally, any other particle; see below), any scenario where the \Referee{Milky Way} features a steep density profile (due to e.g. adiabatic contraction) may be firmly ruled out.

%--------------------------------------------------------------------------
 \section{Different annihilation channels}
 \label{secAnnihilationChannels}
%--------------------------------------------------------------------------

From the point of view of particle physics, dark matter annihilation directly into electron-positron pairs is arguably not the most natural channnel.
In most models, dark matter annihilates into heavier products, and then these particles produce lower-energy electrons and positrons as secondaries.

In this section, we investigate how the upper limits to the annihilation cross-section and the inner slope of the density profile depend on the annihilation channel.
More precisely, we consider different source functions, replacing the Dirac delta in equation~(\ref{eqInjectionSpectrum}) by the appropriate injection spectrum.
We used the electron-positron fluxes at production computed by \citet{Cirelli+11}, including electroweak corrections \citep{Ciafaloni+11}, for all the leptonic channels, as well as for annihilation into top and bottom quarks.

%__________________________________
\begin{figure*}
\includegraphics[width=.33\textwidth]{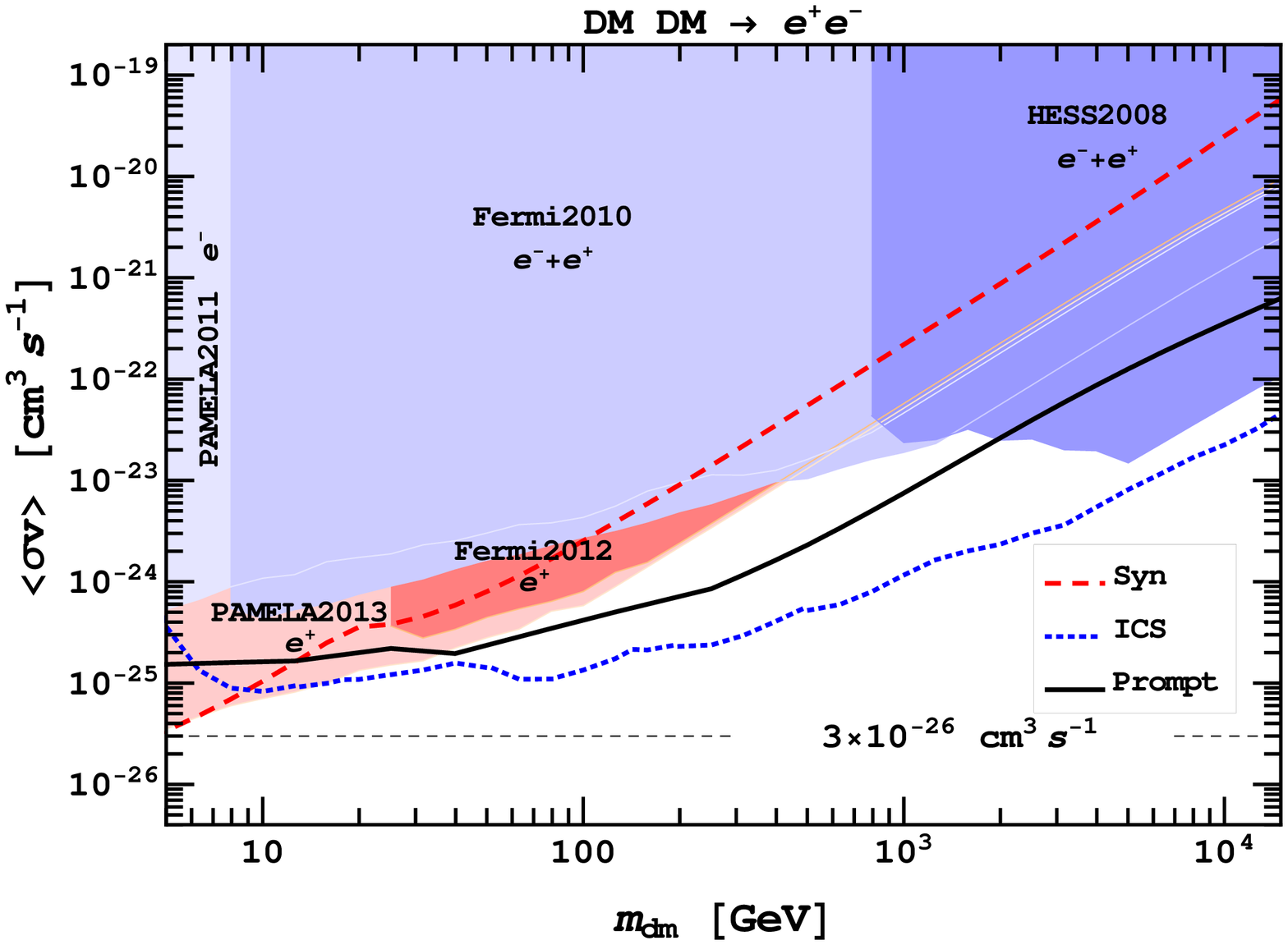}
\includegraphics[width=.33\textwidth]{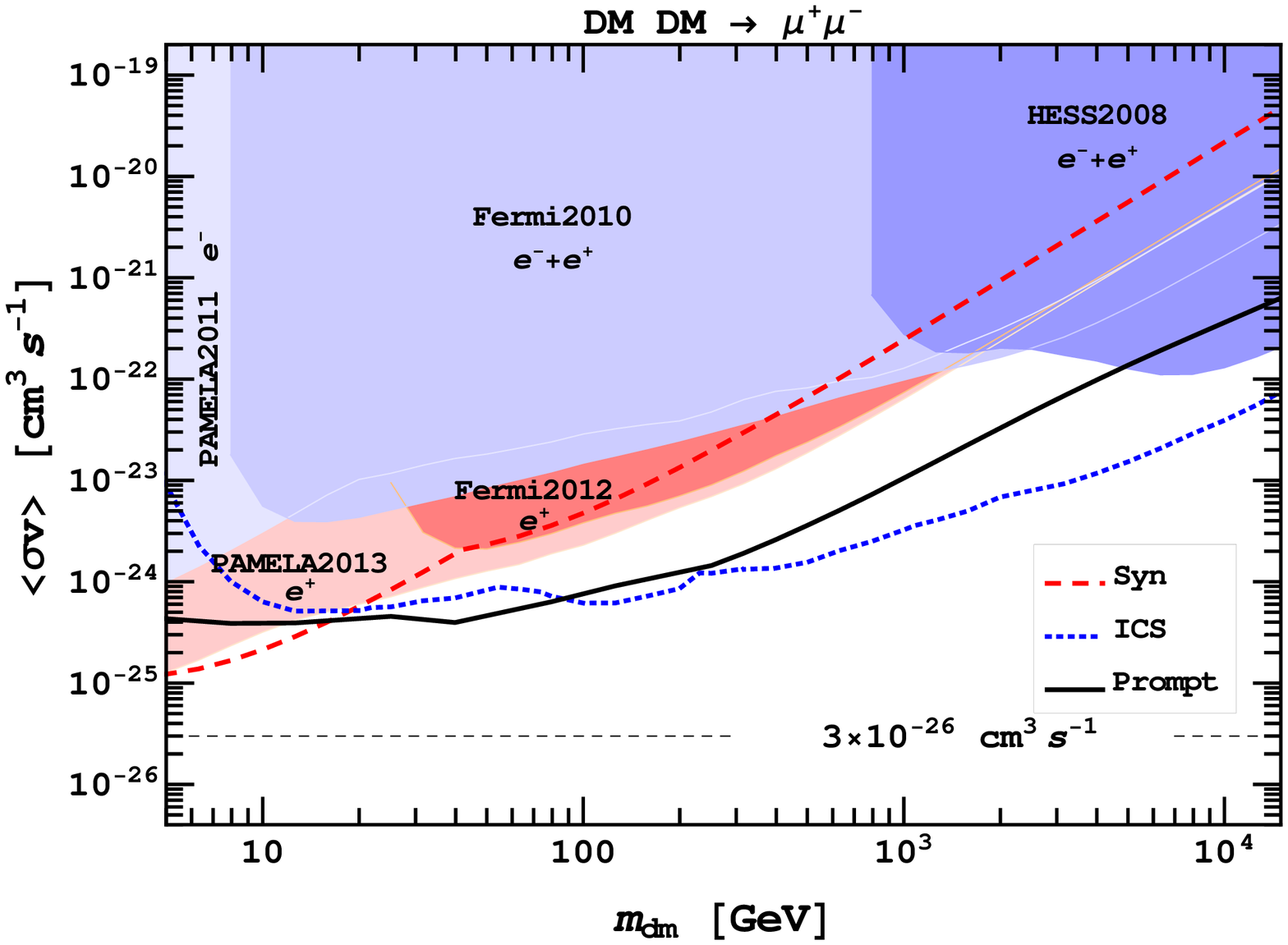}
\includegraphics[width=.33\textwidth]{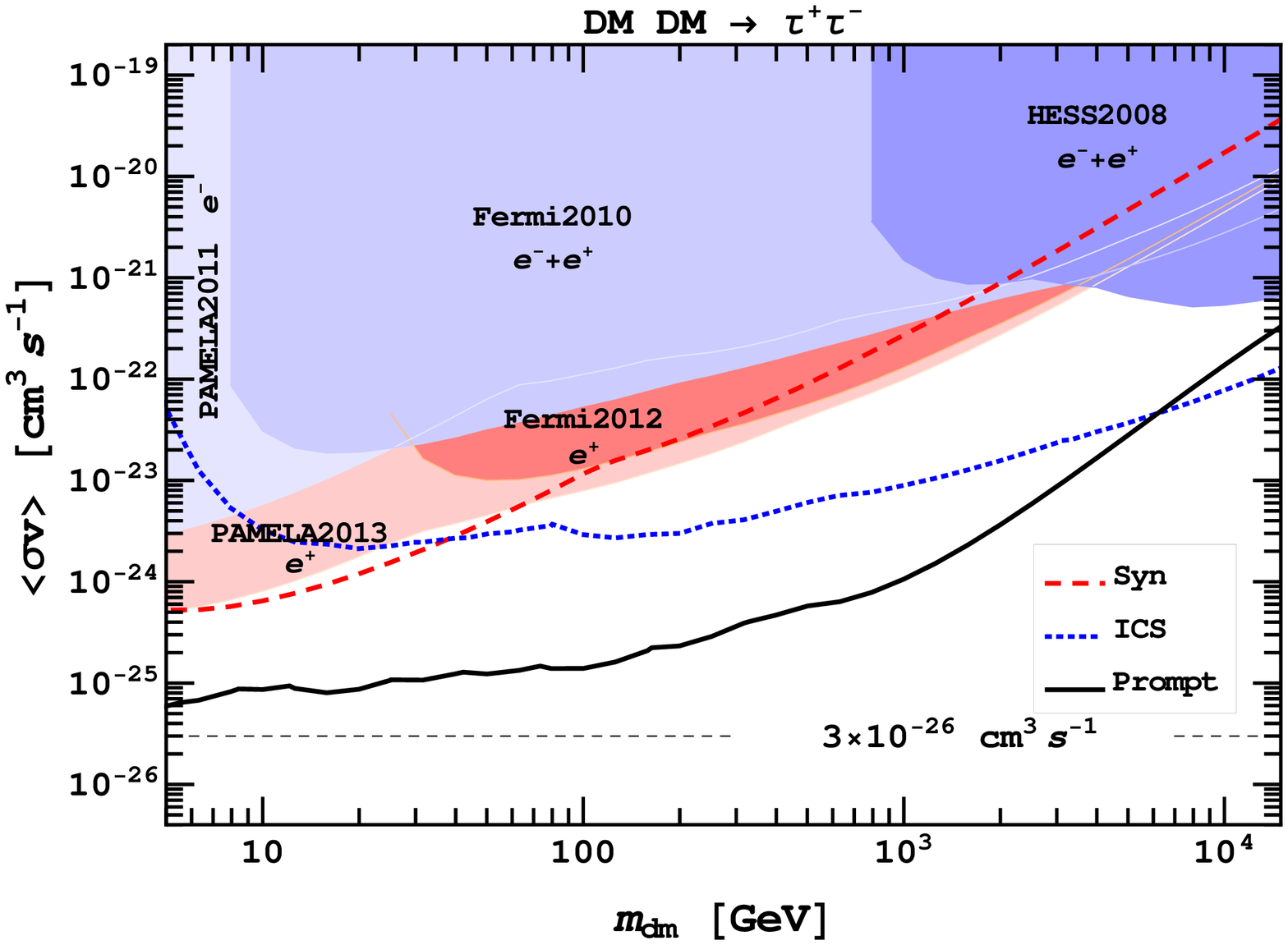}

\includegraphics[width=.33\textwidth]{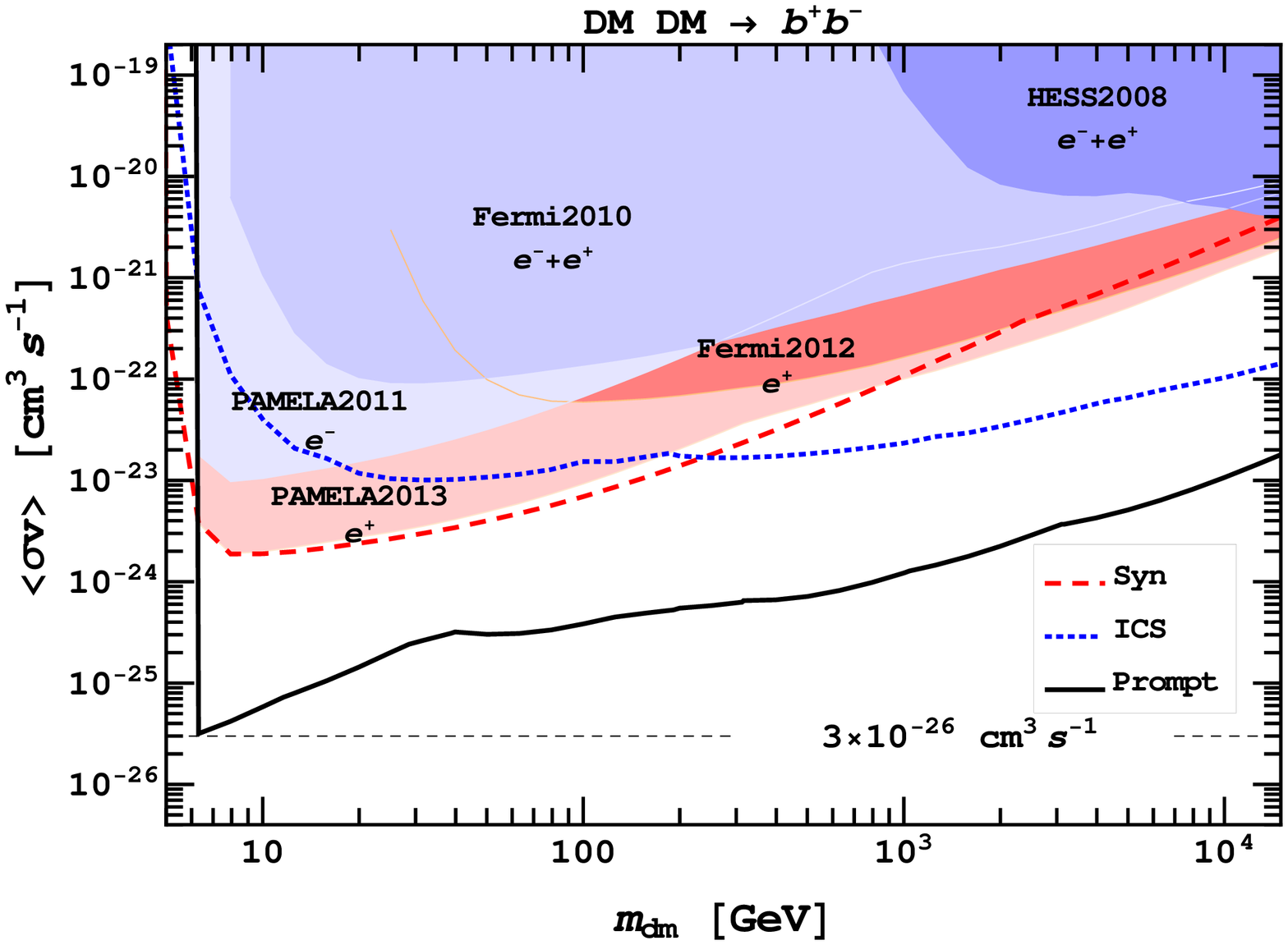}
\includegraphics[width=.33\textwidth]{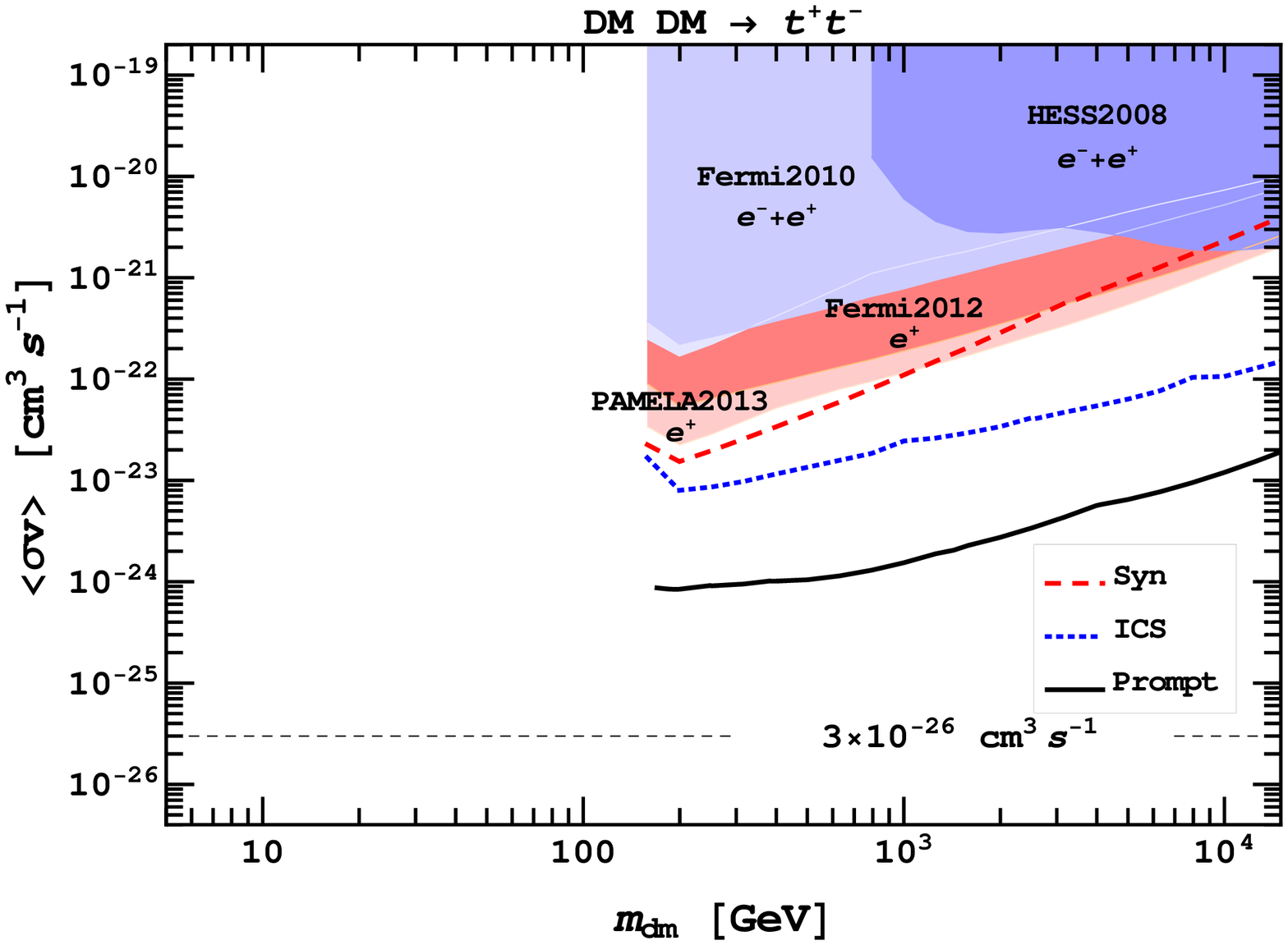}
\caption
{
Constraints on the annihilation cross-section for different annihilation channels.
} 
\label{figChannels_sigma}
\end{figure*}
%__________________________________

The upper limits on the annihilation cross-section into each channel are plotted in Figure~\ref{figChannels_sigma}.
Not surprisingly, the results for the electron channel (including a detailed model of the injection spectrum) are very similar to Figure~\ref{figCanonical}, assuming a Dirac delta.
The use of equation~(\ref{eqInjectionSpectrum}) seems thus perfectly justified in the present context.

The constraints obtained for dark matter annihilation into muon-antimuon pairs are somewhat weaker (by a factor of a few, especially at the interesting regime of low dark matter masses) due to the softer injection spectrum, but they are otherwise analogous to the results obtained for the electron channel (in terms of the relative importance of the different physical processes as well as the dependence of each constraint on the injection energy).
The main difference is that, in agreement with previous studies \citep[e.g.][]{Ackermann+12Con}, some modelling of the astrophysical background would be required in order to rule out the thermal cross-section.

For annihilation into $\tau$ particles, the upper limits imposed by ICS, synchrotron emission and the local positron spectrum are even less stringent.
However, in this case, a larger fraction of the initial energy is promptly radiated as gamma rays \citep{Cirelli+11}, providing constraints comparable to those obtained for direct annihilation into electrons and positrons.
%In particular, the canonical cross-section for a thermal relic is excluded for $m_{\rm dm}<3$~GeV.%

Similar conclusions may be reached for annihilation into top or bottom quarks.
For these channels, constraints from the electron-positron population are not particularly severe, but a large number of photons are produced during the hadronic cascade, and therefore prompt emission of gamma rays provides the tightest upper limits for any dark matter mass.

%__________________________________
\begin{figure*}
\includegraphics[width=.33\textwidth]{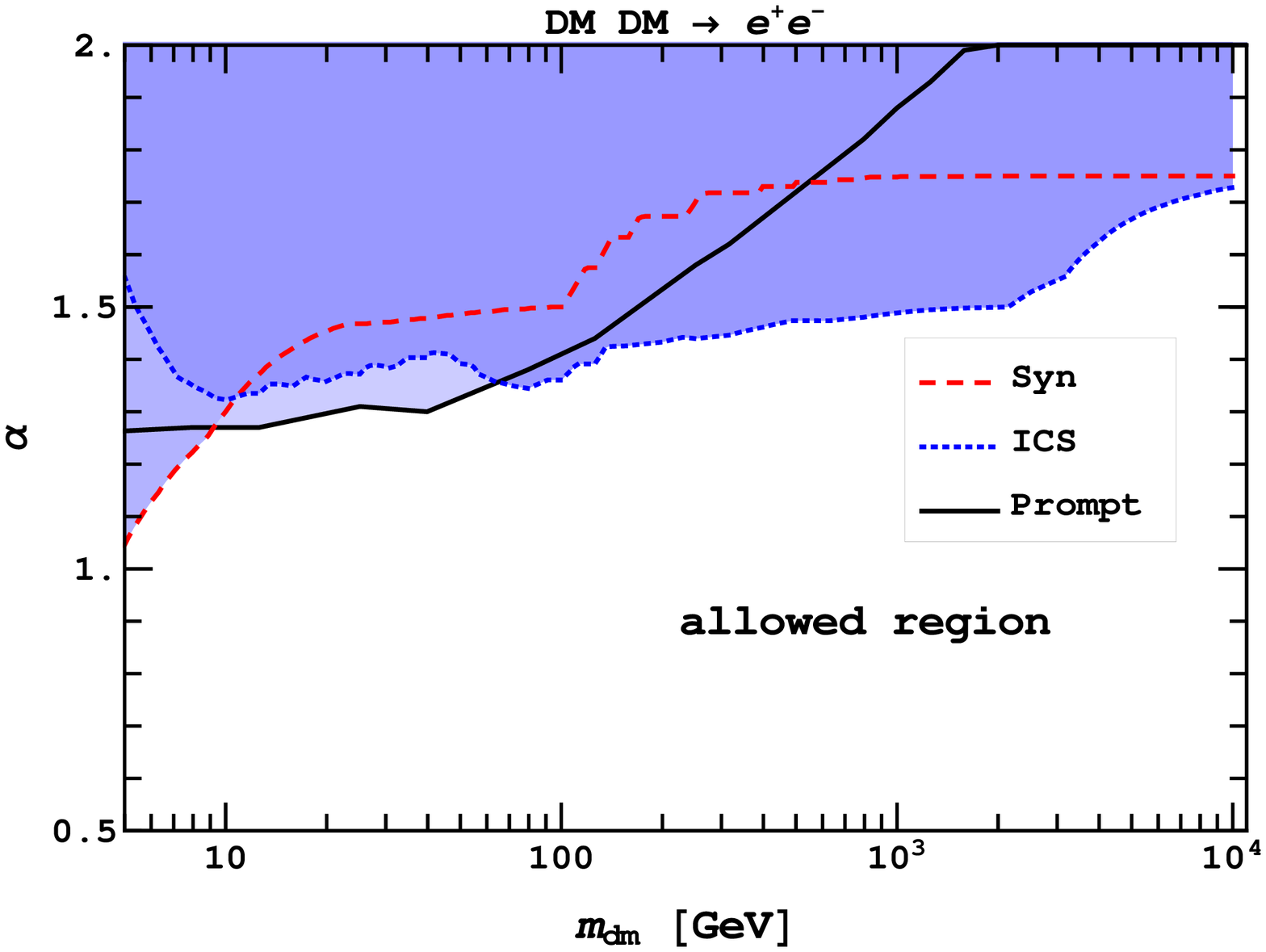}
\includegraphics[width=.33\textwidth]{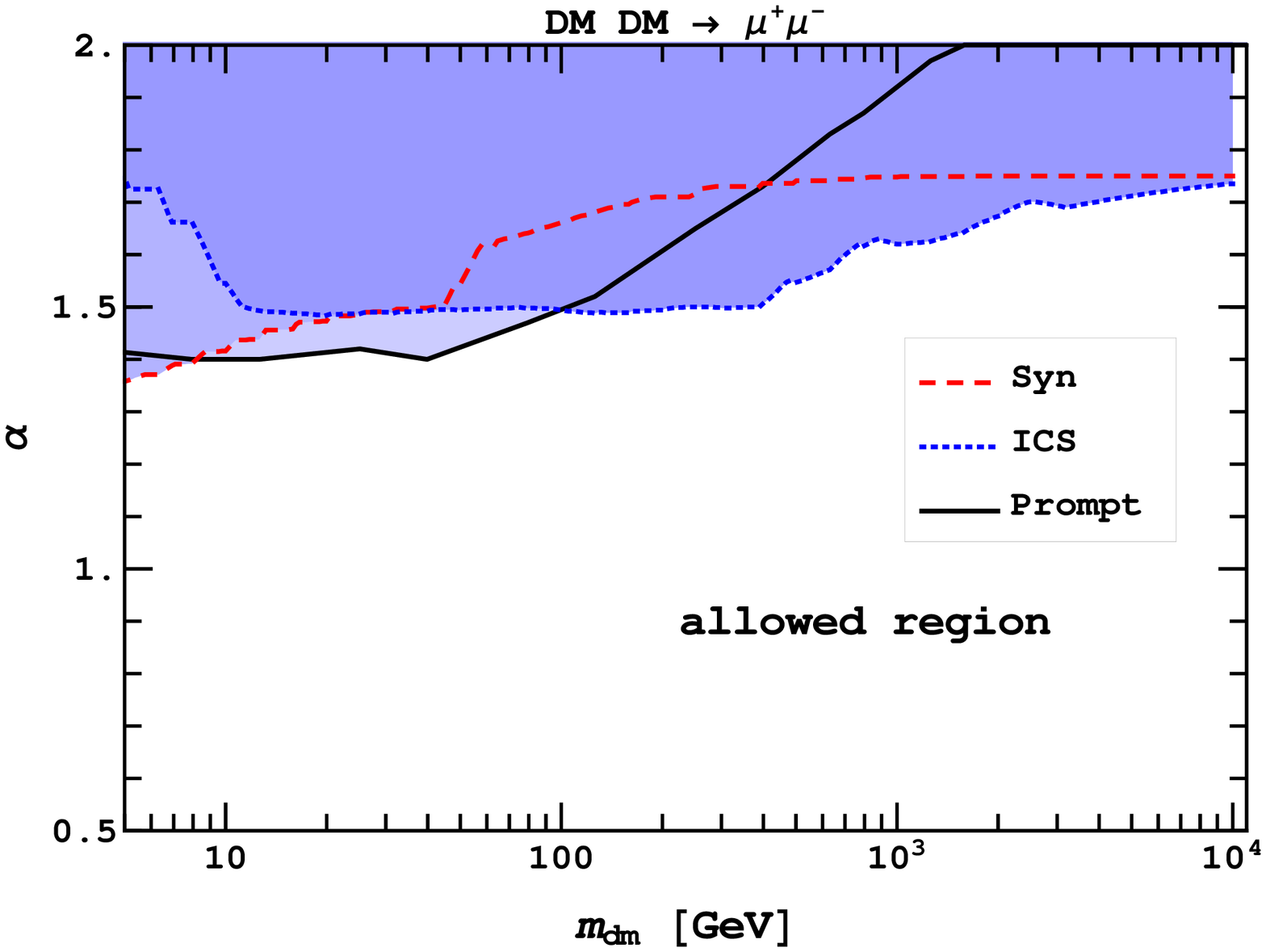}
\includegraphics[width=.33\textwidth]{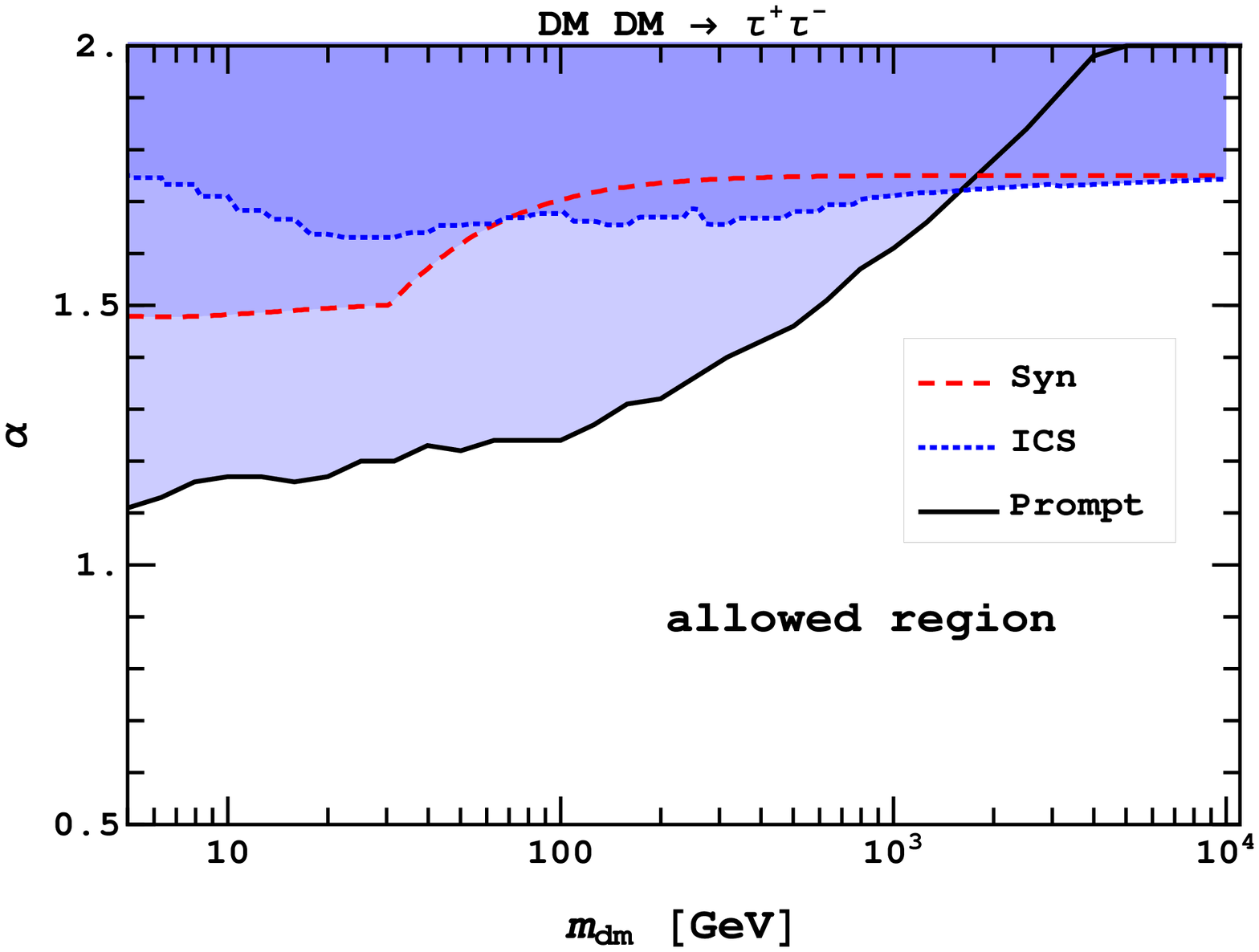}

\includegraphics[width=.33\textwidth]{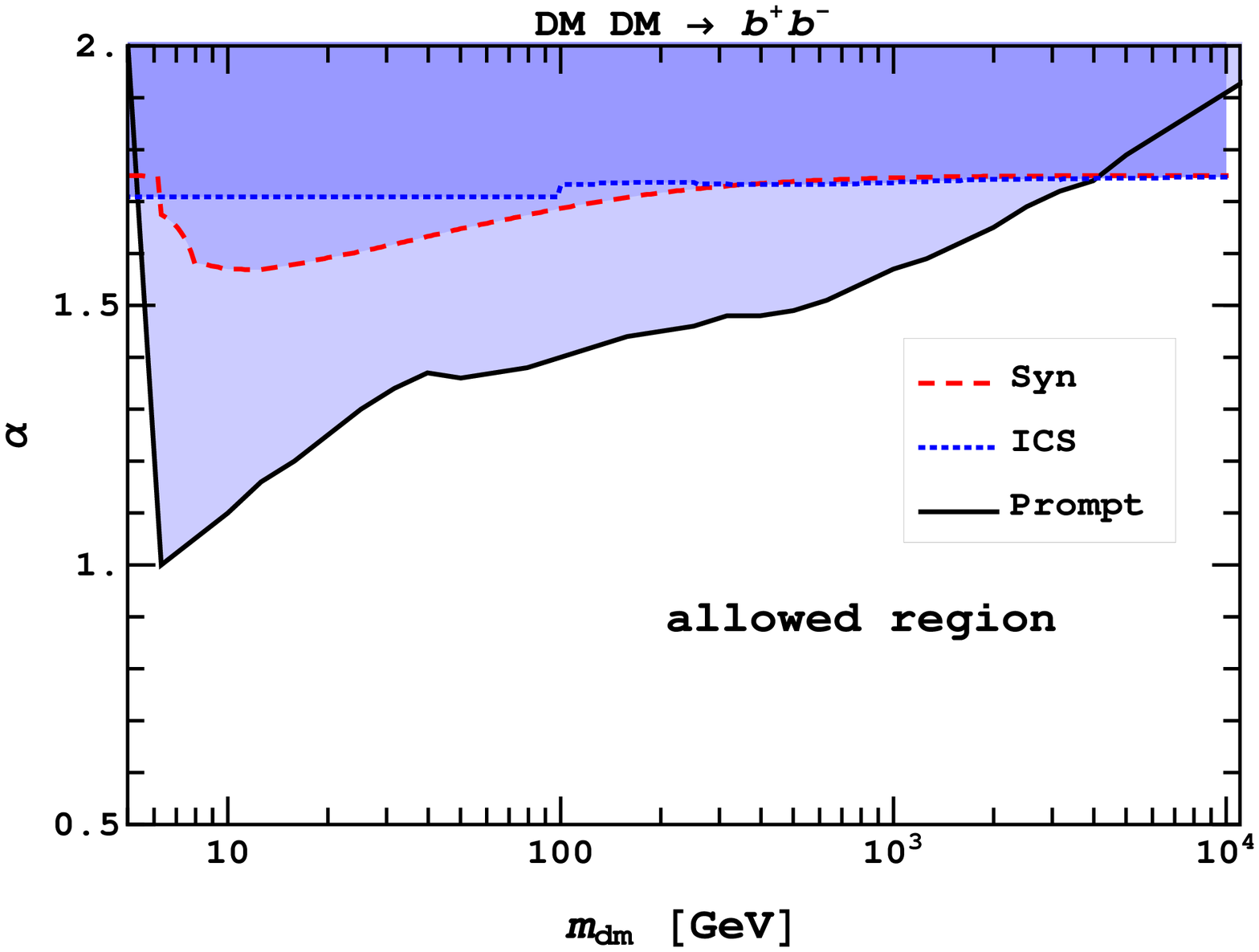}
\includegraphics[width=.33\textwidth]{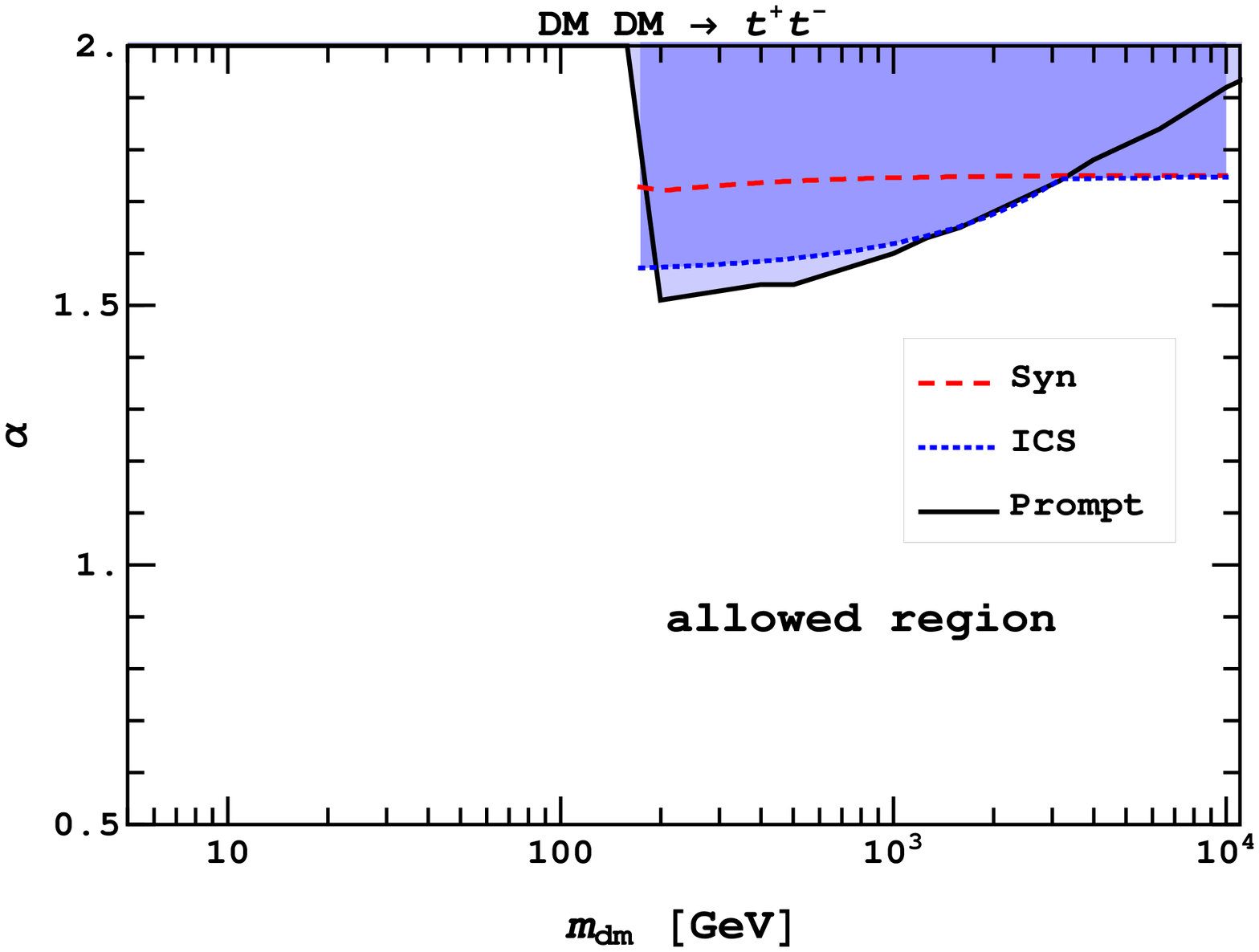}
\caption
{
Constraints on the inner logarithmic slope of the density profile for different annihilation channels, assuming $\langle \sigma v \rangle_{e^\pm}=3 \times 10^{-26}$~cm$^3$~s$^{-1}$.
} 
\label{figChannels_alpha}
\end{figure*}
%__________________________________

Let us now consider the constraints obtained on the inner logarithmic slope $\alpha$ of the dark matter density profile of the Milky Way, assuming that dark matter is a standard thermal relic.
The results, analogous to those presented in Section~\ref{secAlpha}, are shown in Figure~\ref{figChannels_alpha} for each of the selected annihilation channels.
A logarithmic slope steeper than $\alpha \sim 1.5$ is firmly ruled out for $m_{\rm dm} \la 1$~TeV for any annihilation channel, and tighter constraints can be imposed for lower masses in most cases.

%--------------------------------------------------------------------------
 \section{Summary and conclusions}
 \label{secConclusions}
%--------------------------------------------------------------------------

We have investigated the \Referee{constraints} on the dark matter annihilation cross-section into electron-positron pairs by comparing the predictions of an analytic model of particle propagation with a multi-wavelength set of observational data obtained from the literature.
We have compared the expected emission from synchrotron radiation, inverse Compton scattering and final-state radiation within the Milky Way with 18 maps of the sky at different frequencies: the Haslam radio map at 408 MHz, the 7-year data from the Wilkinson Microwave Anisotropy Probe (WMAP) in its 5 bands (23 GHz, 33 GHz, 41 GHz, 61 GHz, and 94 GHz), and gamma-ray maps from the Fermi Large Area Telescope (LAT) binned in 12 different channels (from 0.3 to 300 GeV).
A straightforward statistical criterion has been followed in order to mask the most obvious astrophysical signals (i.e. the emission from the Galactic disc and prominent point sources), and observational upper limits are derived from the remaining spherically-symmetric component.
In addition, we have also imposed that the predicted abundance of electrons and positrons in the solar neighbourhood does not exceed the measurements by PAMELA, HESS and Fermi.
Our main results can be summarized as follows:
\begin{enumerate}
%  \item The constraints from synchrotron and inverse Compton emission are always weaker than those from final-state radiation from the Galactic centre and the positron spectrum in the solar neighbourhood.
% This result is valid for all the values of the magnetic field, diffusion coefficient, and models of the interstellar radiation field that we have considered.
 \item If the density profile of the Milky Way halo is steep ($\alpha>1$), the exact value of inner logarithmic slope plays a crucial role in the upper limit on the annihilation cross-section.
The adiabatic contraction scenario is hardly consistent with any dark matter candidate lighter than $\sim 100$~GeV and $\langle \sigma v \rangle_{e^\pm} = 3 \times 10^{-26}$~cm$^3$~s$^{-1}$.
 \item If the density profile of the Milky Way halo is relatively shallow ($\alpha<1$), the upper limit on the cross-section is set by the local positron spectrum for low values of the injection energy.
Combining both types of messenger (photons and positrons) is thus of the utmost importance in this case.
Considering the positron spectrum separately makes possible to rule out cross-sections above $3 \times 10^{-26}$~cm$^3$~s$^{-1}$ for dark matter particles lighter than a few GeV.
 \item If the primary product of dark matter annihilation is any particle other than the electron, similar constraints can be obtained, although in this case prompt emission of gamma rays becomes the most relevant process.
%  \item \Referee{These results may vary within a factor of a few, depending on the details of particle propagation and the intensity of the magnetic field. Both effects are more important at the high-mass end}.
\end{enumerate}

Let us conclude by noting that the current upper limits are close to -- or have just reached -- the expected annihilation cross-section for a thermal relic,
and similar constraints (sometimes even stronger, and very robust against uncertainties) have also been obtained from the analysis of the CMB, the gamma-ray emission from nearby dwarf galaxies, and direct detection experiments.
Therefore, a better understanding of the production of positrons and gamma rays by astrophysical sources should thus lead to the detection of an indirect signal from dark matter annihilation in the Milky Way, providing at the same time an exquisite probe of the distribution of dark matter in the innermost regions of the Galactic halo.
Otherwise, the most straightforward versions of the leptophillic dark matter scenario would be ruled out completely.

%__________________________________
 \section*{Acknowledgments}

%__________________________________

Funding for this work has been provided by the DFG Research Grant AS\,312/1-1 (Germany).
MW also acknowledges support from Br\"{u}ckenprogramm of Universit\"{a}t Potsdam (Germany) and would like to thank V.~M\"{u}ller and J.C.~Mu\~{n}oz-Cuartas for useful discussions, as well as F.~Breitling and J.~Klar for their help with programming questions.
YA is supported by grant AYA2010-21887-C04-03 from the former \emph{Ministerio de Ciencia e Innovaci\'on} (Spain), as well as the \emph{Ram\'{o}n y Cajal} programme (RyC-2011-09461), now managed by the \emph{Ministerio de Econom\'{i}a y Competitividad} (fiercely cutting back on the Spanish scientific infrastructure).
We thank the participants of the ``Dark matter meetings'' organized by the \emph{Instituto de F\'{i}sica Te\'{o}rica} at the UAM for several interesting discussions; in particular, it is a pleasure to thank M.~Peir\'{o} and B.~Zaldivar for pointing us toward the {\sc pppc4dmid} suite.
Finally, we would like to thank the anonymous referee for a careful, constructive reading of the manuscript, that has led to significant improvements with respect to the original version.

%%%%%%%%%%%%%%%%%%%%%%%%%%%%%%%%%%%%%%%%%%%%%%%%%%%%%%%%%%%%%%%%%%%%%%%%%%%%%%%
 \bibliographystyle{mn2e}
 \bibliography{references}
%%%%%%%%%%%%%%%%%%%%%%%%%%%%%%%%%%%%%%%%%%%%%%%%%%%%%%%%%%%%%%%%%%%%%%%%%%%%%%%

% \newpage 

\appendix

%__________________________________
\section{Observed intensities}
\label{secTables}

%__________________________________
\begin{figure*}

\includegraphics[width=5.5cm]{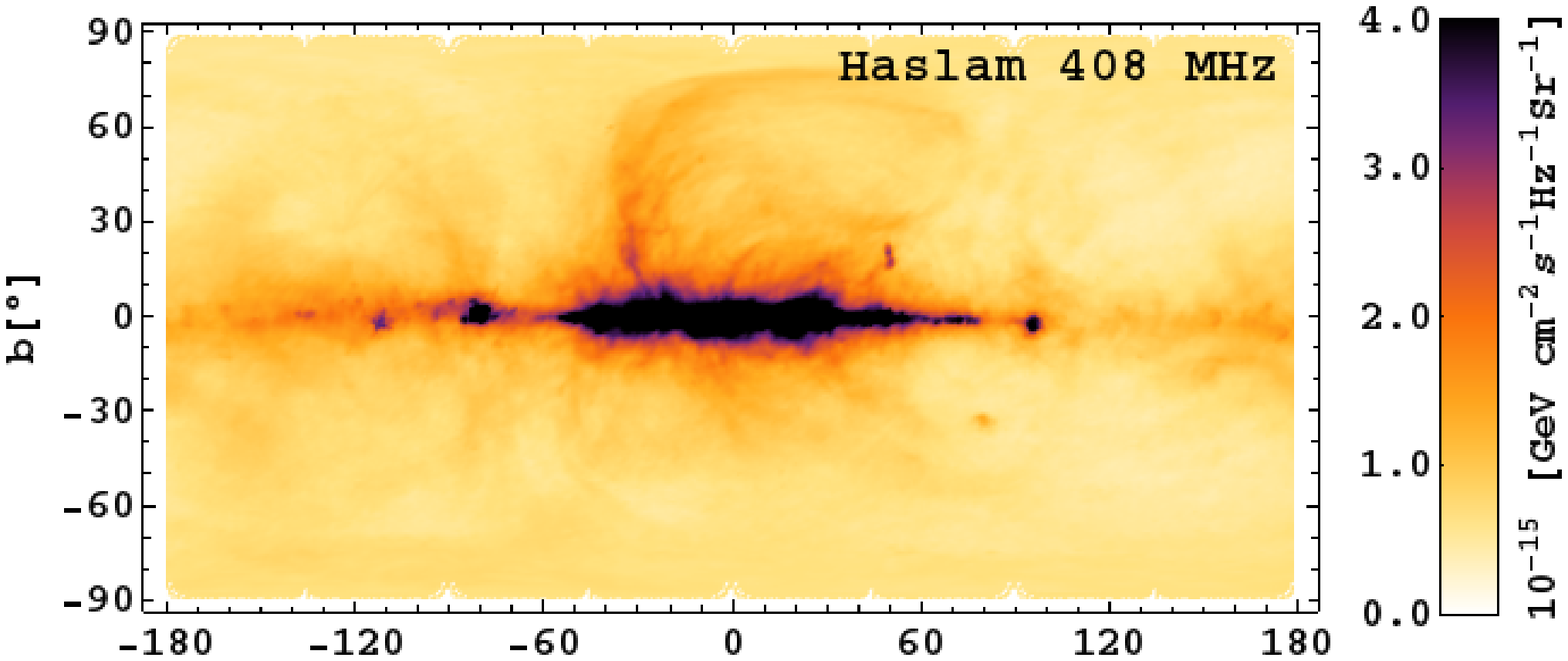}\hfill
\includegraphics[width=5.5cm]{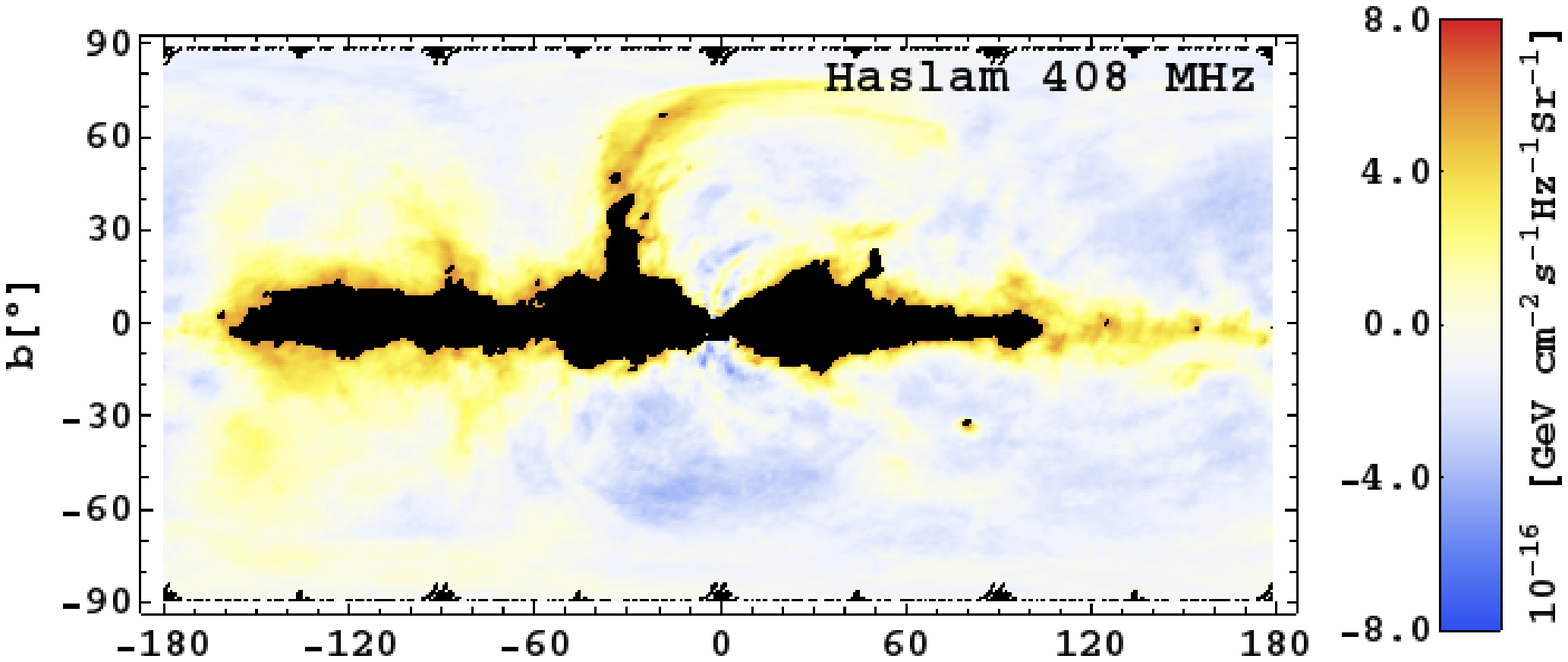}\hfill
\includegraphics[width=5.3cm]{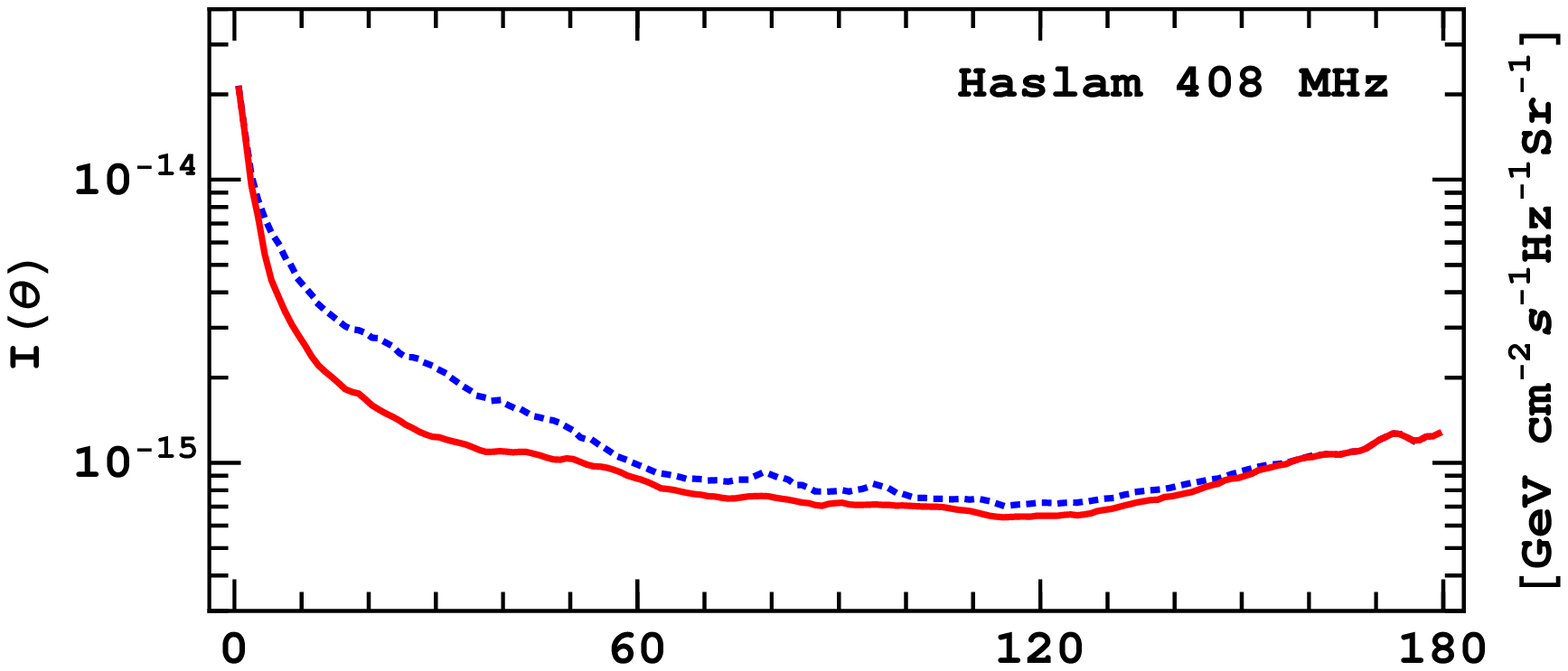}\hfill

\includegraphics[width=5.5cm]{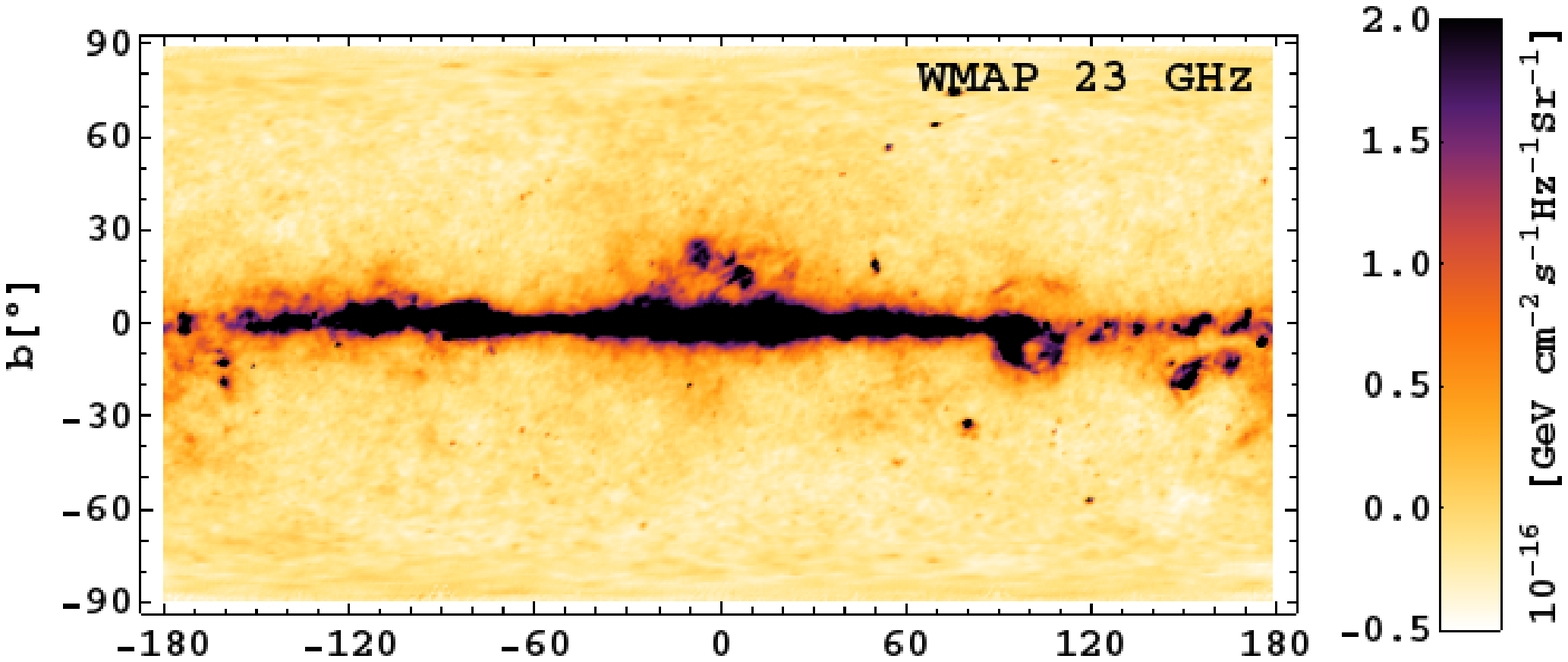}\hfill
\includegraphics[width=5.5cm]{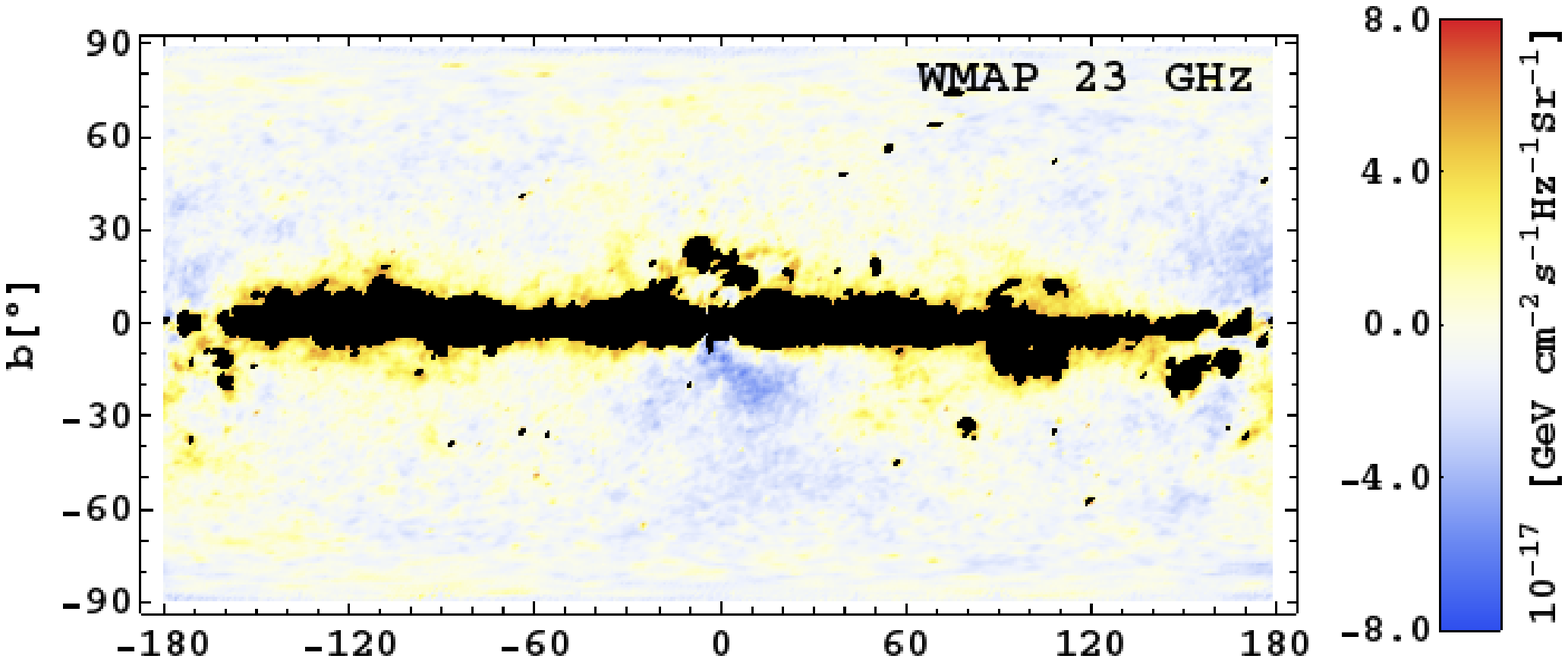}\hfill
\includegraphics[width=5.3cm]{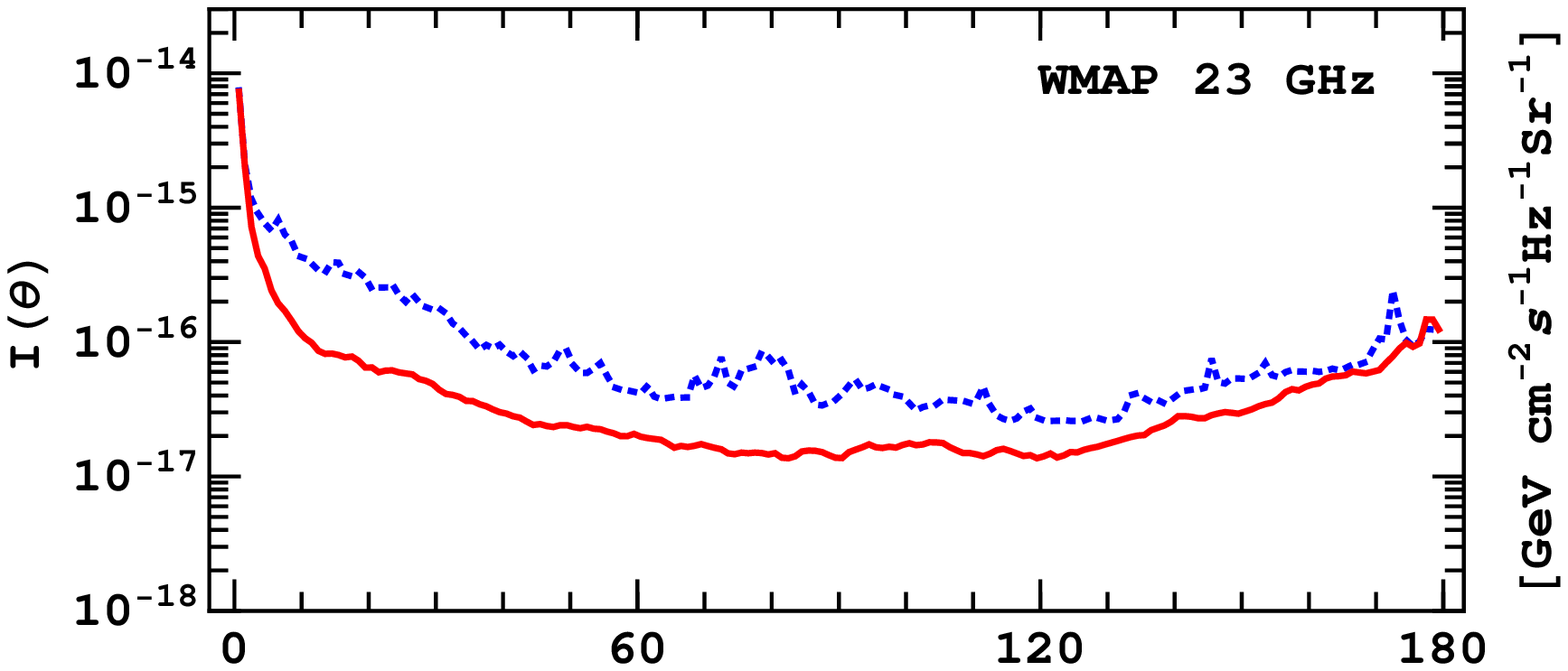}\hfill

\includegraphics[width=5.5cm]{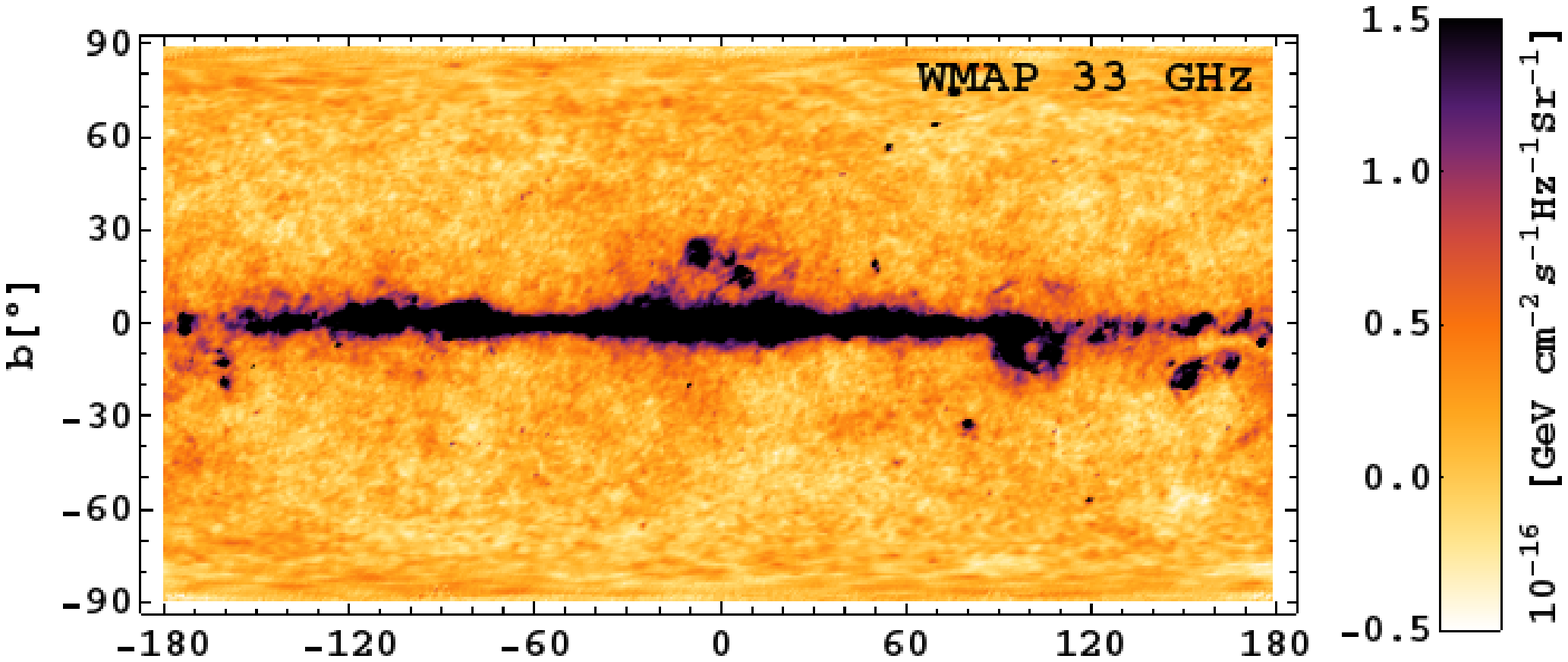}\hfill
\includegraphics[width=5.5cm]{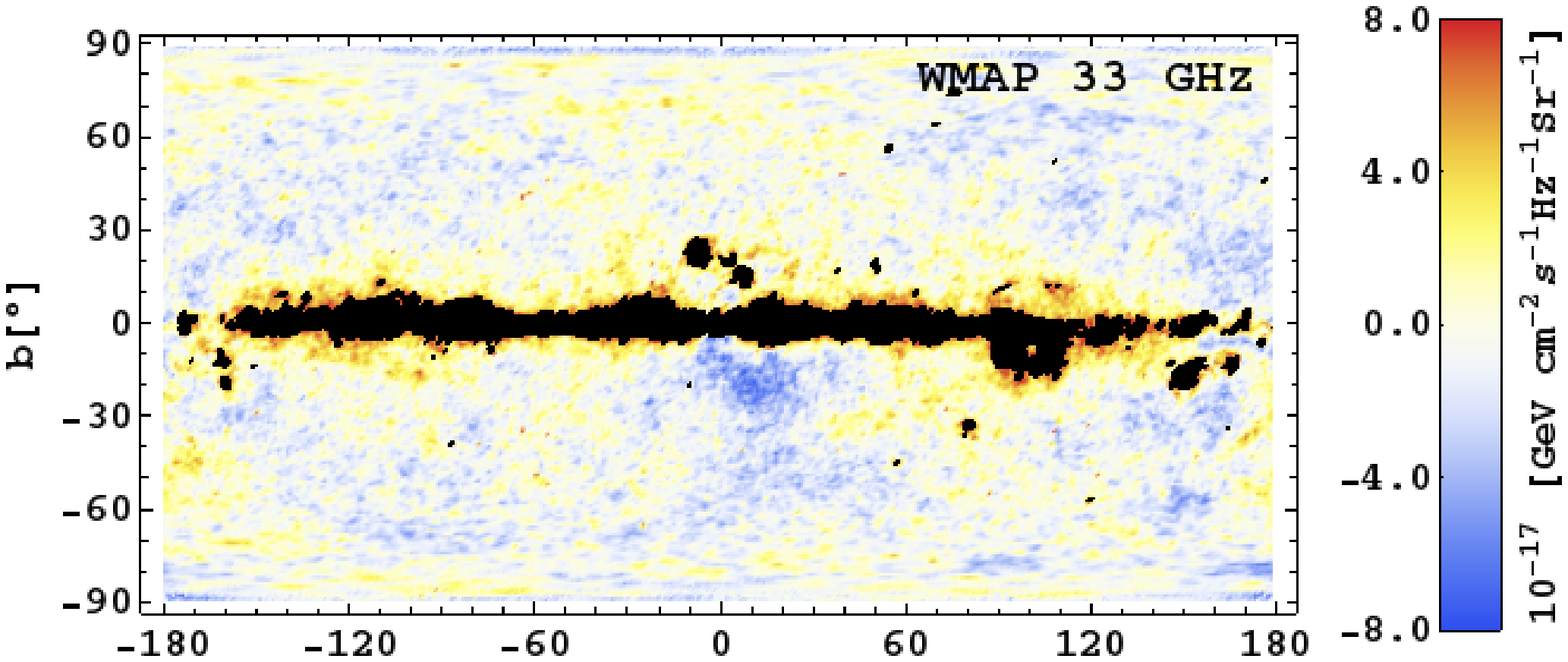}\hfill
\includegraphics[width=5.3cm]{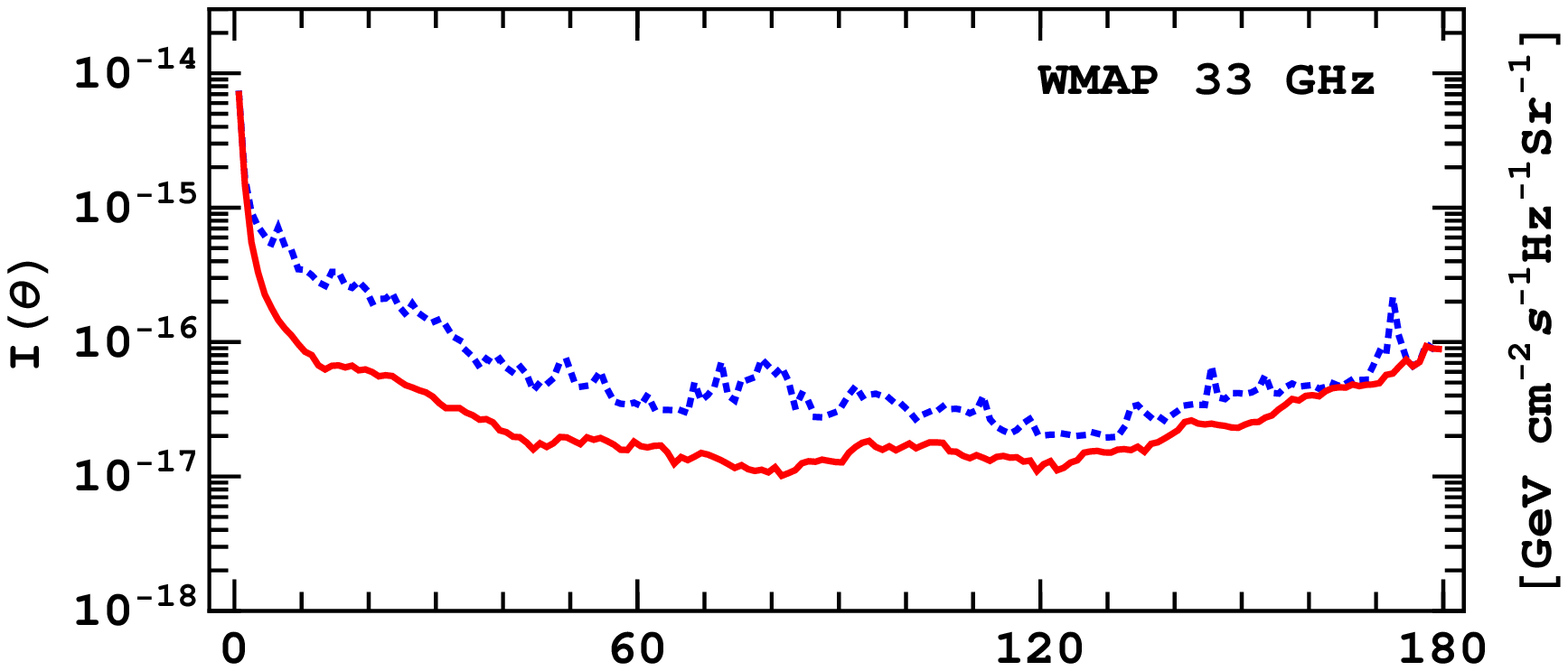}\hfill

\includegraphics[width=5.5cm]{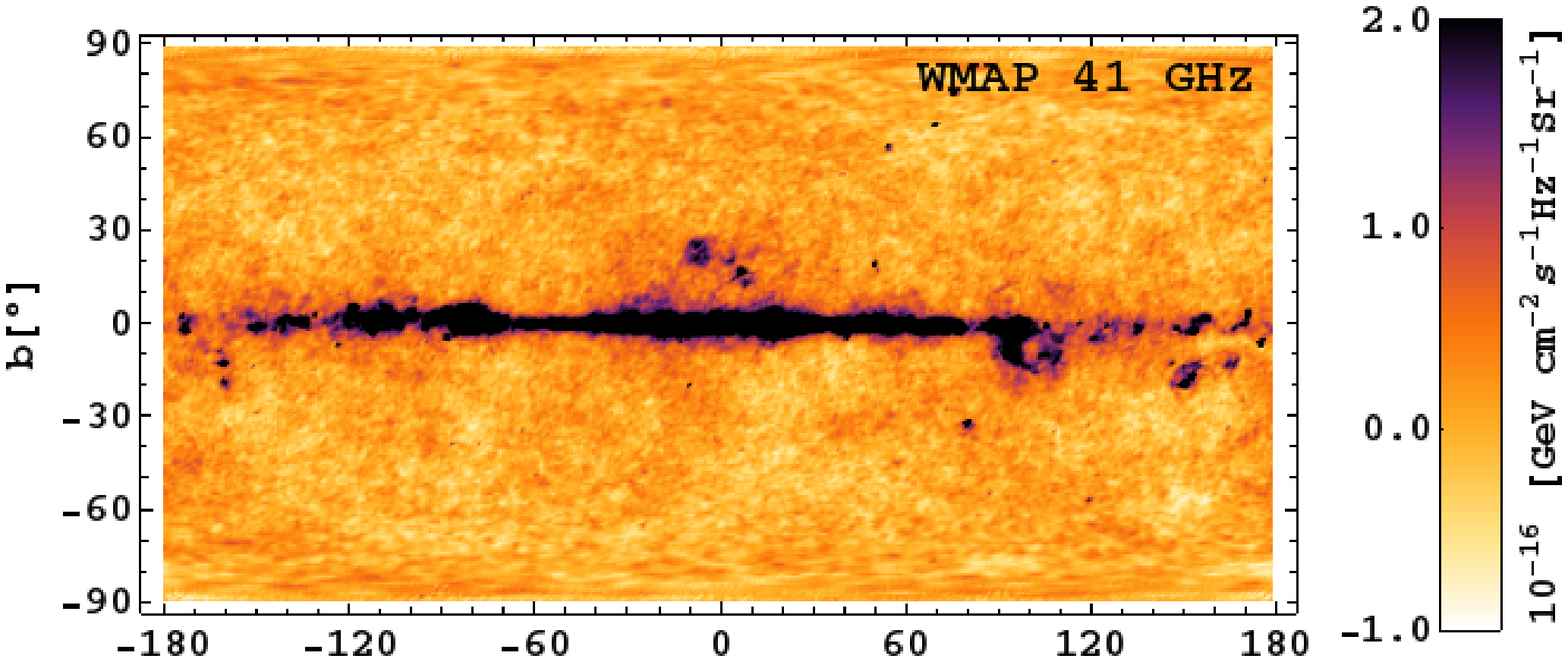}\hfill
\includegraphics[width=5.5cm]{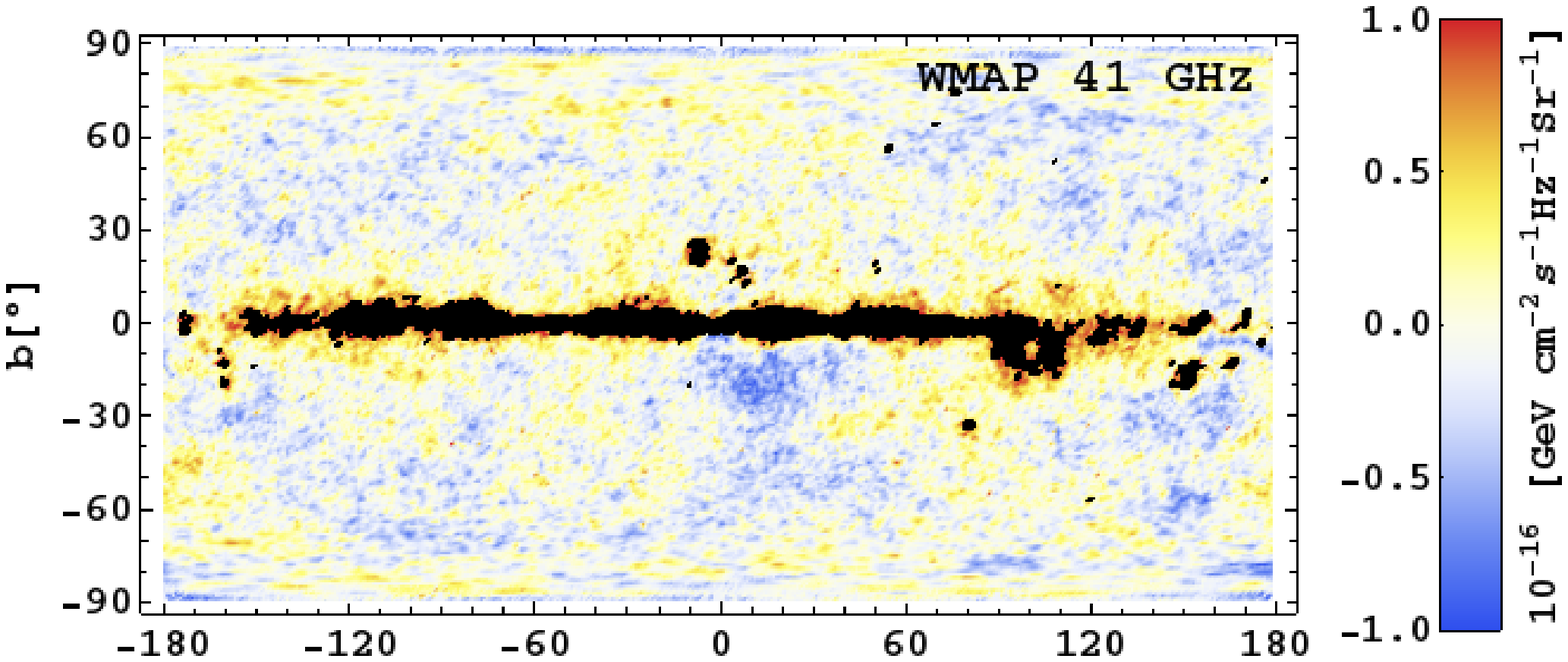}\hfill
\includegraphics[width=5.3cm]{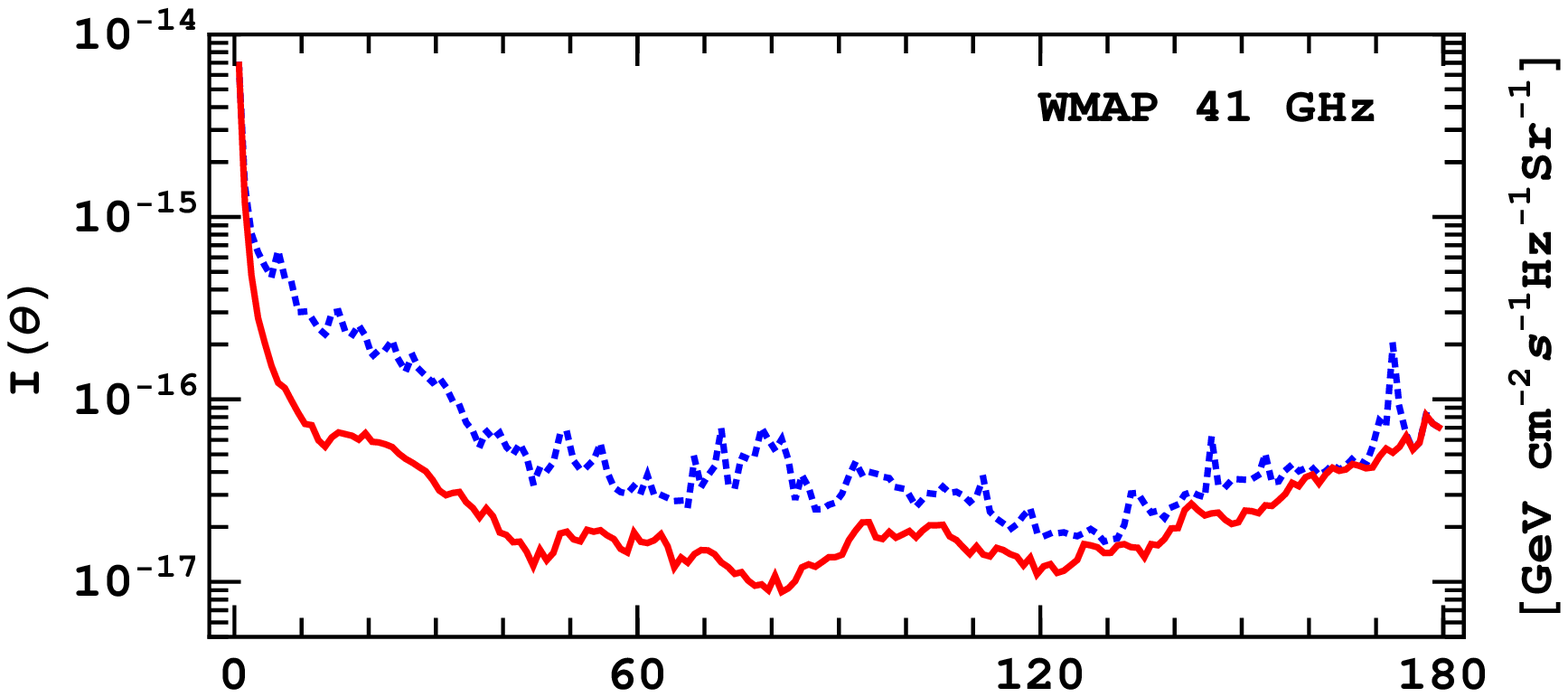}\hfill

\includegraphics[width=5.5cm]{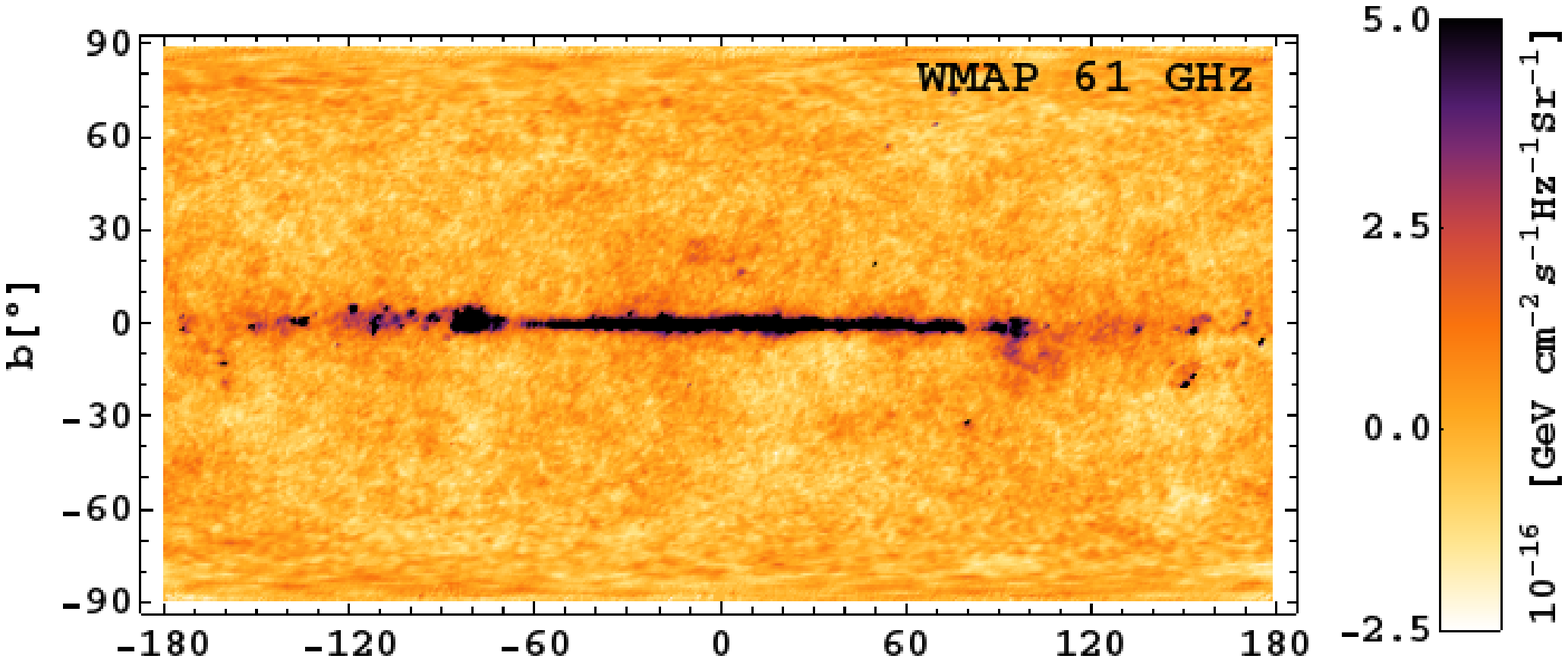}\hfill
\includegraphics[width=5.5cm]{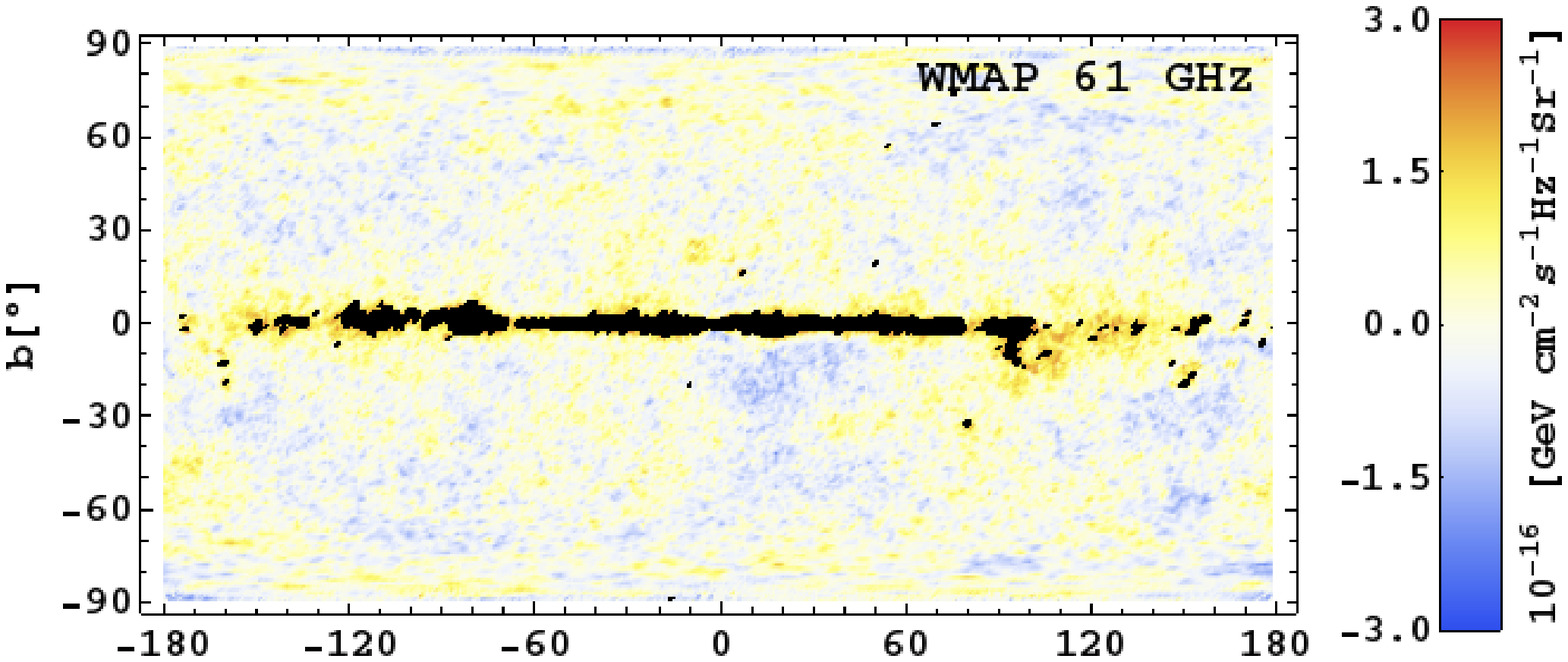}\hfill
\includegraphics[width=5.3cm]{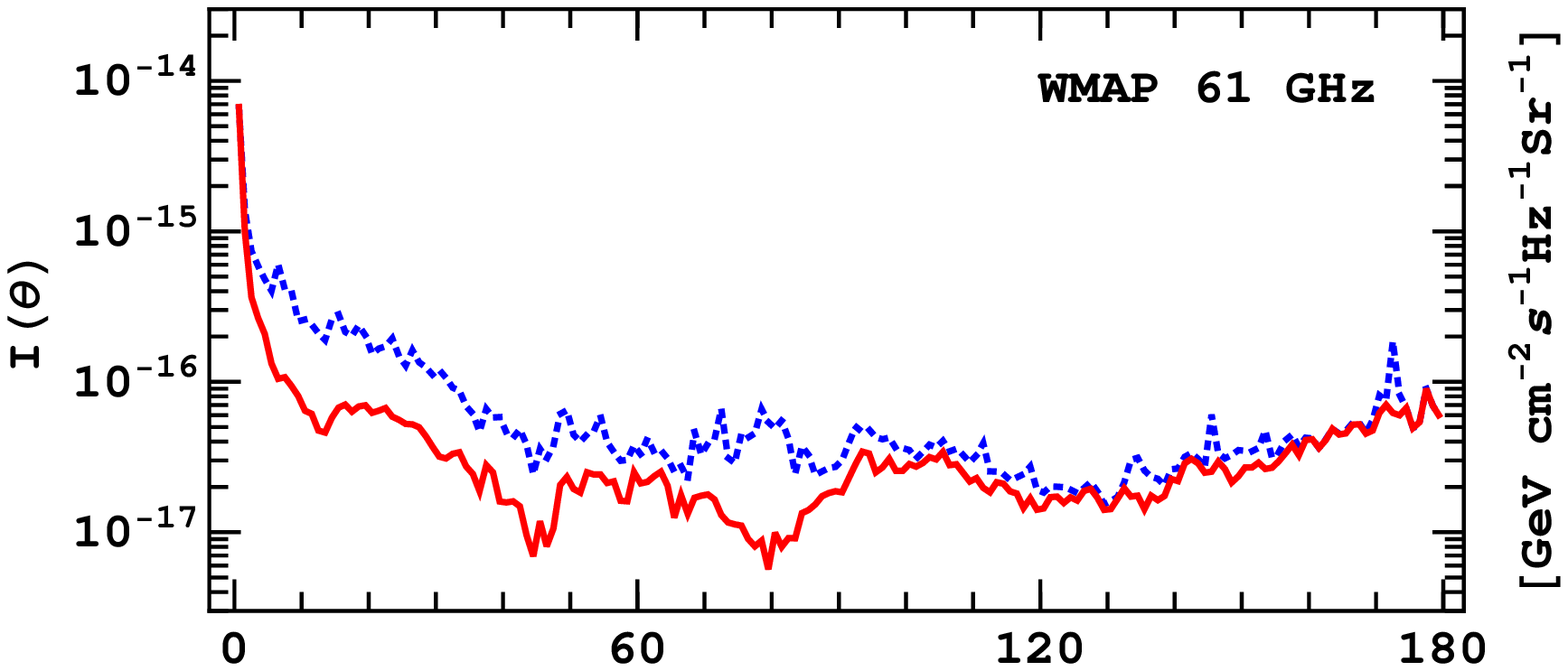}\hfill

\includegraphics[width=5.5cm]{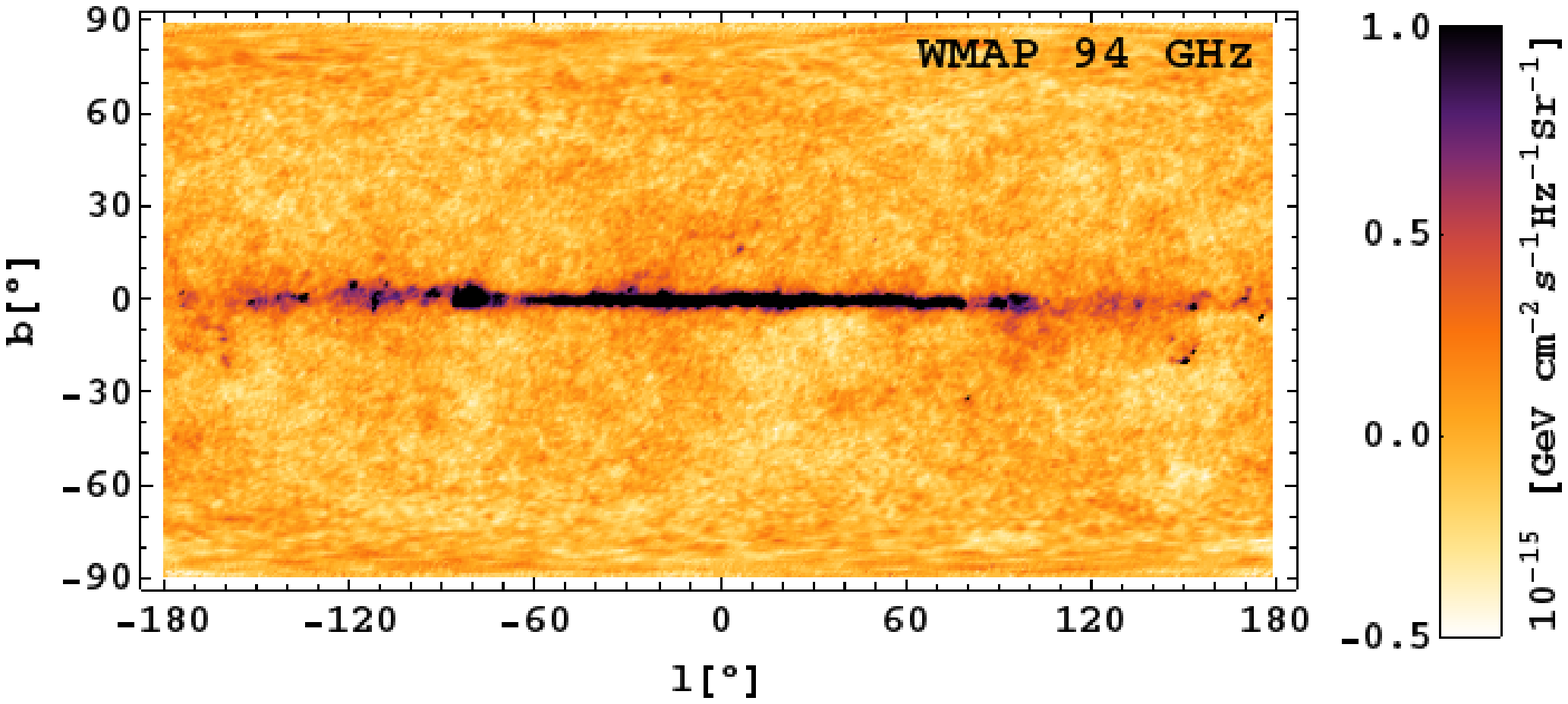}\hfill
\includegraphics[width=5.5cm]{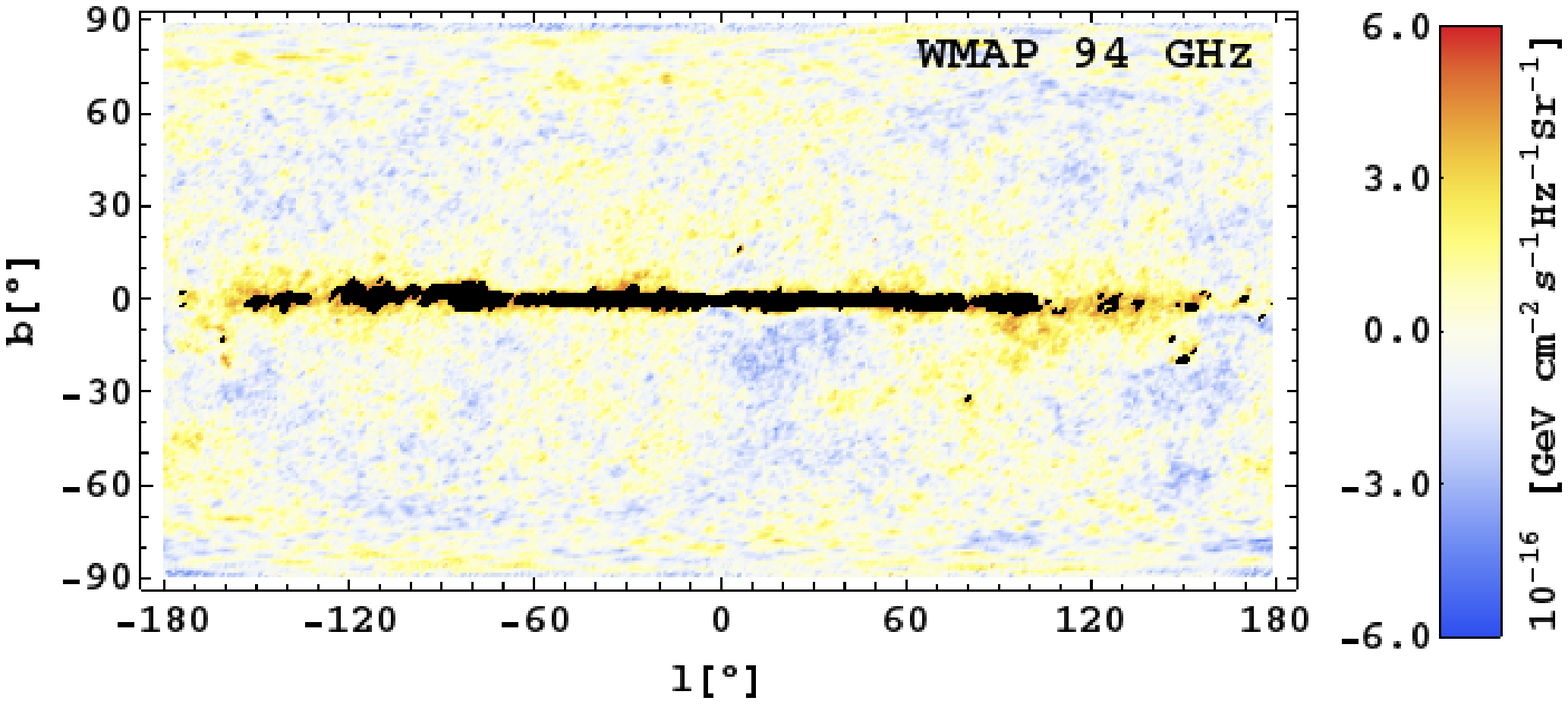}\hfill
\includegraphics[width=5.3cm]{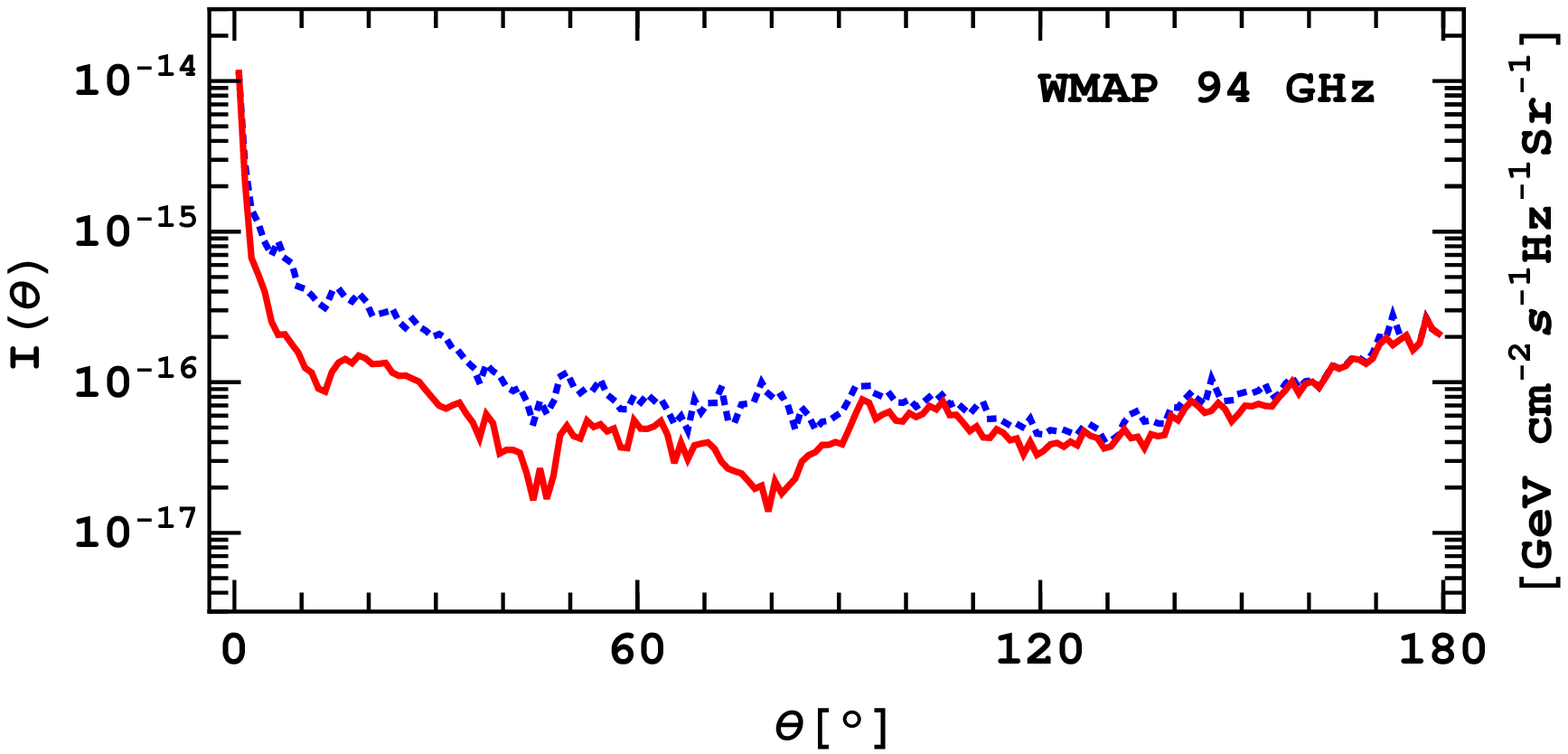}\hfill

\caption
{
Haslam and WMAP intensity maps $I(l,b)$ in Galactic coordinates (left), masked residual maps $I(l,b)-I(\theta)$ (middle), and spherically-averaged intensities (right).
Dotted blue lines represent the original mean intensity $I_0 (\theta)$, while solid red lines correspond to the final intensity $I(\theta)$ after discarding the outliers (black areas in the masked residual maps).
} 
\label{figHaslam&WMAPMap}
\end{figure*}
%__________________________________

%__________________________________
\begin{figure*}

\includegraphics[width=5.5cm]{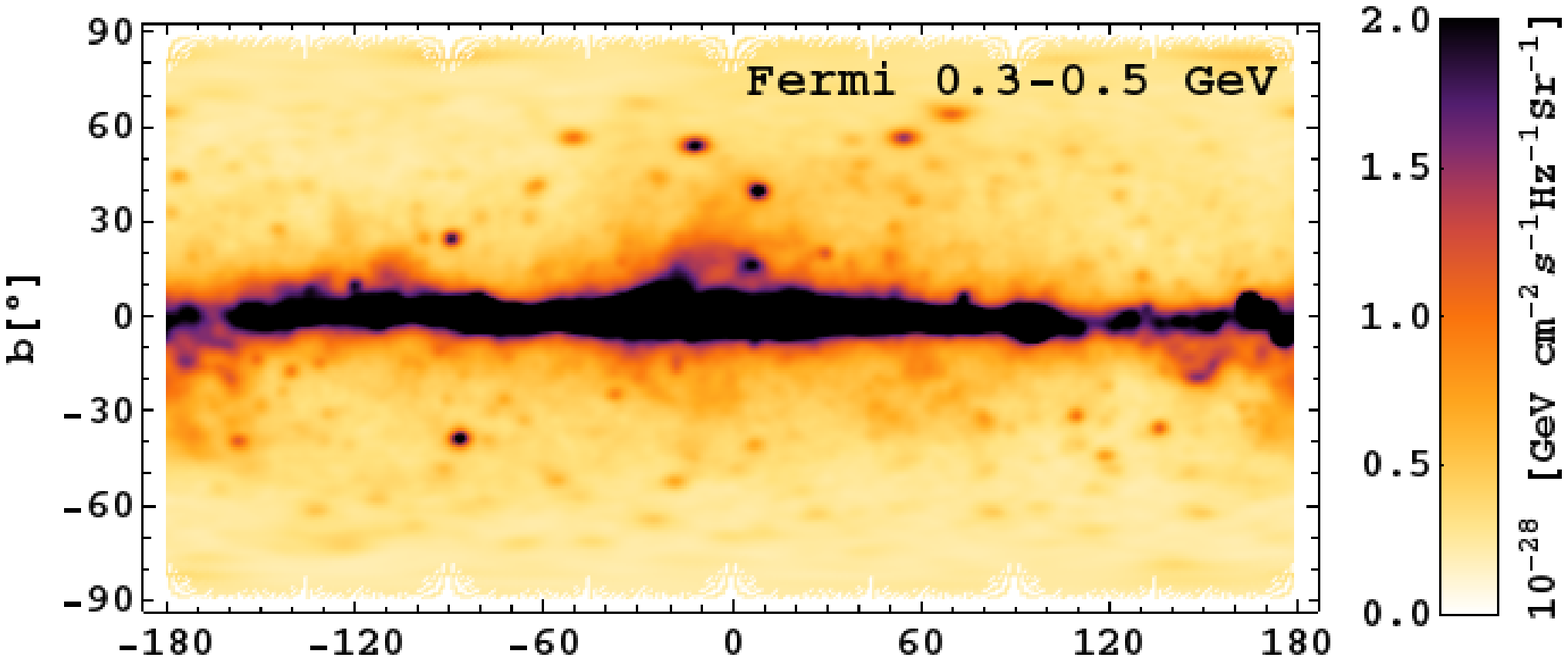}\hfill
\includegraphics[width=5.5cm]{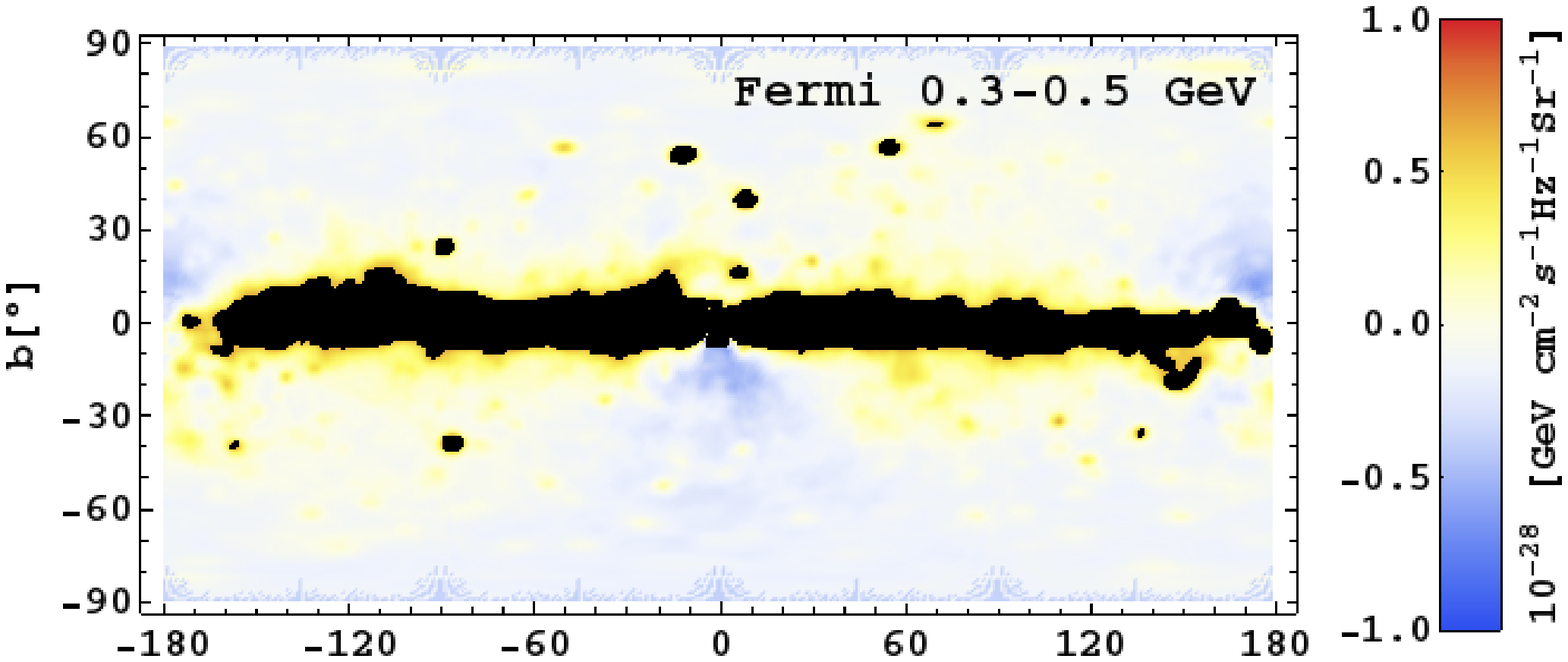}\hfill
\includegraphics[width=5.3cm]{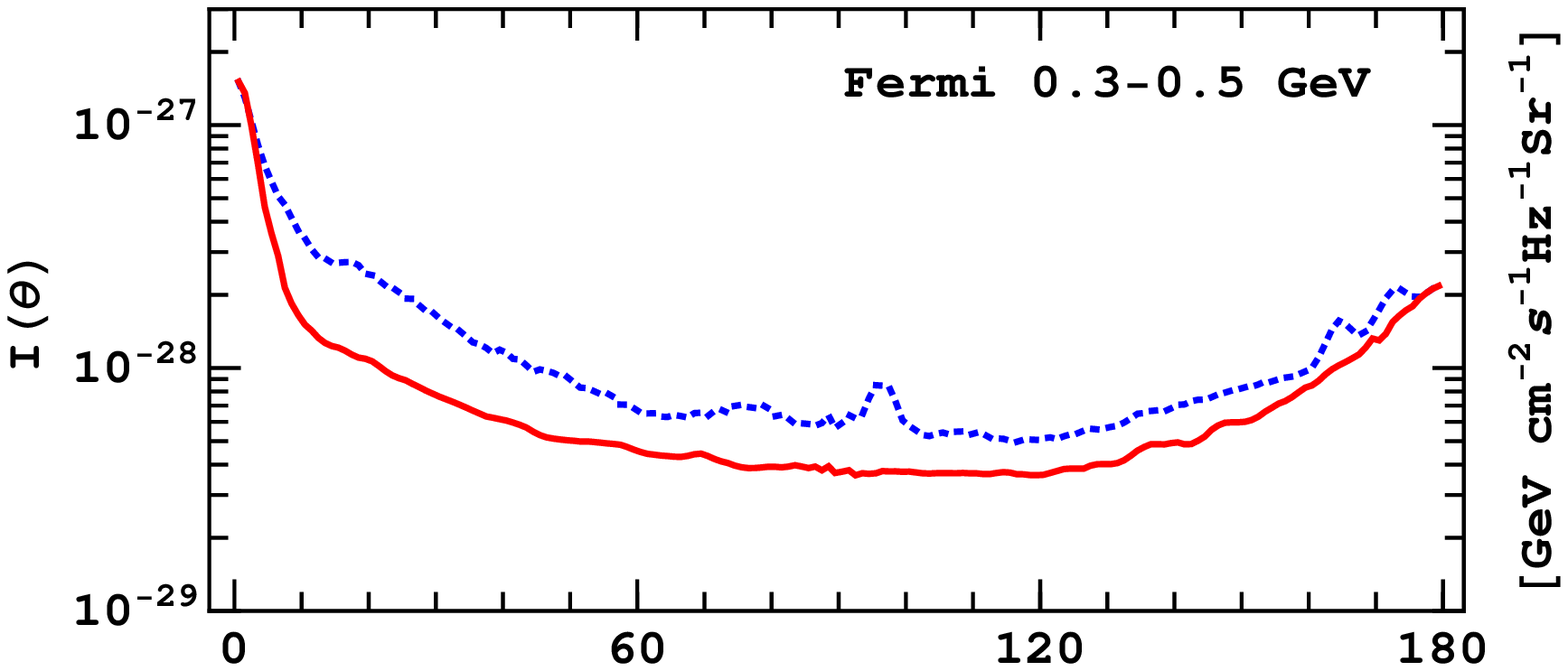}\hfill

\includegraphics[width=5.5cm]{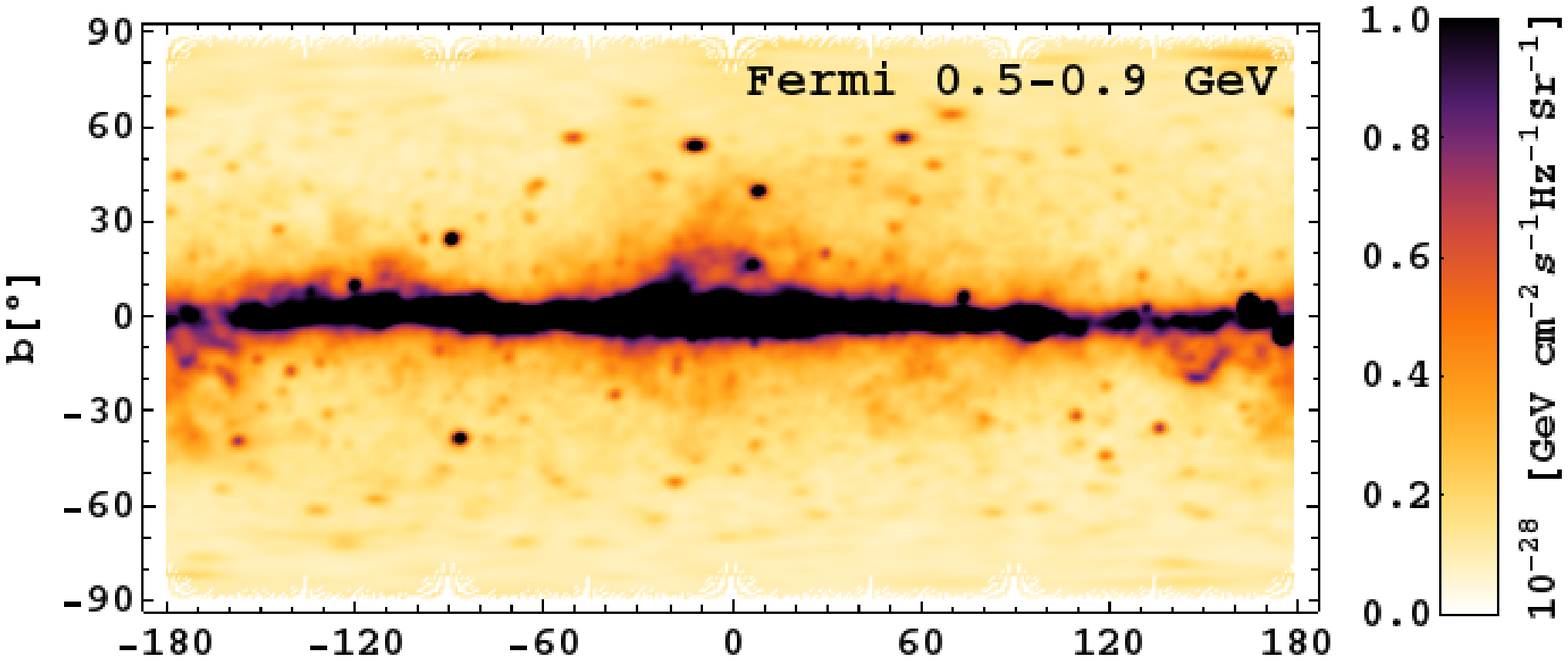}\hfill
\includegraphics[width=5.5cm]{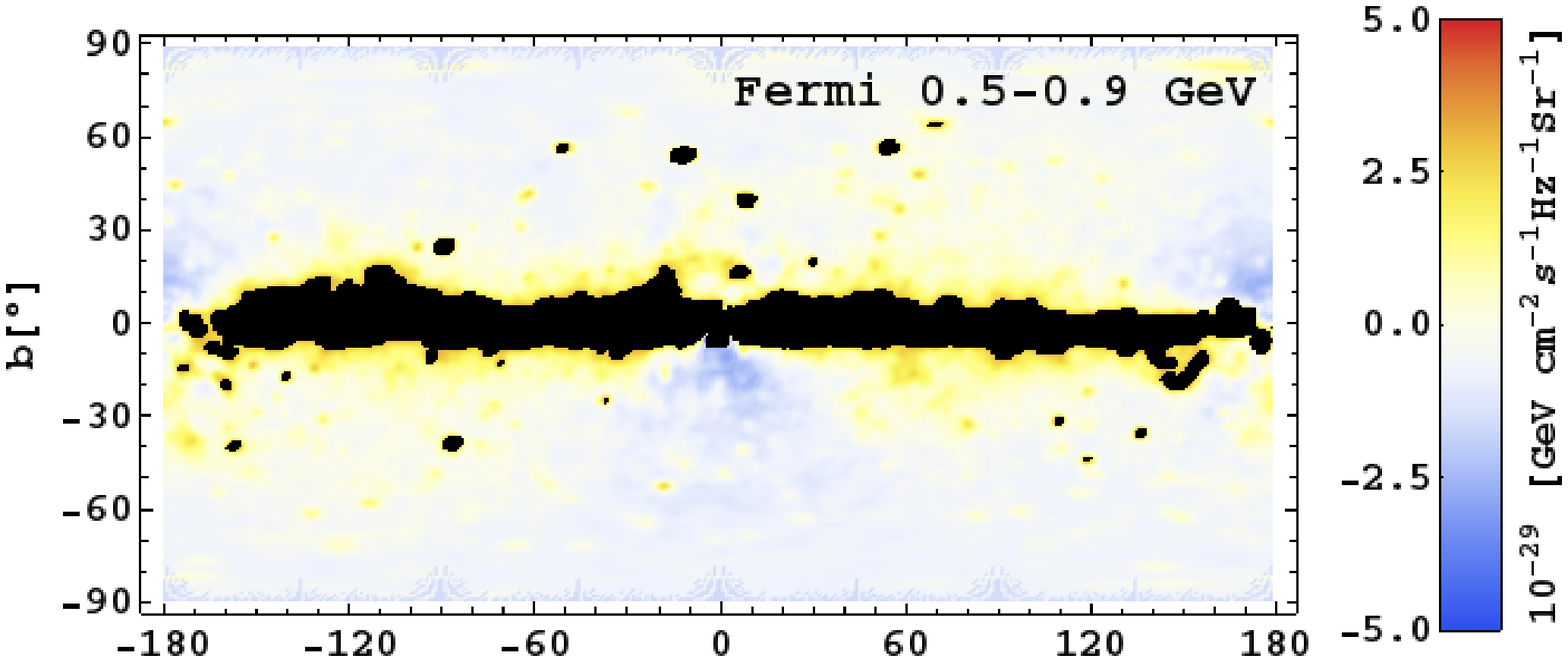}\hfill
\includegraphics[width=5.3cm]{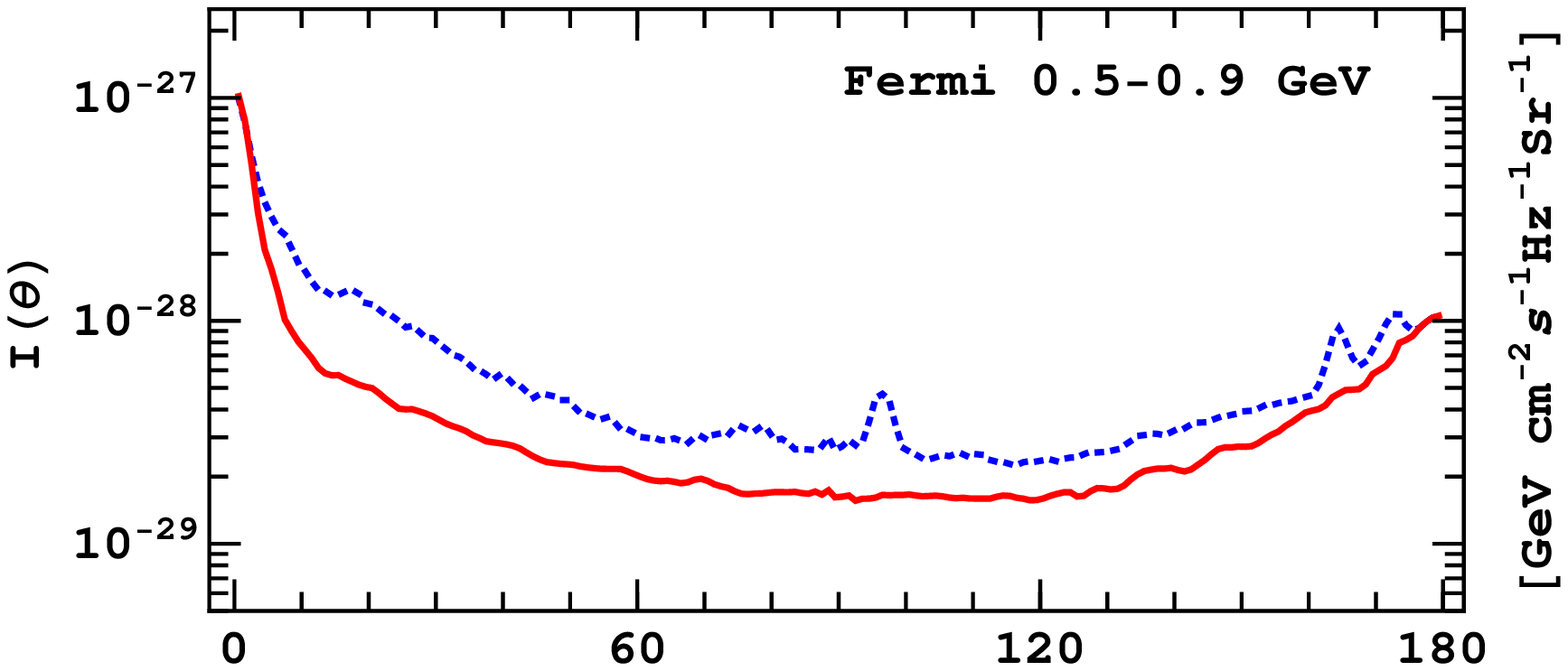}\hfill

\includegraphics[width=5.5cm]{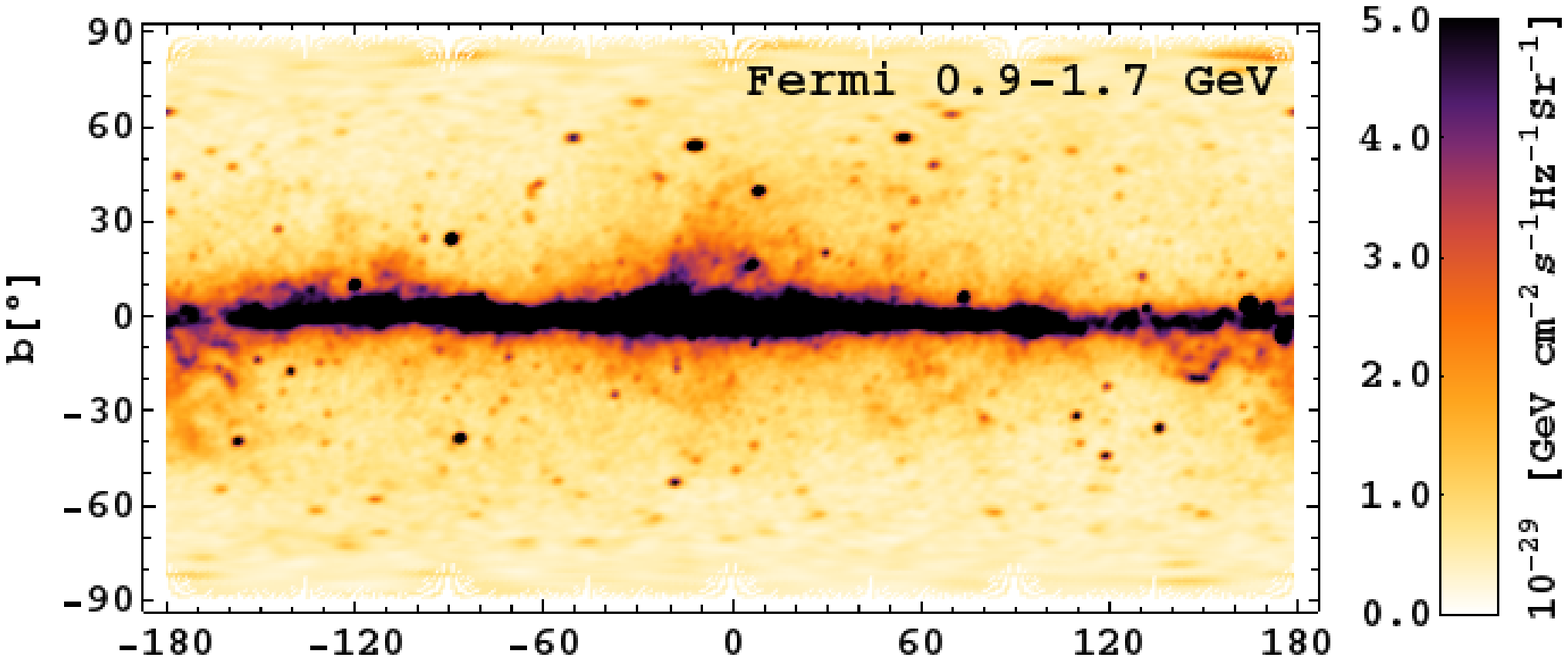}\hfill
\includegraphics[width=5.5cm]{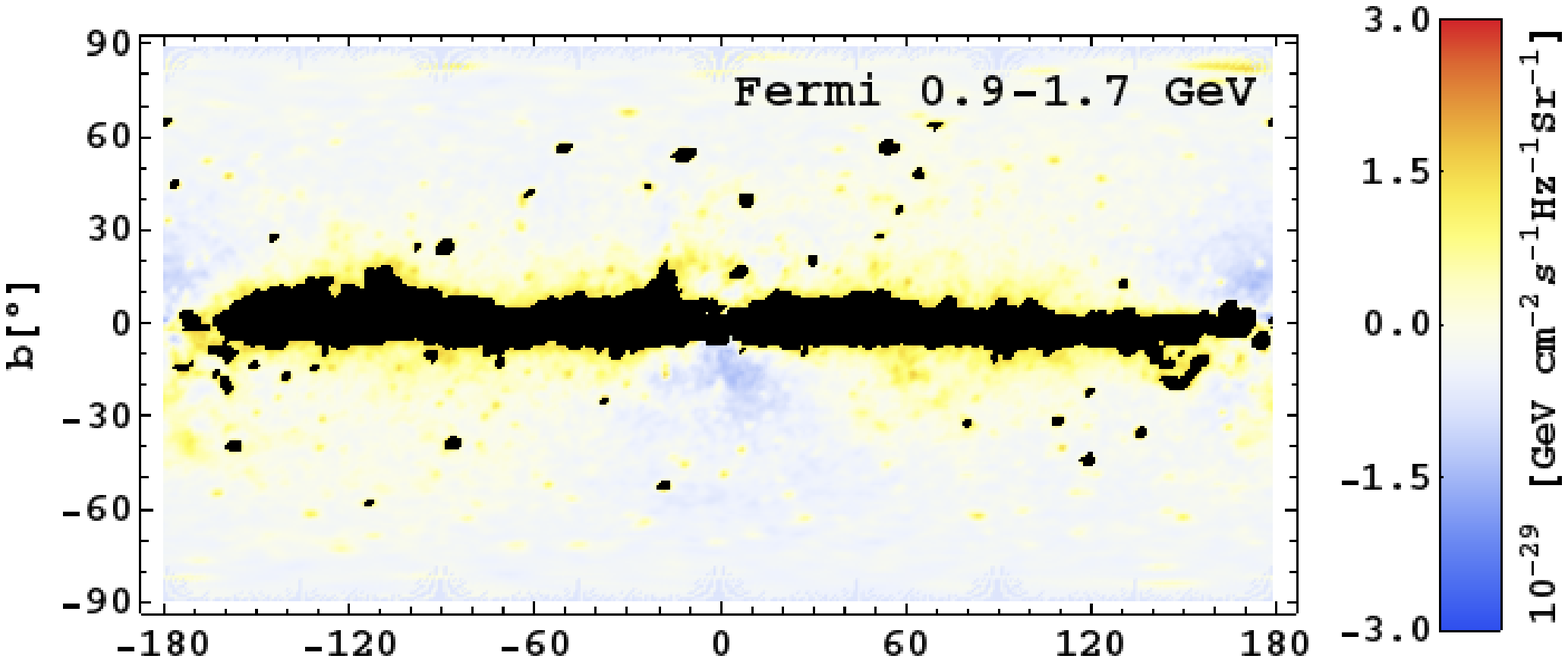}\hfill
\includegraphics[width=5.3cm]{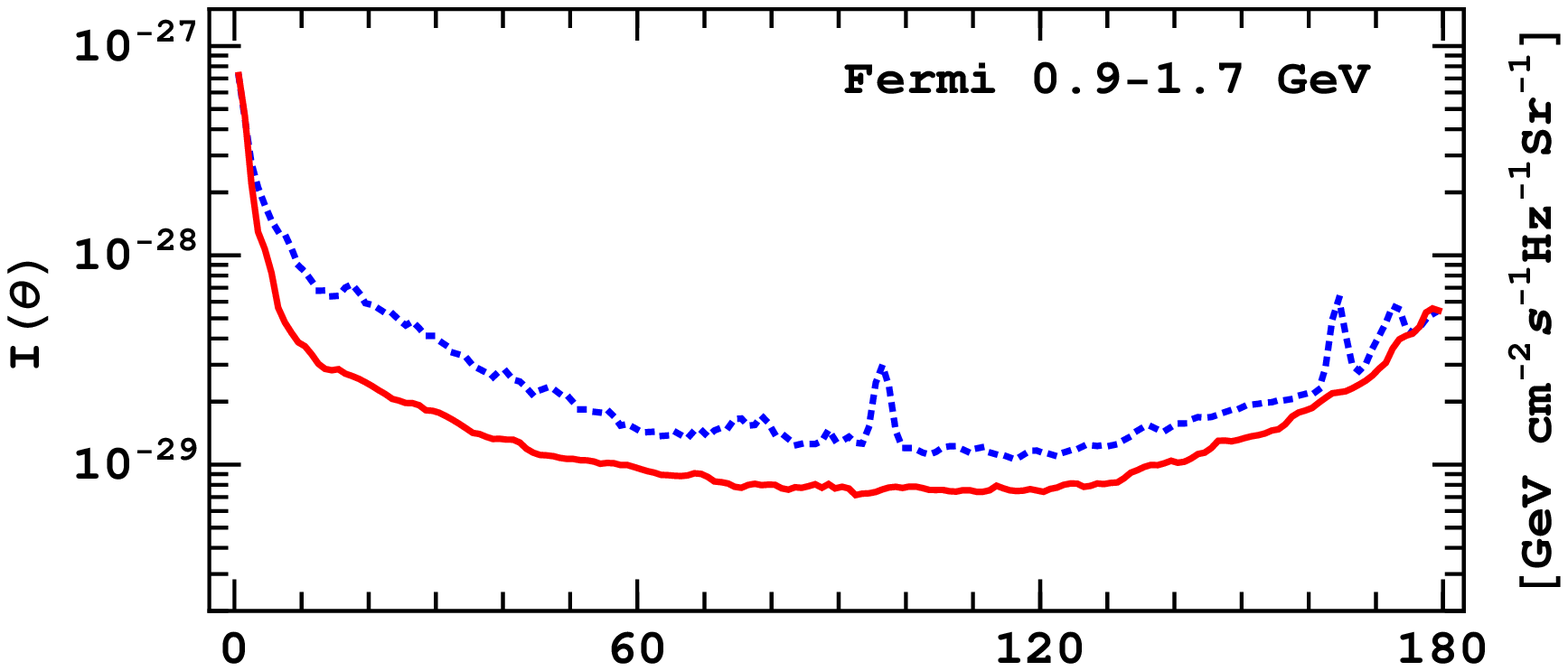}\hfill

\includegraphics[width=5.5cm]{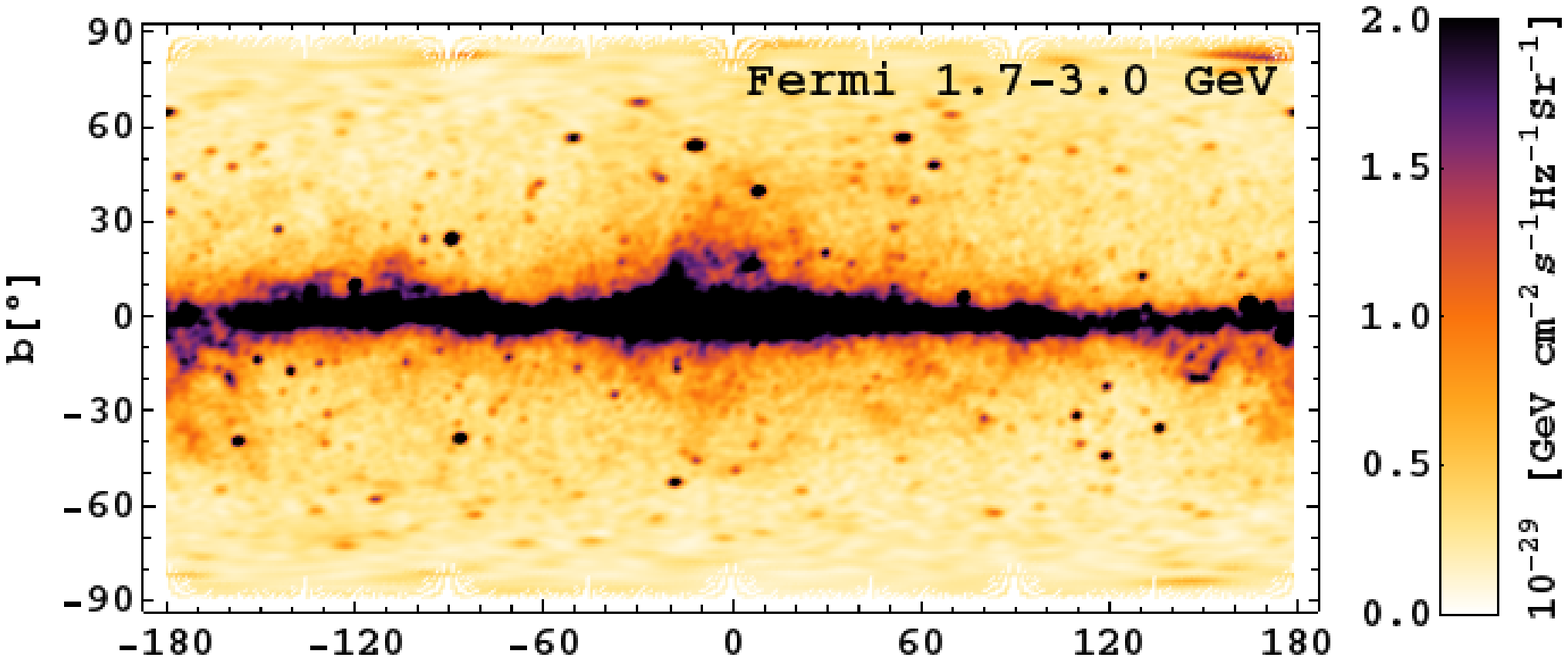}\hfill
\includegraphics[width=5.5cm]{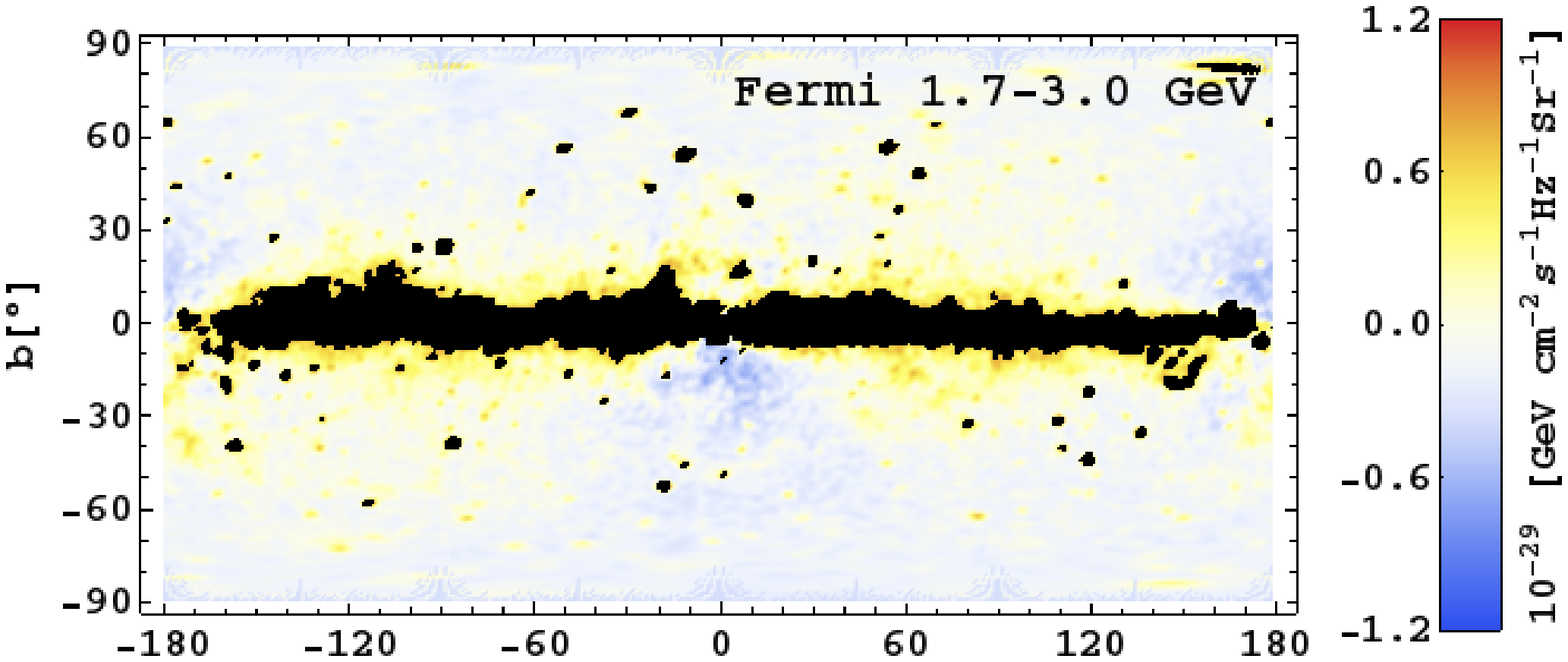}\hfill
\includegraphics[width=5.3cm]{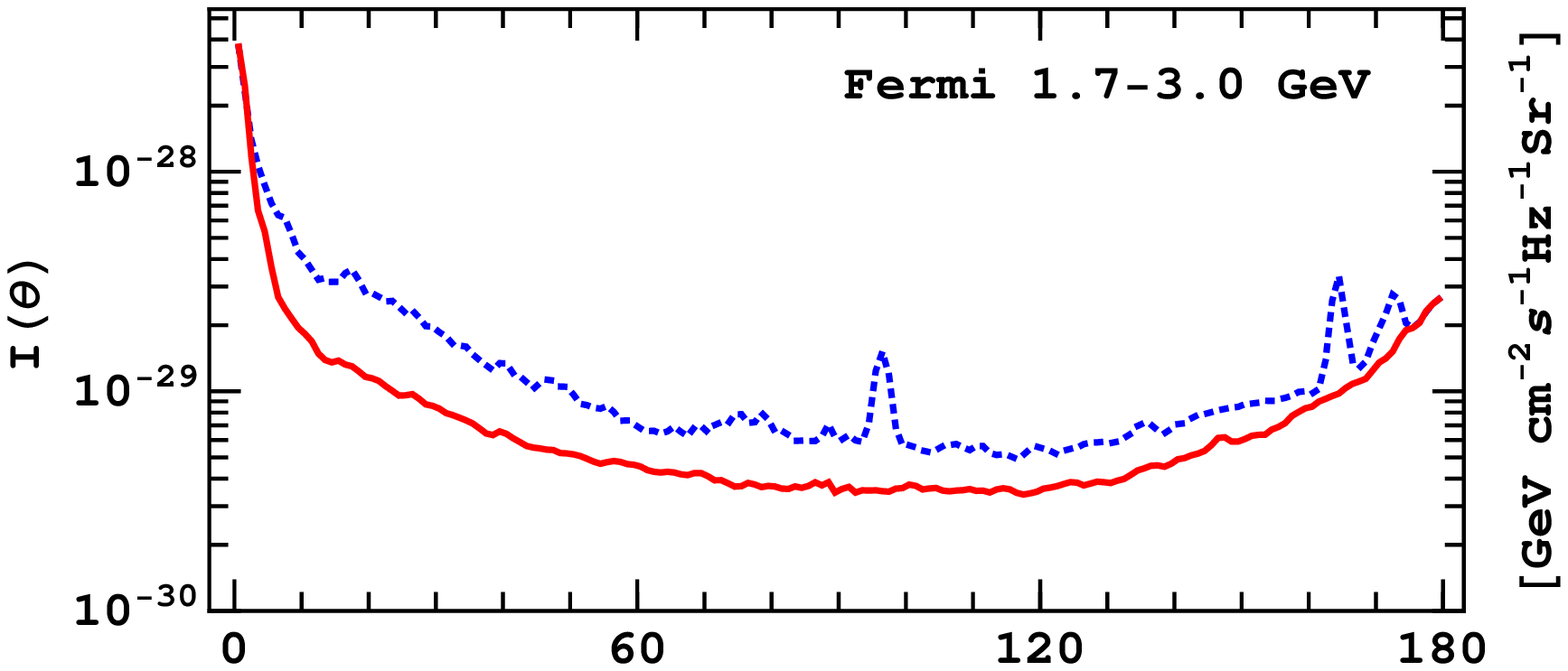}\hfill

\includegraphics[width=5.5cm]{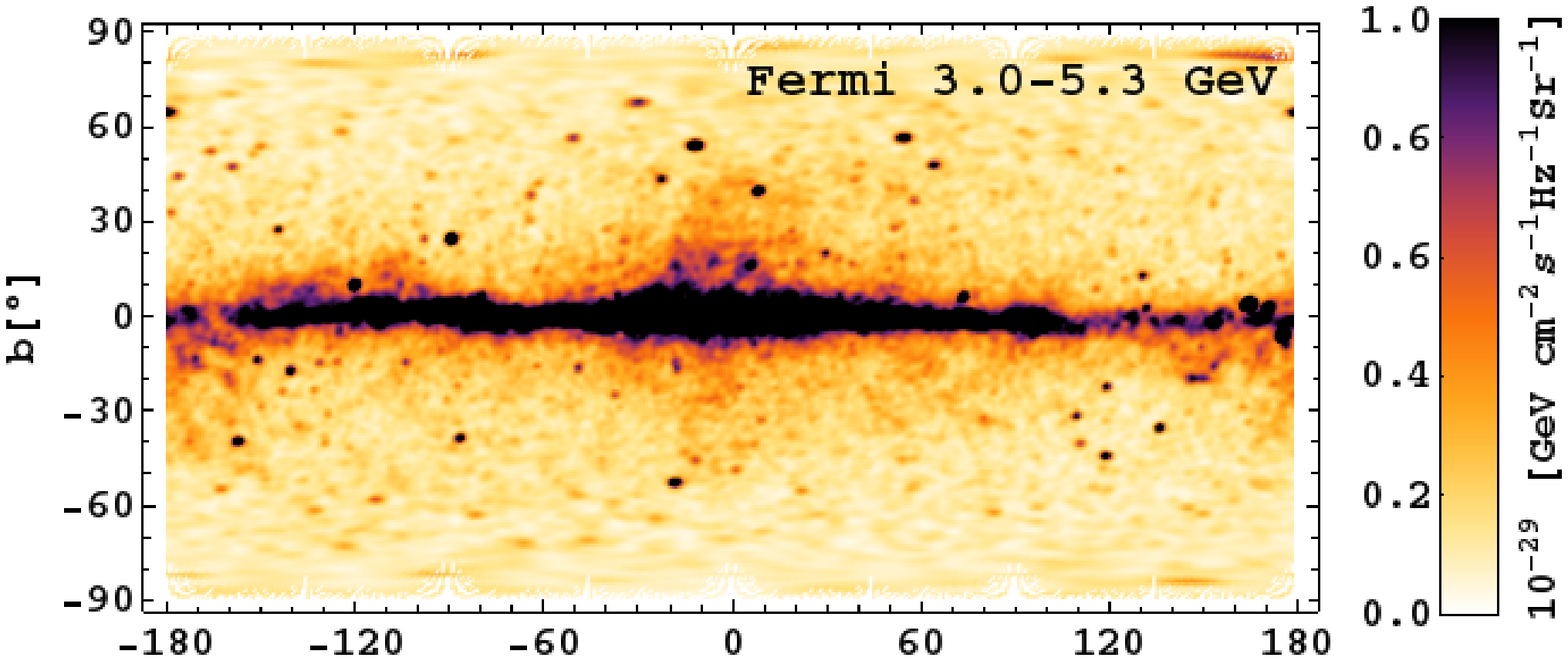}\hfill
\includegraphics[width=5.5cm]{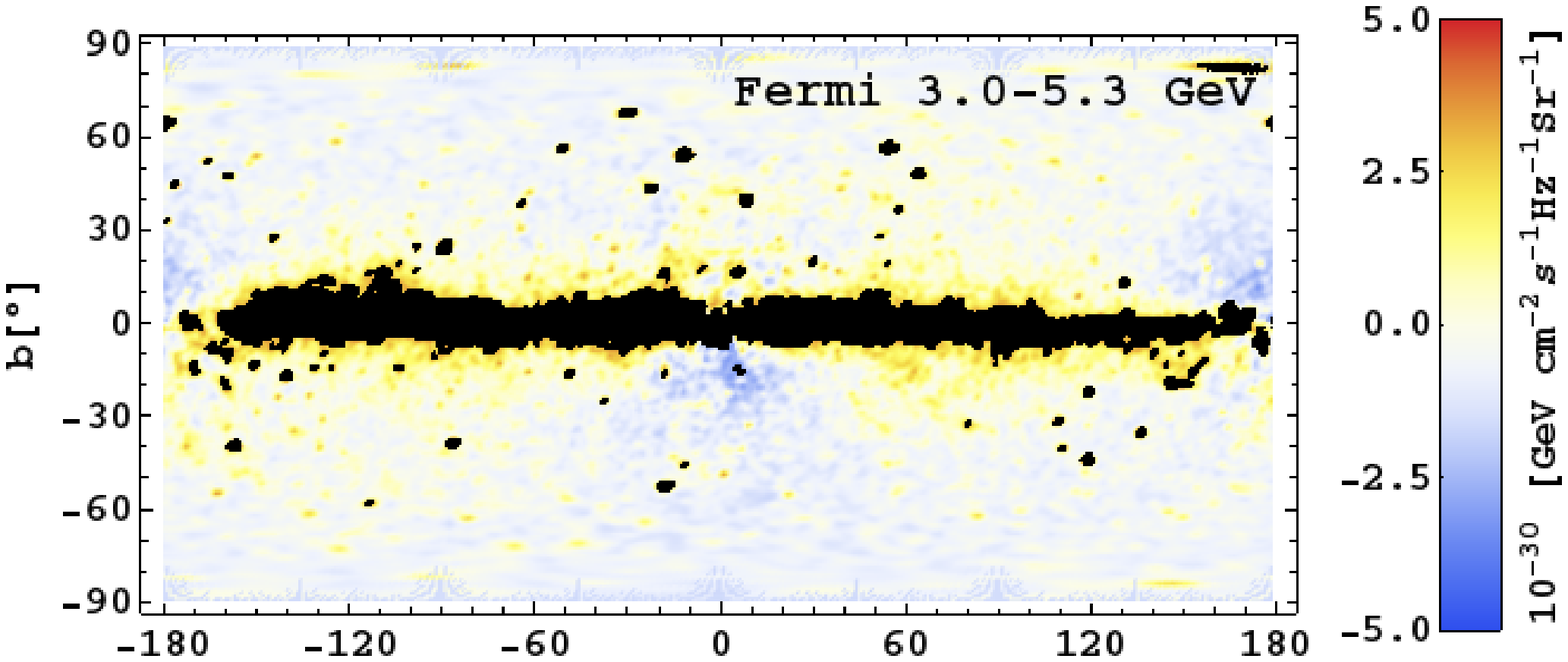}\hfill
\includegraphics[width=5.3cm]{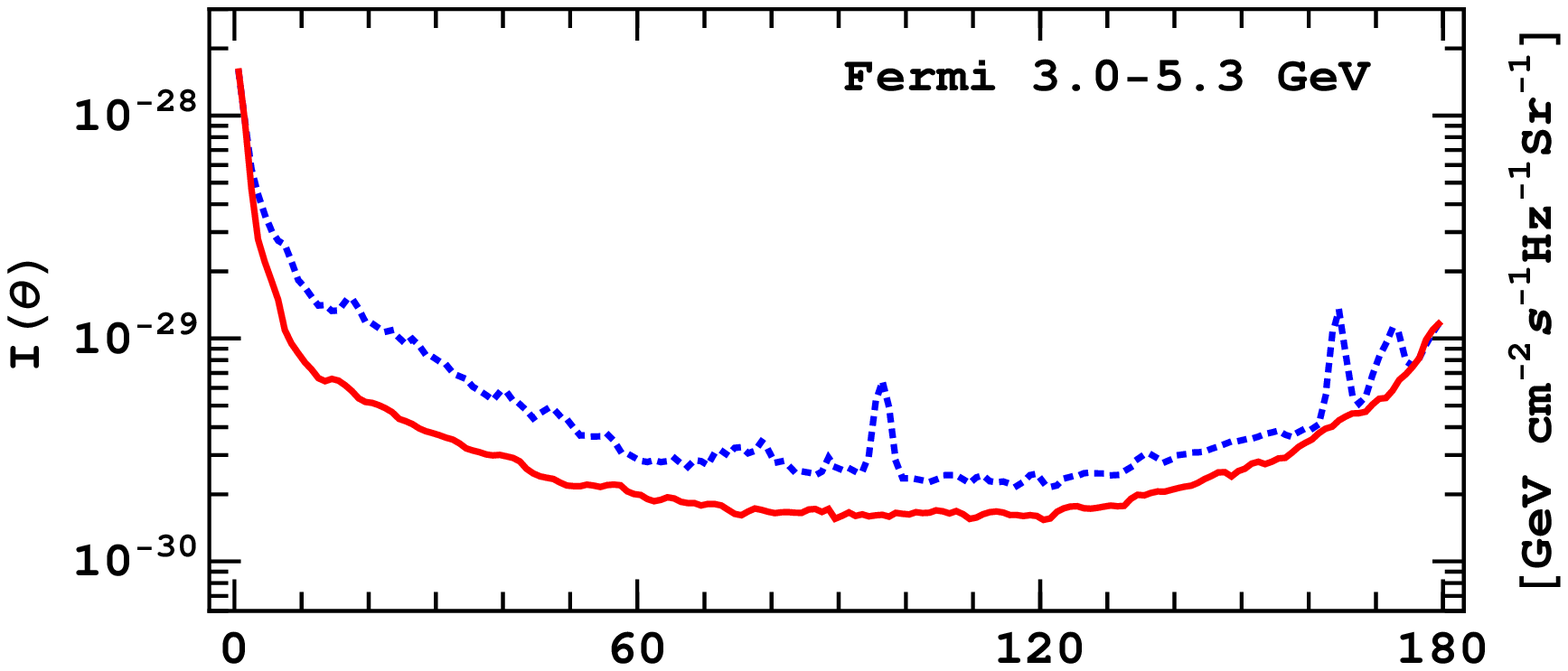}\hfill

\includegraphics[width=5.5cm]{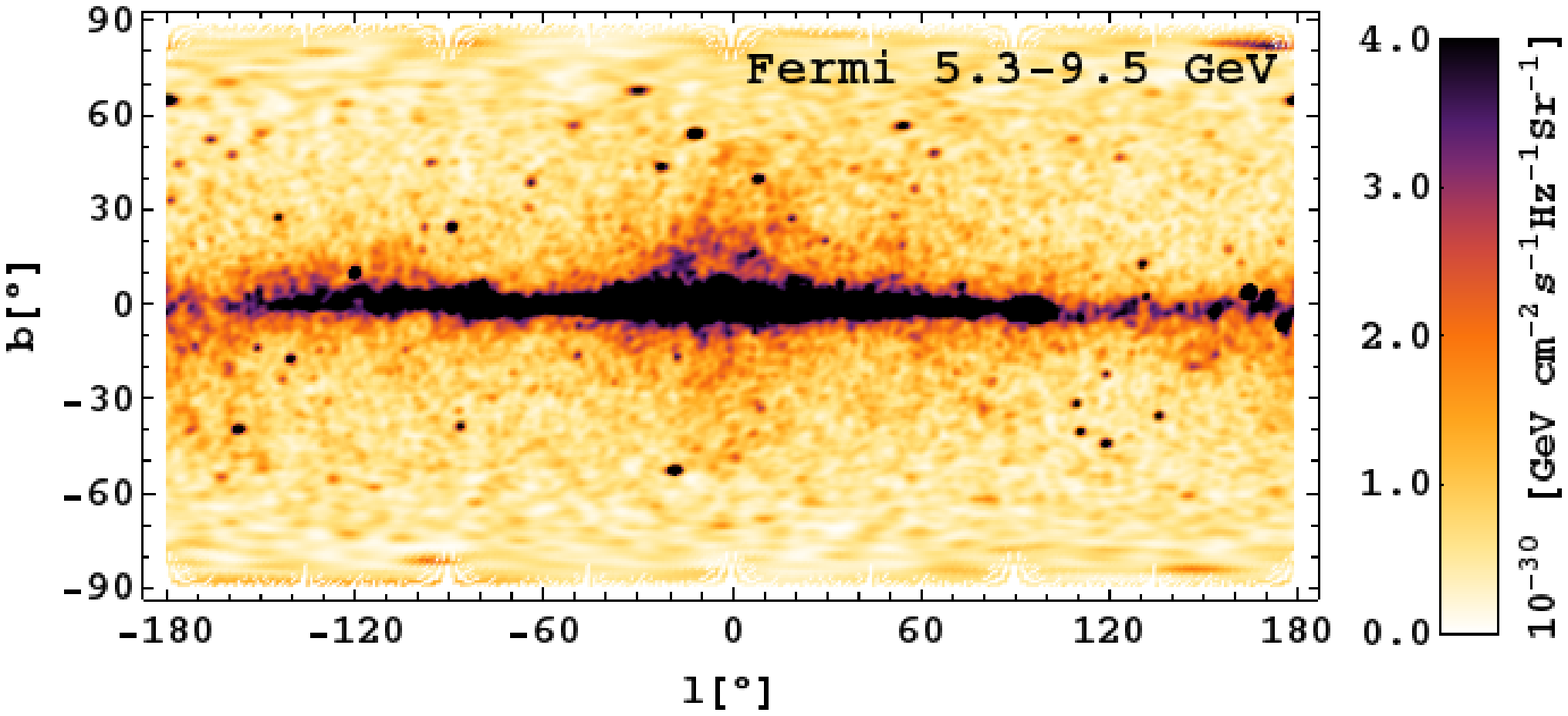}\hfill
\includegraphics[width=5.5cm]{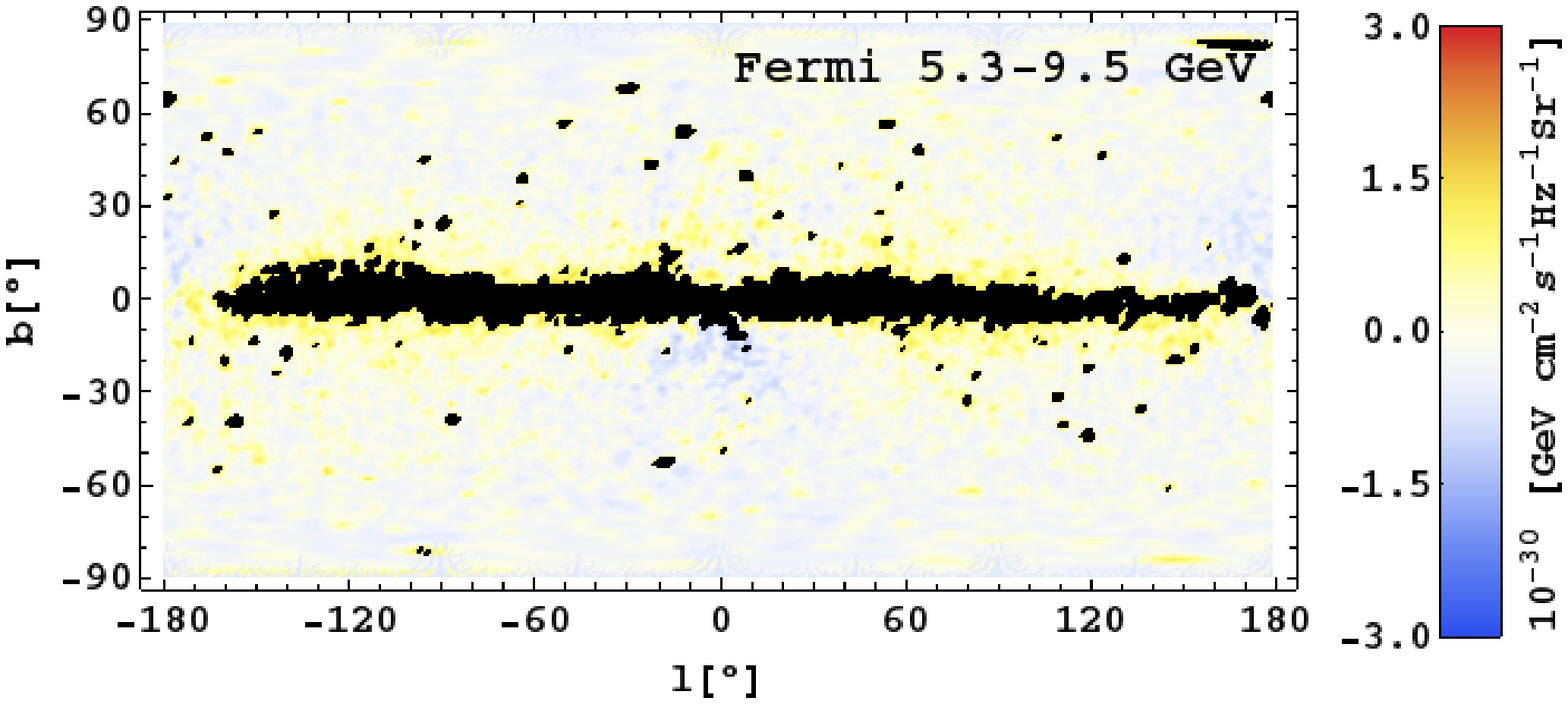}\hfill
\includegraphics[width=5.3cm]{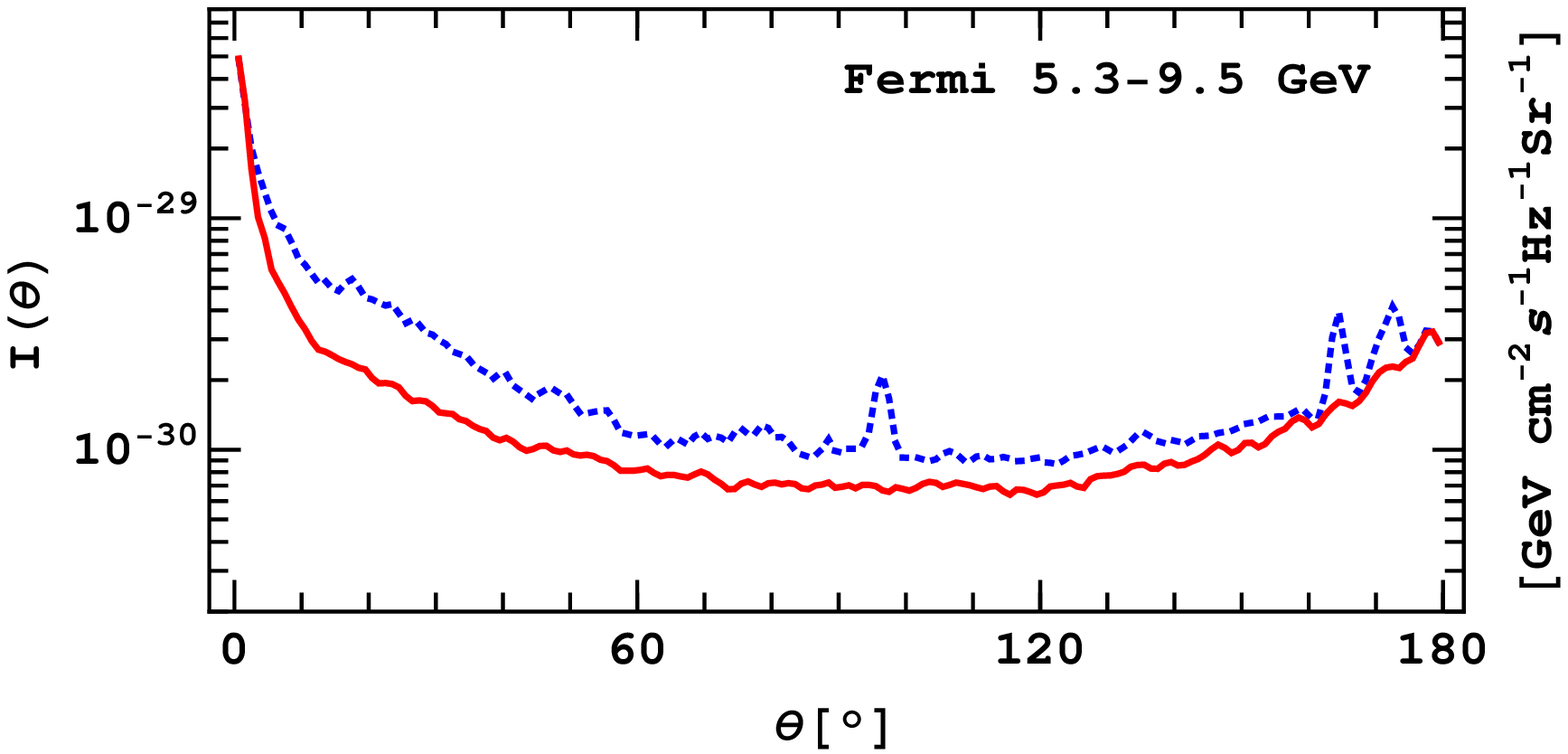}\hfill

\caption
{
Fermi intensity maps $I(l,b)$ in Galactic coordinates (left), masked residual maps $I(l,b)-I(\theta)$ (middle), and spherically-averaged intensities (right).
Dotted blue lines represent the original mean intensity $I_0 (\theta)$, while solid red lines correspond to the final intensity $I(\theta)$ after discarding the outliers (black areas in the masked residual maps).
}
\label{figFermiMap1}
\end{figure*}
%__________________________________

%__________________________________
\begin{figure*}

\includegraphics[width=5.5cm]{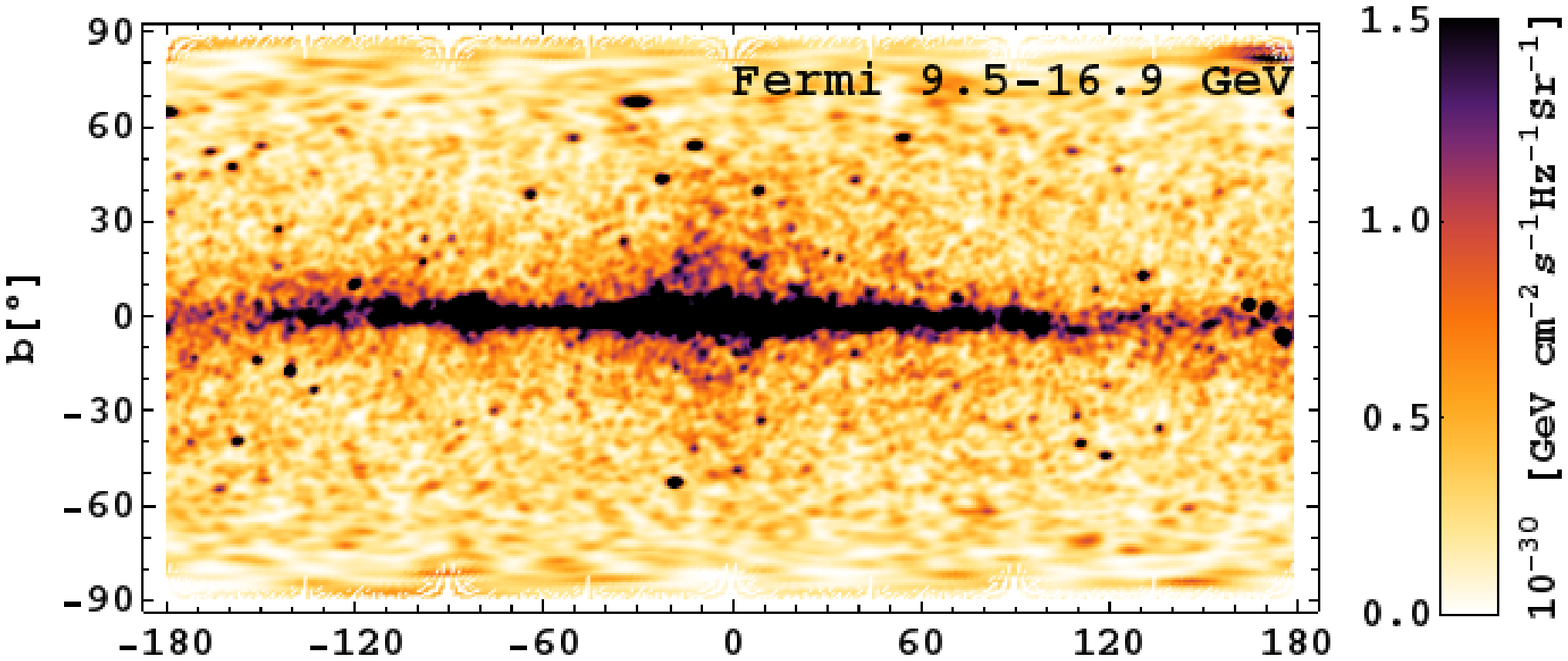}\hfill
\includegraphics[width=5.5cm]{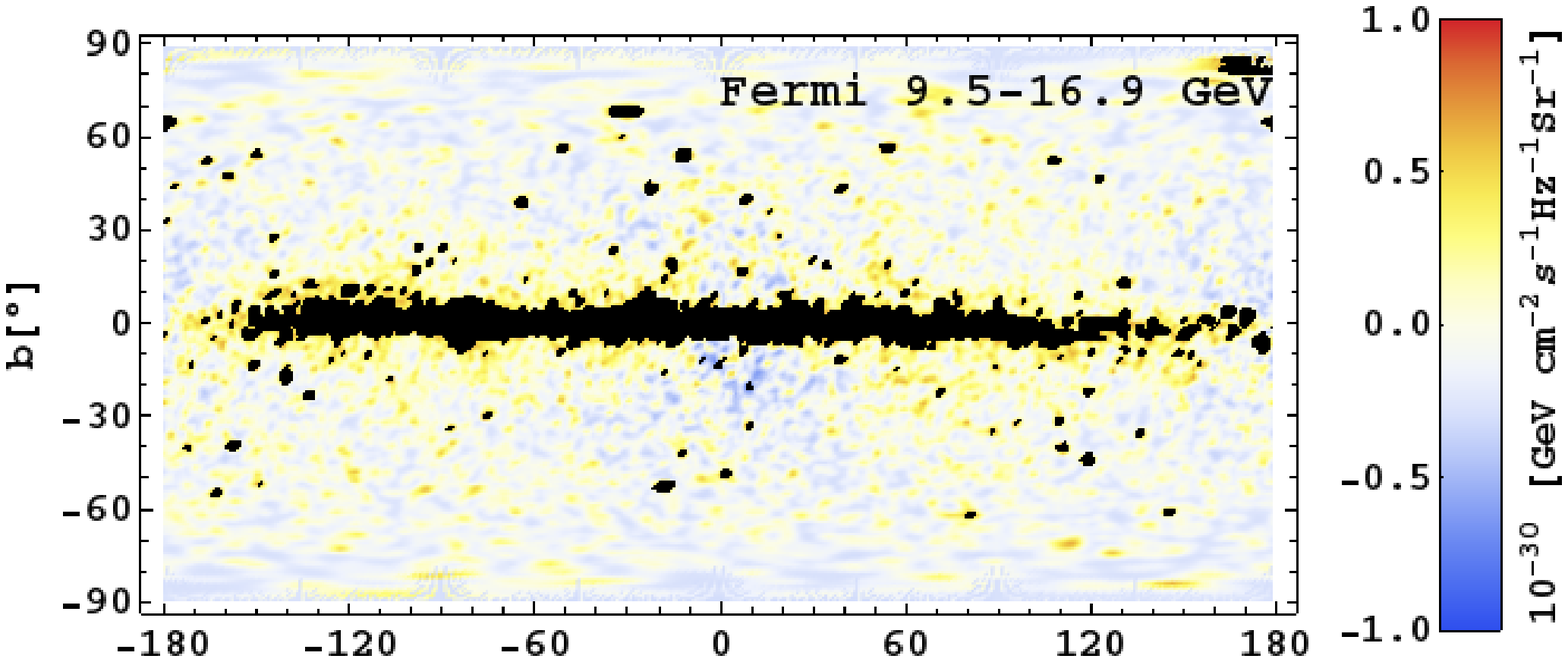}\hfill
\includegraphics[width=5.3cm]{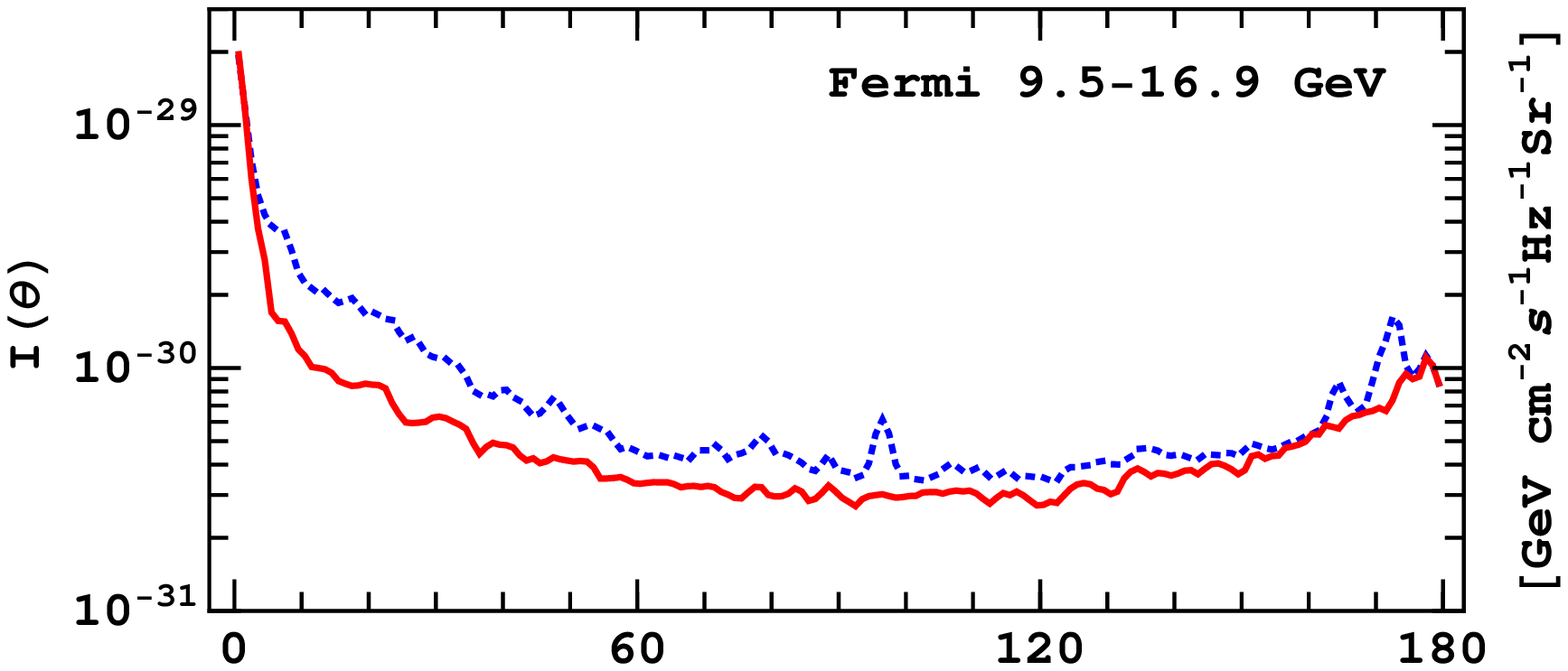}\hfill

\includegraphics[width=5.5cm]{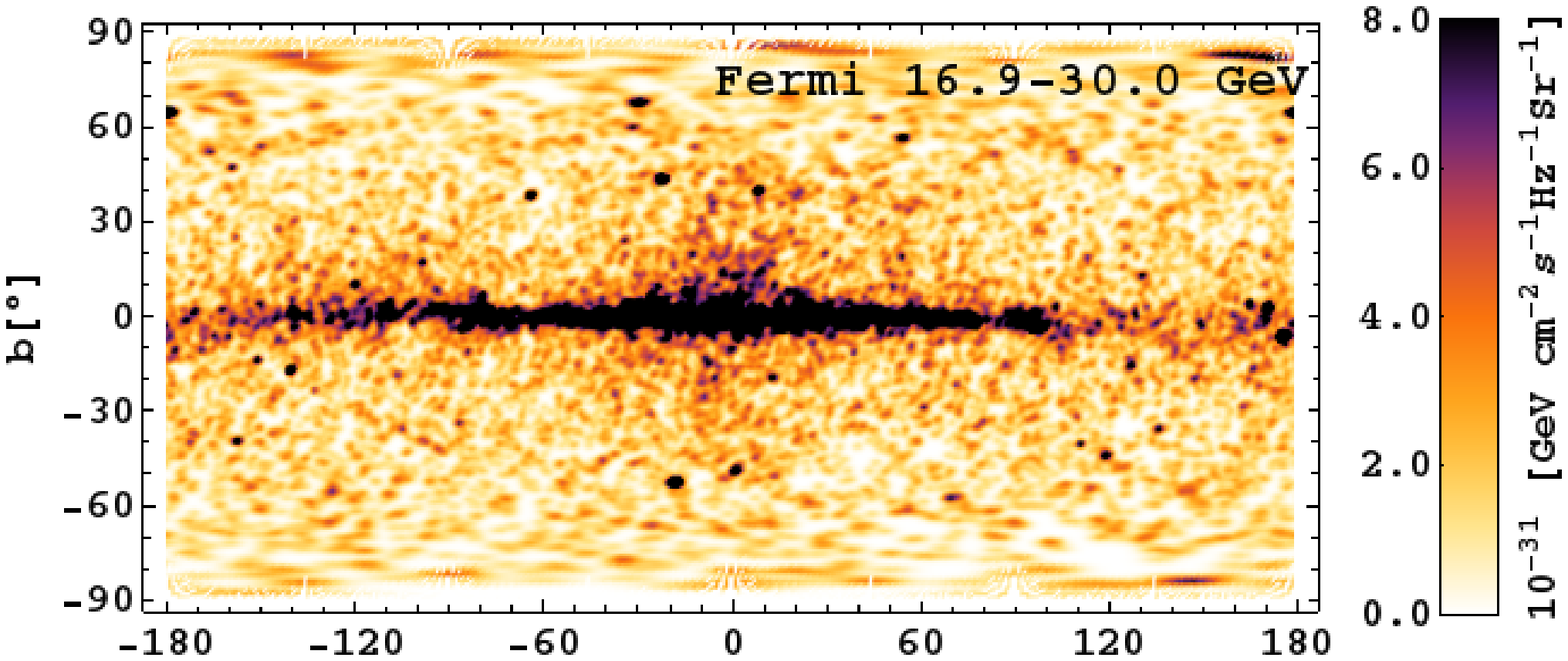}\hfill
\includegraphics[width=5.5cm]{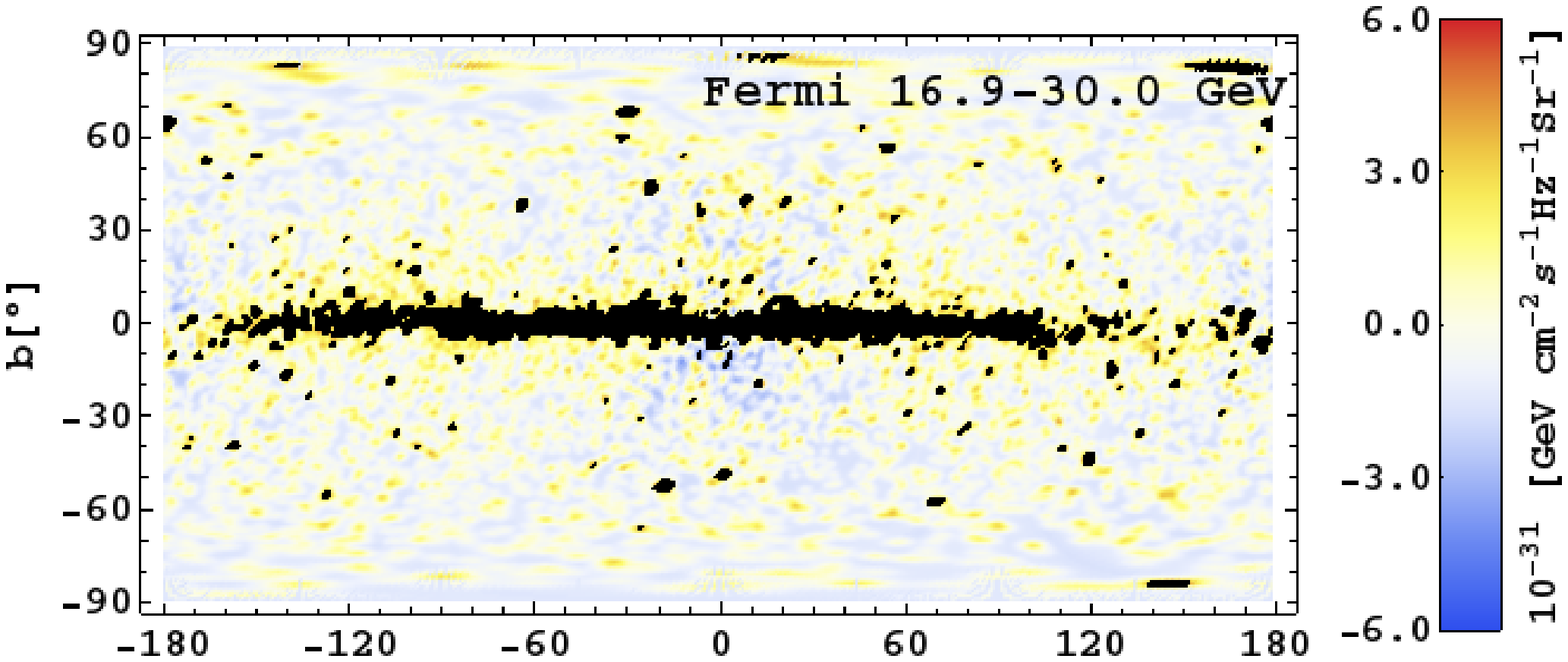}\hfill
\includegraphics[width=5.3cm]{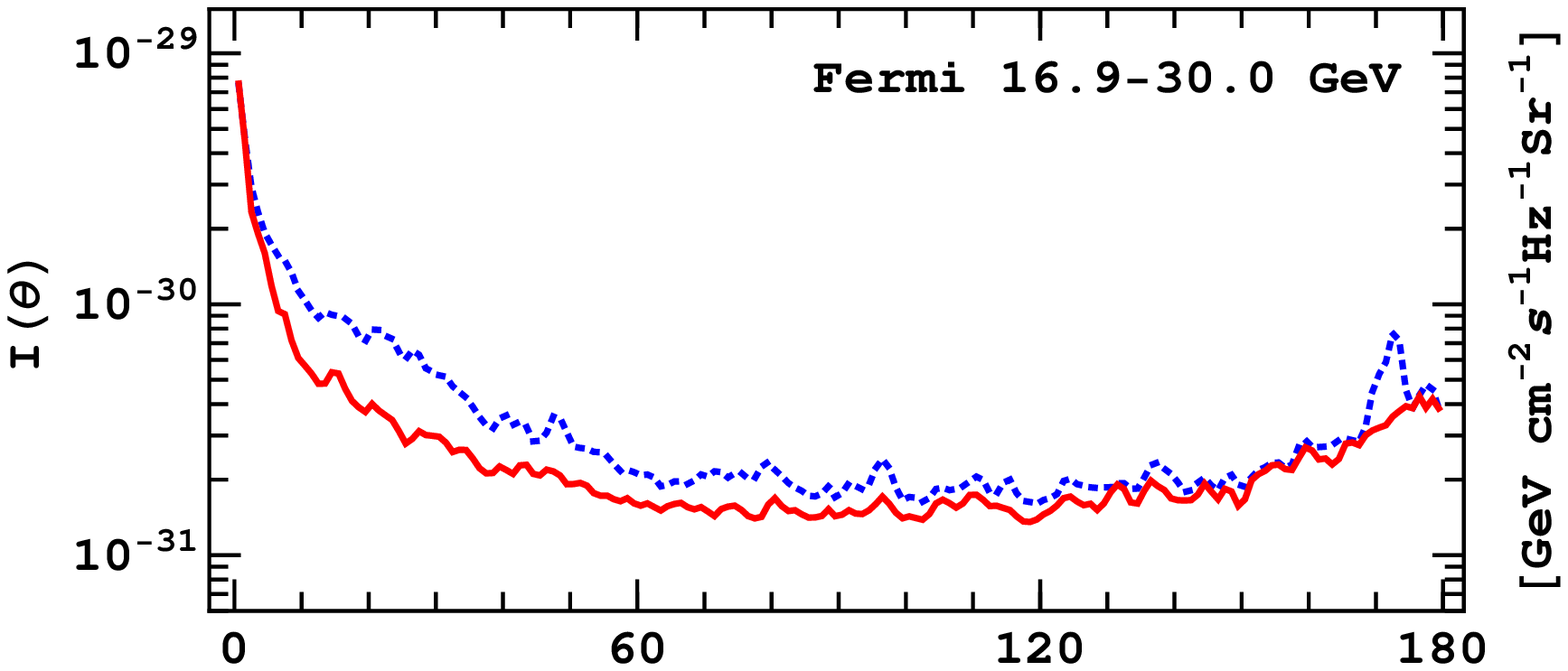}\hfill

\includegraphics[width=5.5cm]{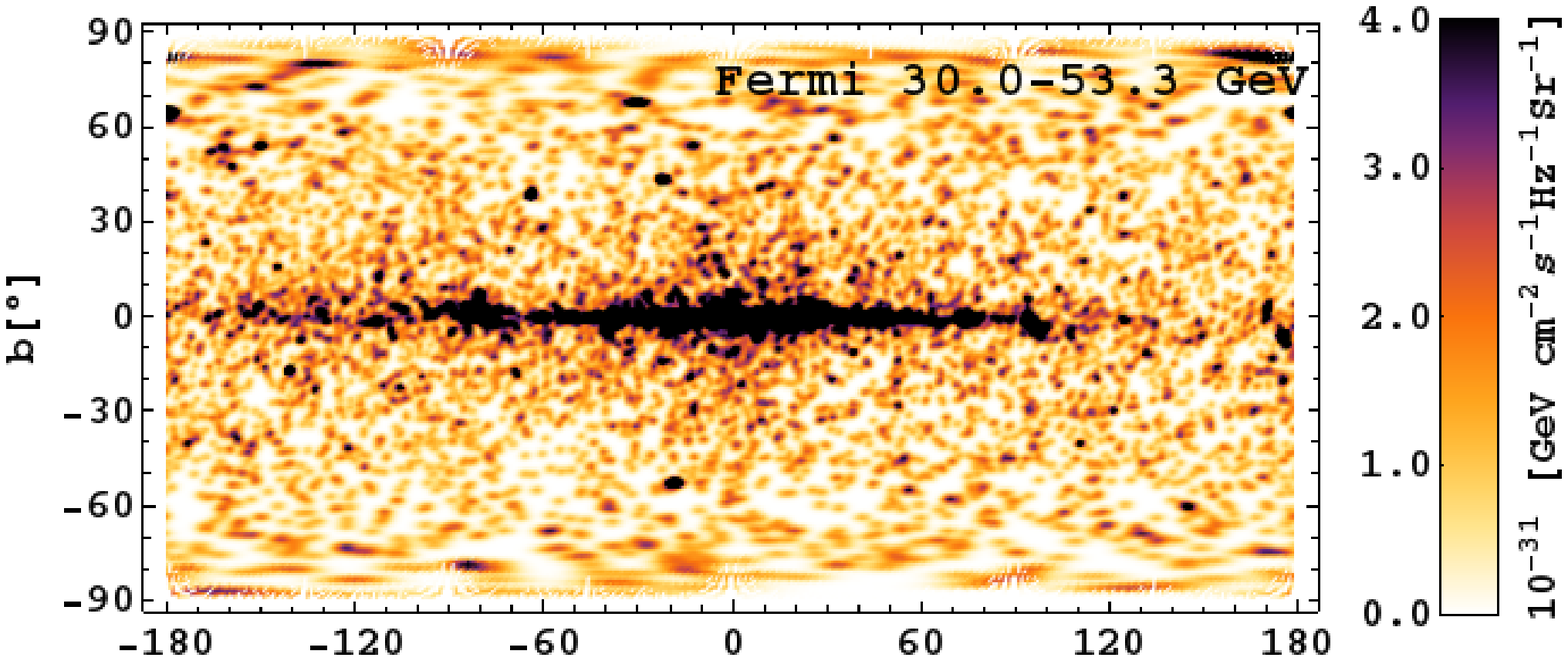}\hfill
\includegraphics[width=5.5cm]{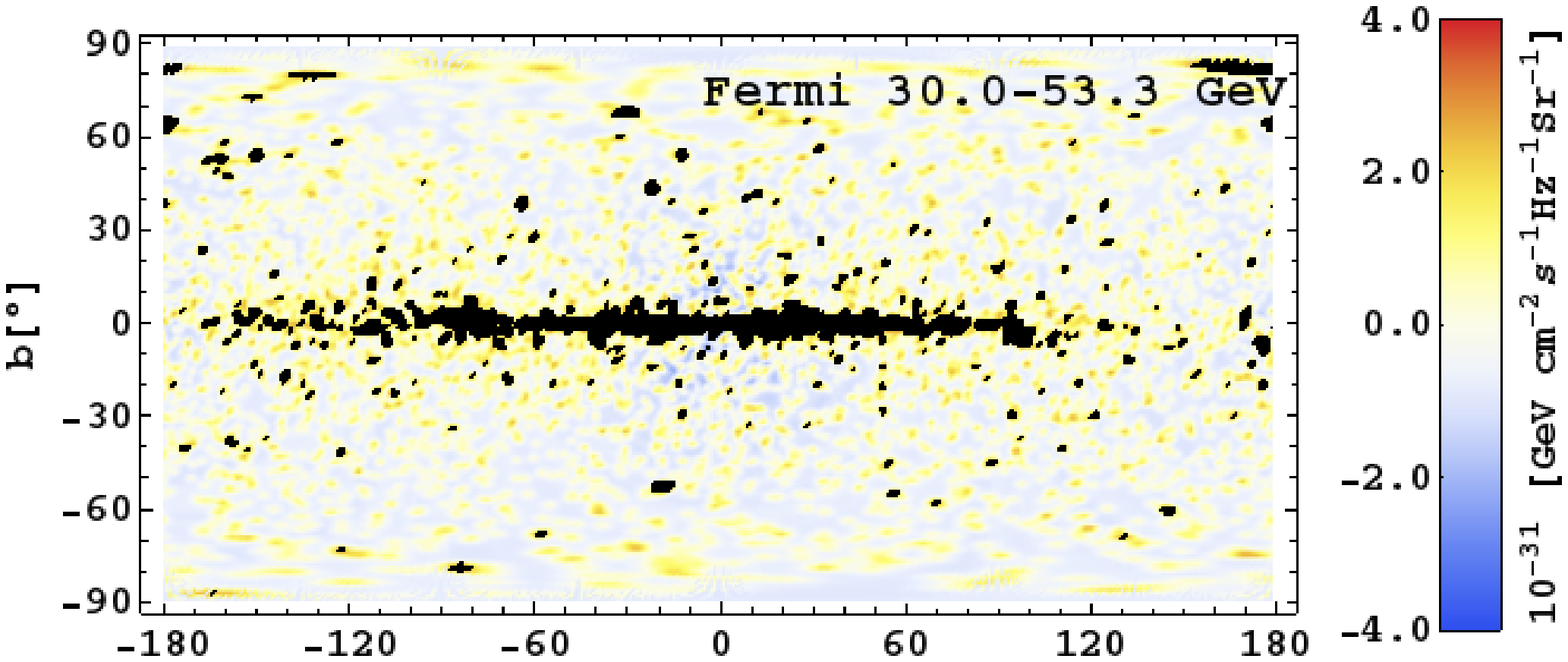}\hfill
\includegraphics[width=5.3cm]{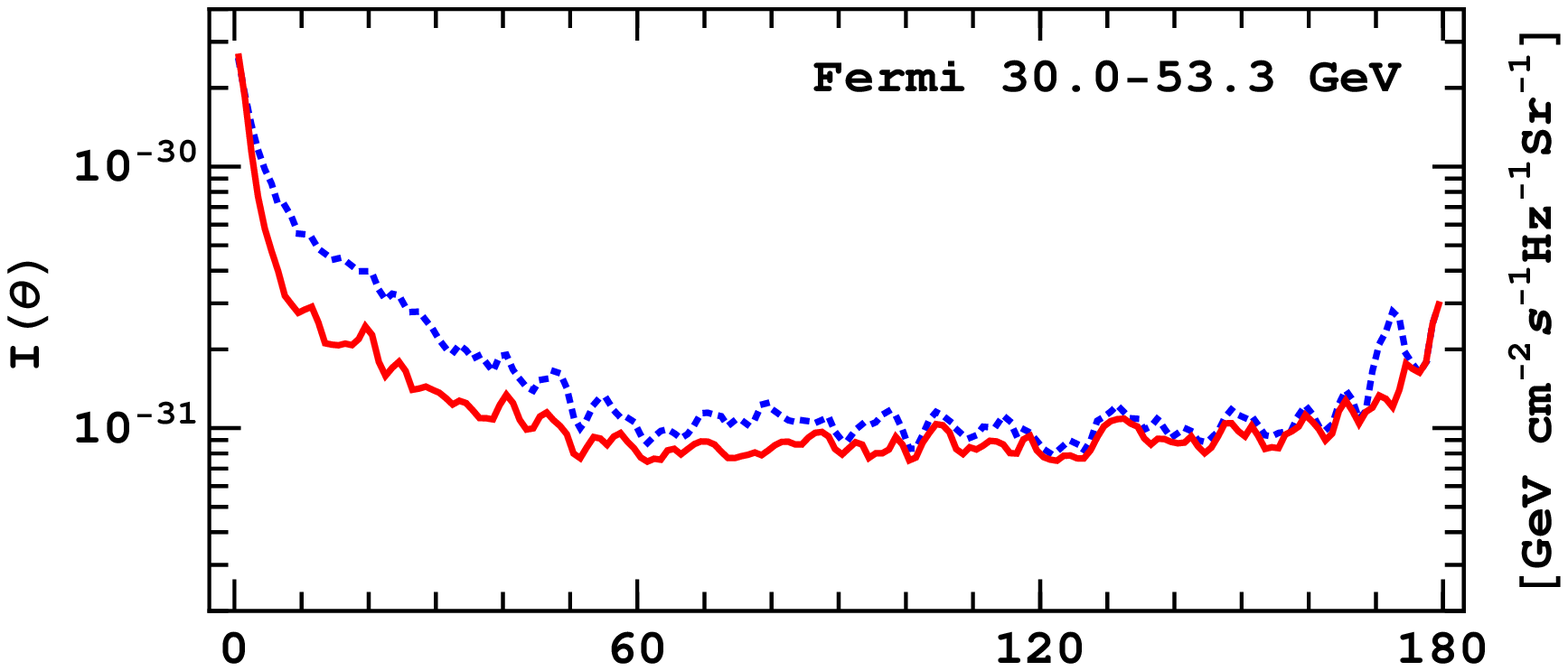}\hfill

\includegraphics[width=5.5cm]{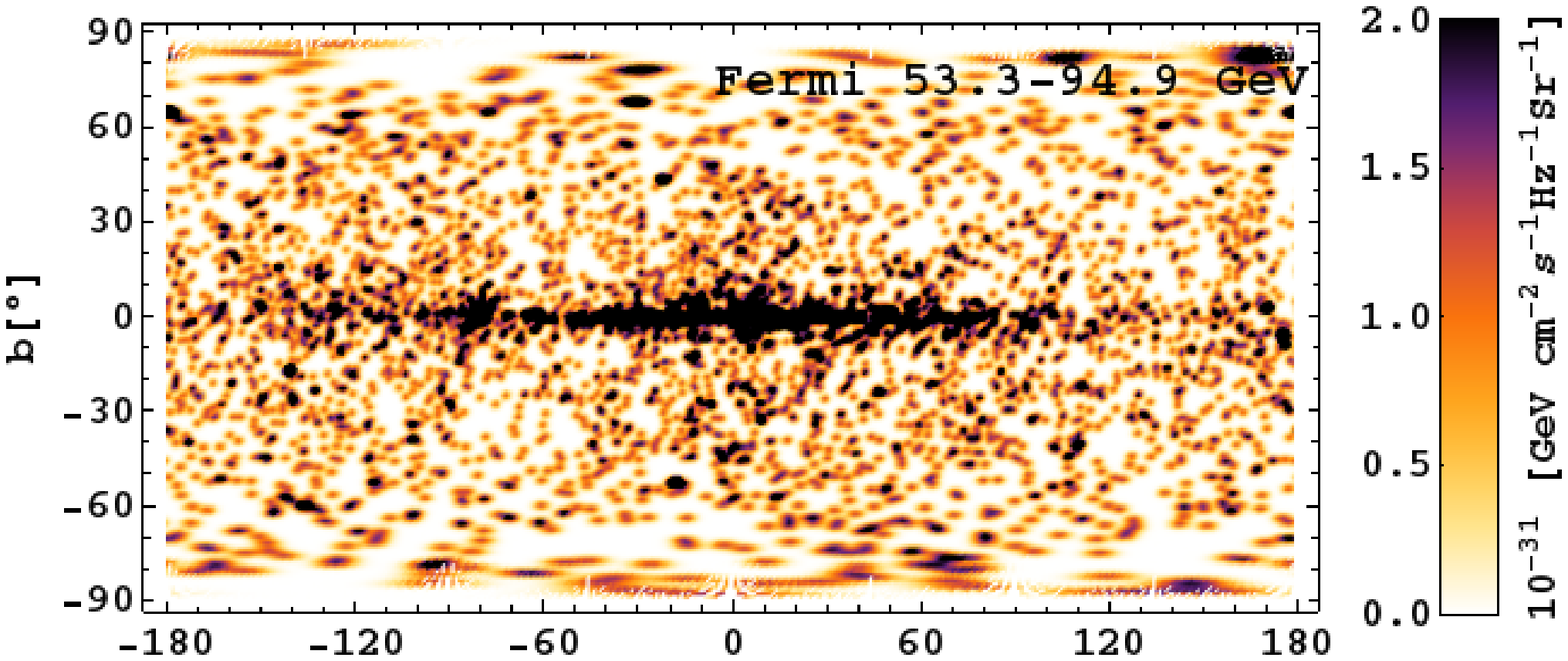}\hfill
\includegraphics[width=5.5cm]{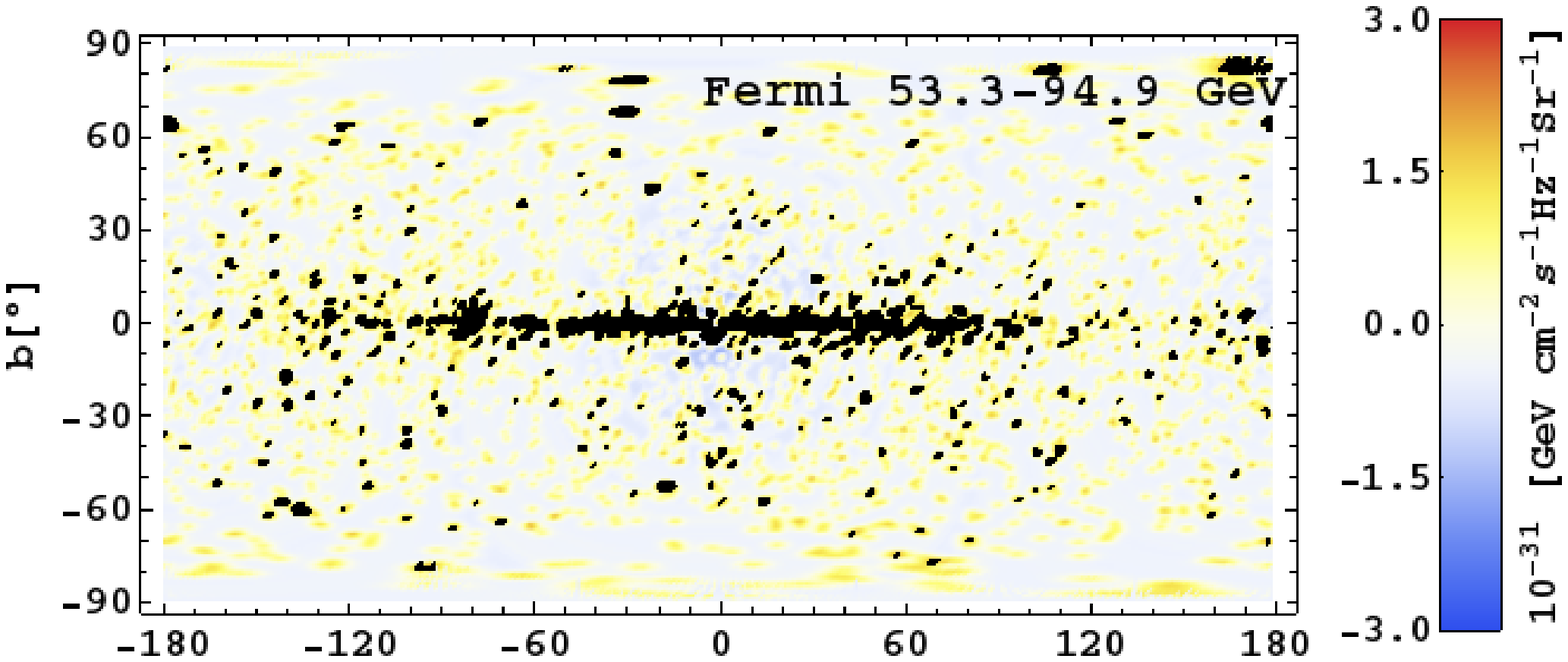}\hfill
\includegraphics[width=5.3cm]{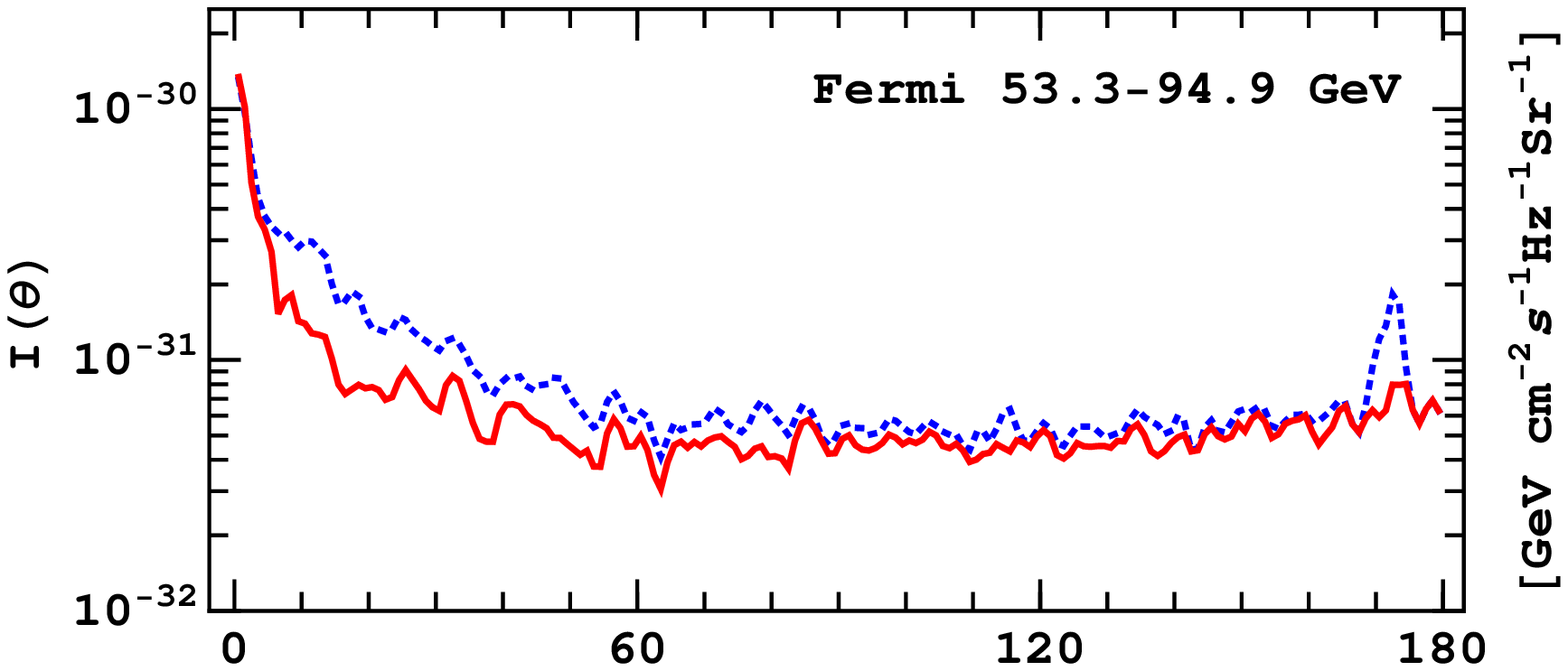}\hfill

\includegraphics[width=5.5cm]{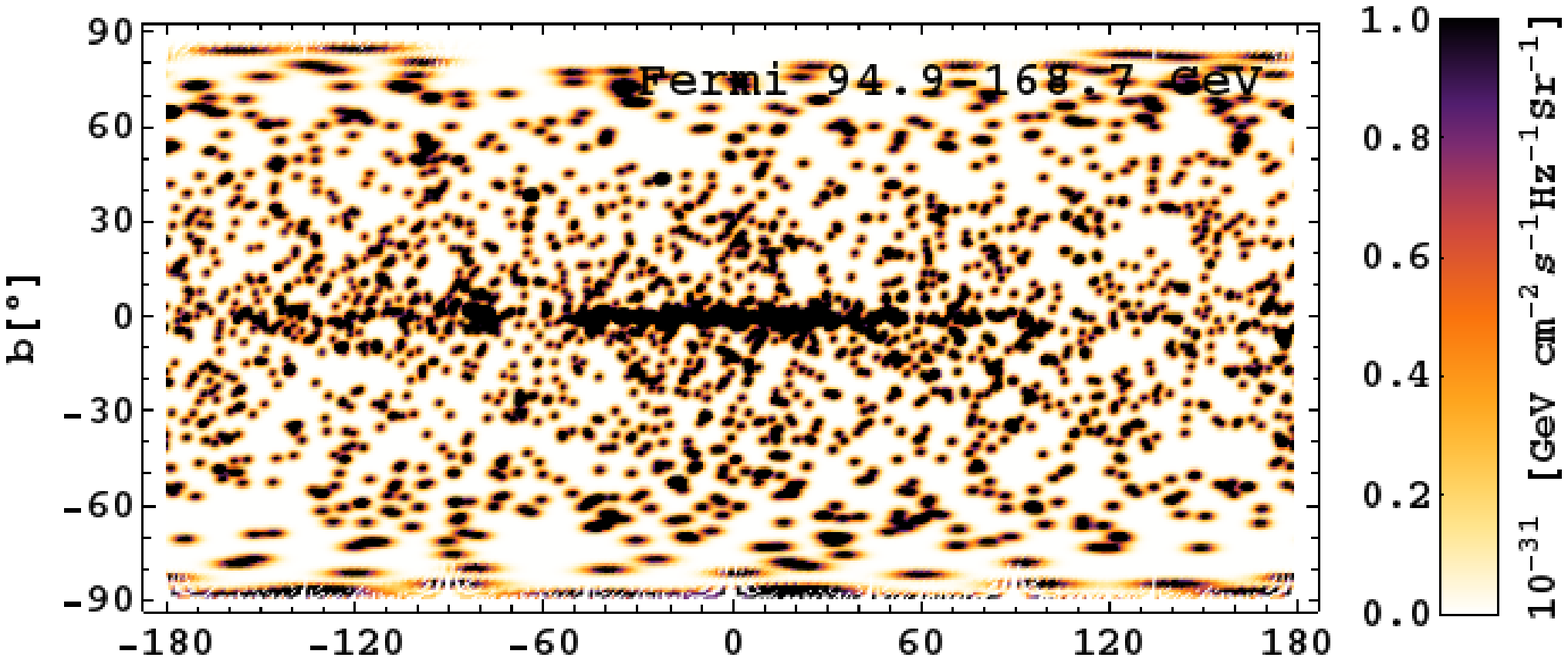}\hfill
\includegraphics[width=5.5cm]{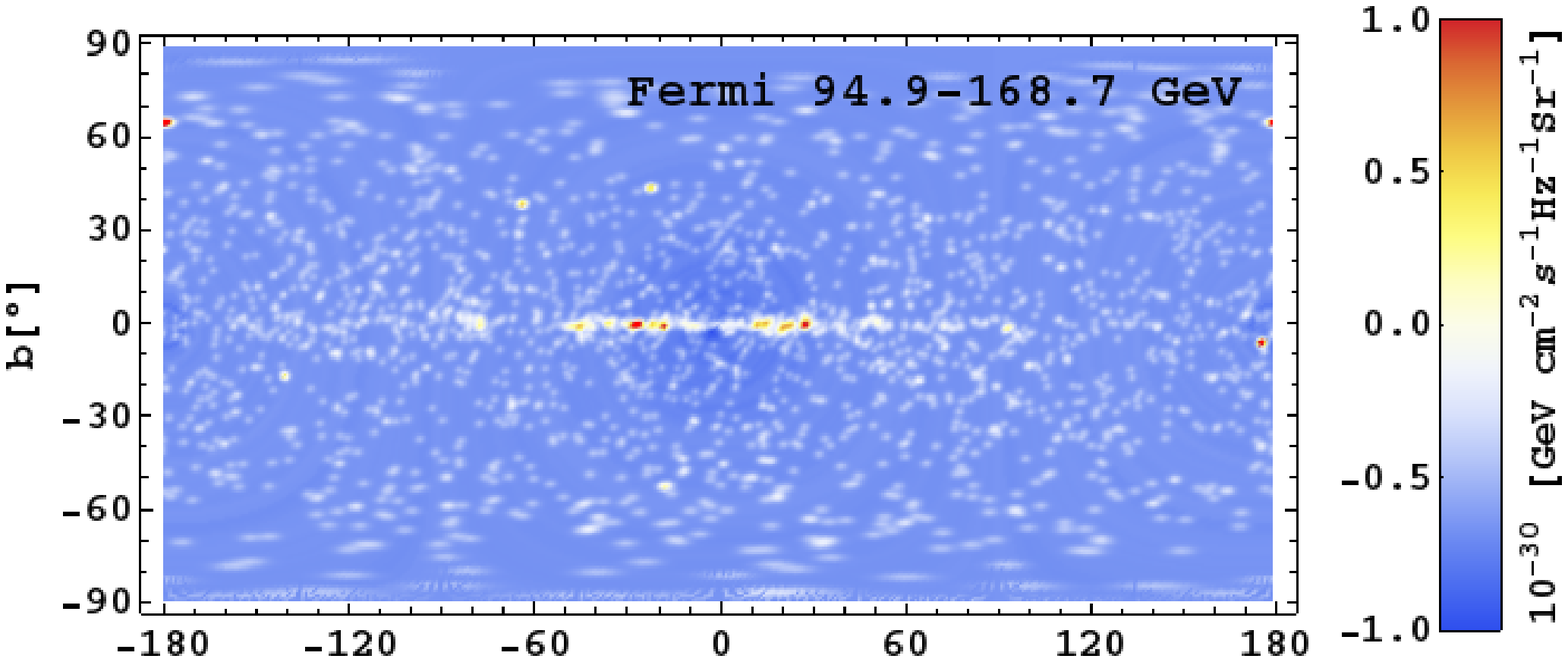}\hfill
\includegraphics[width=5.3cm]{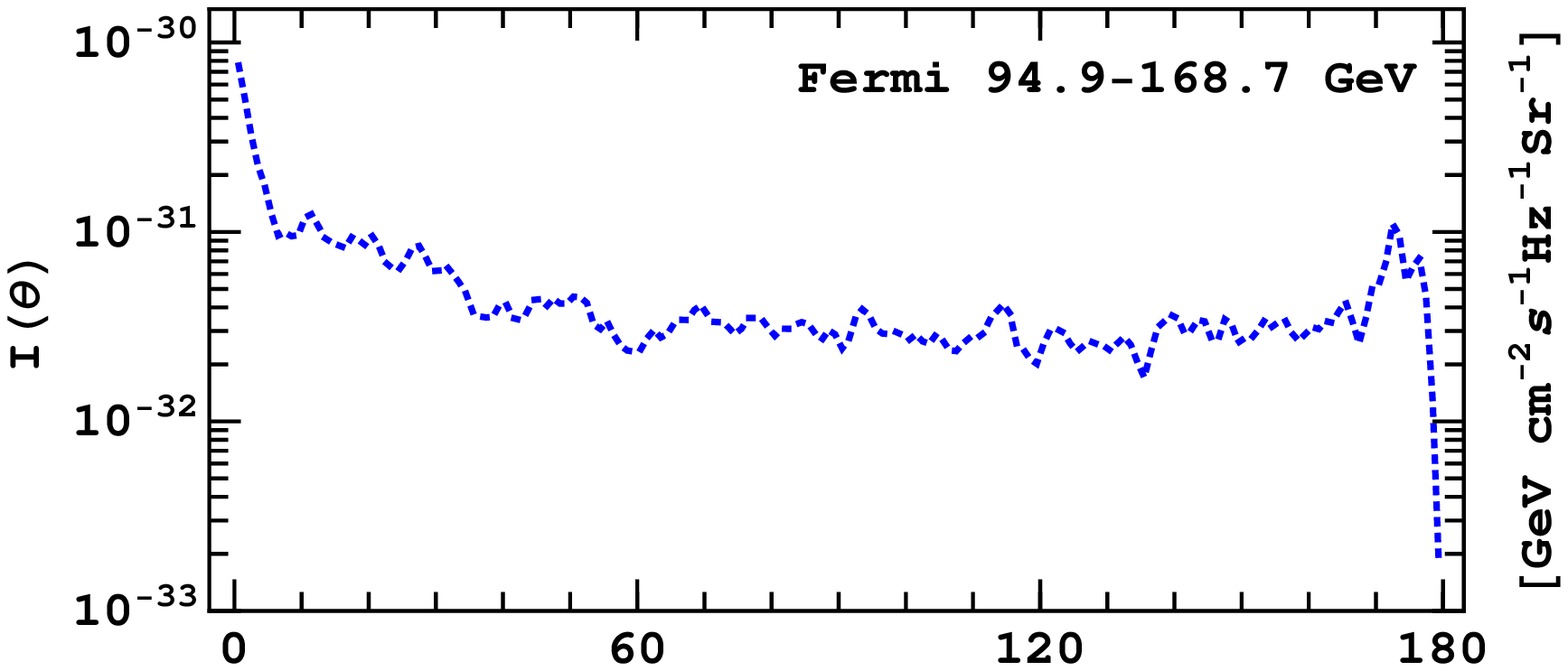}\hfill

\includegraphics[width=5.5cm]{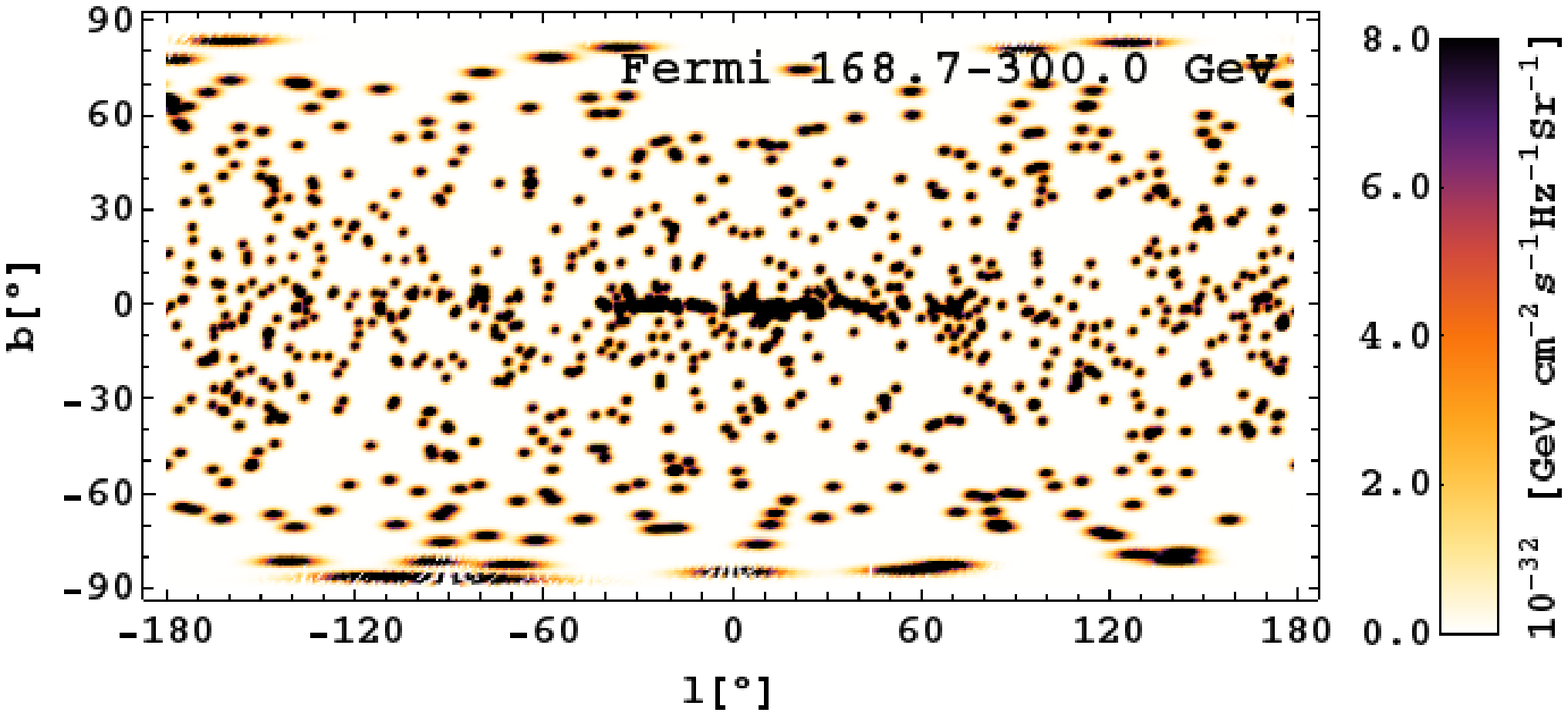}\hfill
\includegraphics[width=5.5cm]{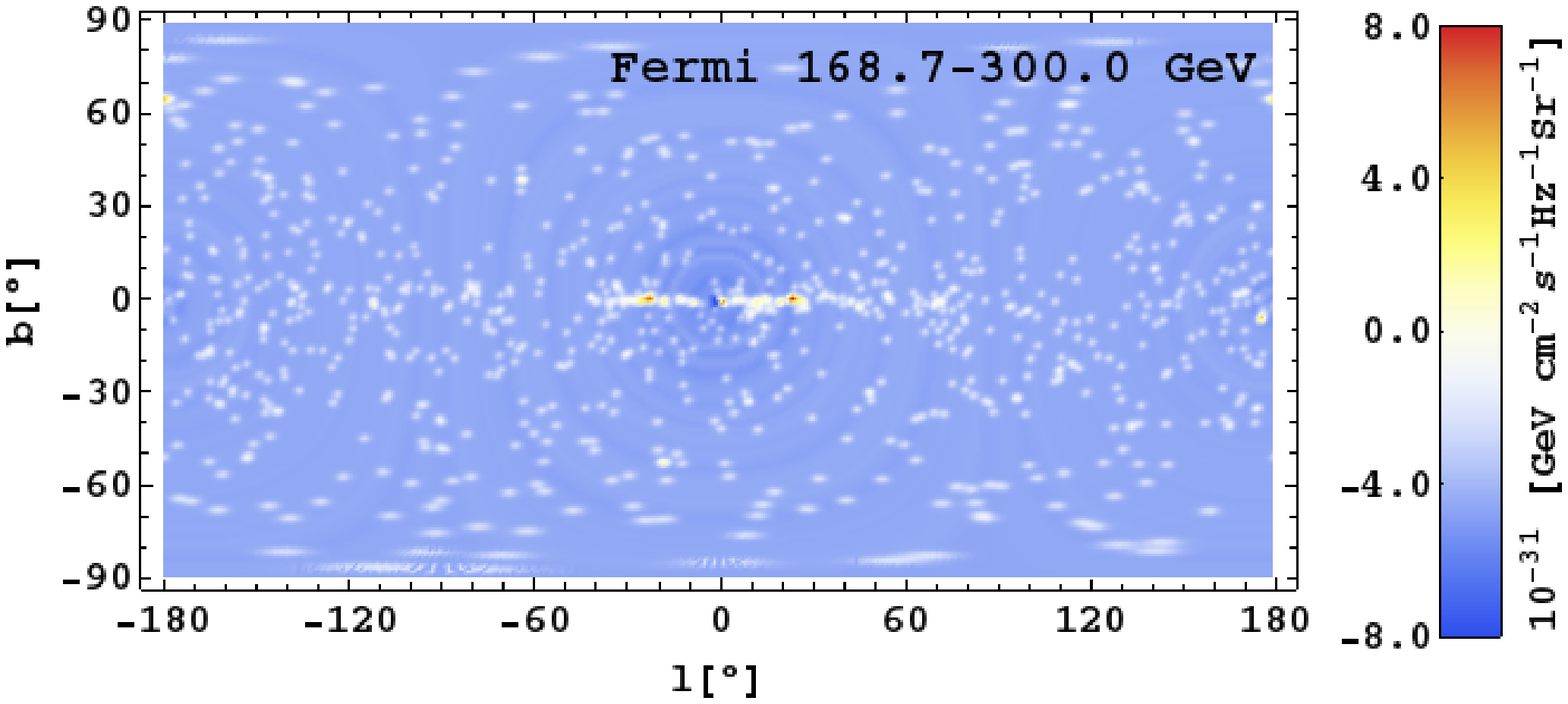}\hfill
\includegraphics[width=5.3cm]{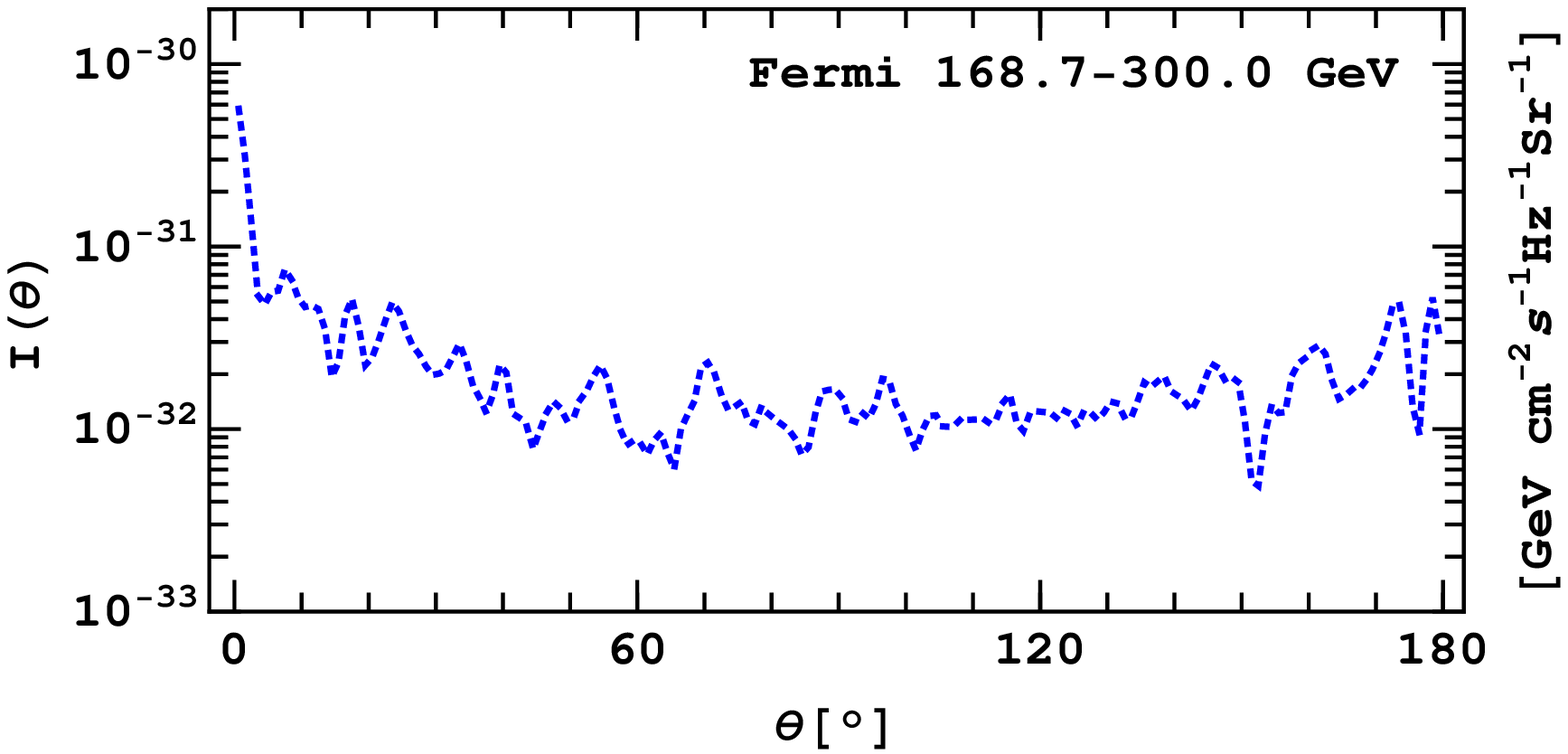}\hfill

\caption
{
Fermi intensity maps $I(l,b)$ in Galactic coordinates (left), masked residual maps $I(l,b)-I(\theta)$ (middle), and spherically-averaged intensities (right).
Dotted blue lines represent the original mean intensity $I_0 (\theta)$, while solid red lines correspond to the final intensity $I(\theta)$ after discarding the outliers (black areas in the masked residual maps).
For the last two bands, we opted to use the original average intensity $I_0(\theta)$ without applying any mask.
}
\label{figFermiMap2}
\end{figure*}
%__________________________________

%__________________________________
\begin{table*}
\begin{center}
\begin{tabular}{cccccccccccccccc}

\\\hline
&\multicolumn{3}{c}{Haslam} &\multicolumn{12}{c}{WMAP} \\
&\multicolumn{3}{c}{408 MHz}
&\multicolumn{3}{c}{23 GHz} 
&\multicolumn{3}{c}{33 GHz} 
&\multicolumn{3}{c}{41 GHz} 
&\multicolumn{3}{c}{61 GHz}	
\\\hline
\multicolumn{1}{c}{$\theta$}
&\multicolumn{1}{c}{$I_0 (\theta)$} &\multicolumn{1}{c}{$I(\theta)$} &\multicolumn{1}{c}{$\sigma (\theta)$}
&\multicolumn{1}{c}{$I_0 (\theta)$} &\multicolumn{1}{c}{$I(\theta)$} &\multicolumn{1}{c}{$\sigma (\theta)$}	
&\multicolumn{1}{c}{$I_0 (\theta)$} &\multicolumn{1}{c}{$I(\theta)$} &\multicolumn{1}{c}{$\sigma (\theta)$}
&\multicolumn{1}{c}{$I_0 (\theta)$} &\multicolumn{1}{c}{$I(\theta)$} &\multicolumn{1}{c}{$\sigma (\theta)$}
&\multicolumn{1}{c}{$I_0 (\theta)$} &\multicolumn{1}{c}{$I(\theta)$} &\multicolumn{1}{c}{$\sigma (\theta)$}	\\
\multicolumn{1}{c}{$[^\circ]$}	
&\multicolumn{3}{c}{$\times 10^{-16}$} 
&\multicolumn{3}{c}{$\times 10^{-17}$} 
&\multicolumn{3}{c}{$\times 10^{-17}$} 
&\multicolumn{3}{c}{$\times 10^{-17}$}				
&\multicolumn{3}{c}{$\times 10^{-17}$}							
\\\hline

0.5	&	209.2	&	208.4	&	3.3	&	743.9	&	729.5	&	3.2	&	703.8	&	698.9	&	5.4	&	682.0	&	681.0	&	4.4	&	660.2	&	670.6	&	8.7	\\
1.5	&	147.7	&	145.6	&	3.5	&	210.4	&	202.3	&	3.7	&	167.8	&	148.8	&	4.0	&	147.8	&	119.0	&	5.8	&	137.3	&	98.8	&	11.7	\\
2.5	&	103.5	&	95.2	&	3.4	&	116.2	&	72.0	&	3.3	&	92.3	&	55.2	&	3.9	&	81.1	&	47.9	&	5.4	&	73.9	&	36.5	&	9.3	\\
3.5	&	84.8	&	73.9	&	3.6	&	90.6	&	43.8	&	2.9	&	72.5	&	33.0	&	3.8	&	64.0	&	27.9	&	5.0	&	58.9	&	26.6	&	9.6	\\
4.5	&	73.2	&	54.4	&	2.9	&	76.5	&	35.1	&	3.2	&	61.6	&	22.6	&	4.1	&	54.1	&	20.4	&	5.2	&	48.0	&	20.8	&	10.1	\\
5.5	&	64.9	&	44.1	&	2.5	&	67.6	&	24.2	&	3.3	&	54.0	&	17.9	&	3.5	&	47.0	&	15.3	&	4.3	&	40.3	&	13.2	&	8.6	\\
6.5	&	59.8	&	38.9	&	2.4	&	81.1	&	19.4	&	2.9	&	70.9	&	14.6	&	3.2	&	65.5	&	12.3	&	3.9	&	60.4	&	10.5	&	8.0	\\
7.5	&	53.4	&	34.4	&	2.7	&	63.2	&	17.1	&	2.9	&	52.1	&	12.6	&	2.8	&	46.4	&	11.5	&	4.0	&	41.1	&	10.7	&	7.8	\\
8.5	&	49.4	&	30.9	&	3.2	&	56.4	&	14.4	&	3.1	&	48.1	&	11.2	&	3.5	&	43.7	&	9.8	&	4.2	&	39.9	&	9.4	&	8.2	\\
9.5	&	44.4	&	28.3	&	2.9	&	43.6	&	12.1	&	3.1	&	34.8	&	9.6	&	3.5	&	30.1	&	8.4	&	4.1	&	25.2	&	8.0	&	8.3	\\
10.5	&	41.7	&	26.0	&	2.8	&	41.9	&	10.7	&	3.1	&	34.4	&	8.5	&	3.8	&	30.3	&	7.3	&	4.4	&	25.9	&	6.4	&	8.4	\\
12.5	&	36.3	&	22.1	&	2.7	&	34.2	&	8.6	&	2.9	&	27.9	&	6.7	&	3.3	&	24.5	&	6.0	&	4.2	&	21.0	&	4.7	&	7.5	\\
14.5	&	33.1	&	20.1	&	2.3	&	39.3	&	8.2	&	3.1	&	33.2	&	6.7	&	3.5	&	29.9	&	6.2	&	4.3	&	26.1	&	5.8	&	8.1	\\
16.5	&	30.3	&	18.2	&	2.3	&	32.0	&	7.7	&	3.1	&	26.4	&	6.5	&	3.5	&	23.5	&	6.5	&	4.4	&	21.4	&	7.1	&	8.2	\\
18.5	&	29.4	&	17.6	&	2.3	&	33.2	&	7.3	&	3.2	&	28.2	&	6.2	&	3.5	&	25.5	&	6.0	&	4.4	&	23.7	&	6.8	&	8.5	\\
20.5	&	27.7	&	15.9	&	2.2	&	24.5	&	6.5	&	3.2	&	19.7	&	6.0	&	3.9	&	17.2	&	5.9	&	4.9	&	15.3	&	6.2	&	9.1	\\
25.5	&	23.6	&	13.6	&	1.9	&	19.9	&	5.8	&	3.0	&	16.1	&	4.8	&	3.3	&	14.2	&	4.7	&	4.4	&	12.8	&	5.2	&	8.3	\\
30.5	&	21.3	&	12.3	&	2.1	&	17.8	&	4.4	&	2.1	&	14.7	&	3.5	&	2.8	&	13.1	&	3.2	&	3.7	&	12.0	&	3.2	&	7.6	\\
40.5	&	16.1	&	11.0	&	2.5	&	8.5	&	2.9	&	1.7	&	6.4	&	2.1	&	2.3	&	5.5	&	1.8	&	3.2	&	4.6	&	1.6	&	7.1	\\
50.5	&	13.0	&	10.3	&	2.5	&	6.7	&	2.3	&	1.4	&	5.3	&	1.8	&	2.0	&	4.6	&	1.7	&	2.9	&	4.5	&	1.9	&	6.7	\\
60.5	&	9.7	&	8.7	&	2.2	&	4.2	&	2.0	&	1.6	&	3.4	&	1.7	&	2.3	&	3.1	&	1.7	&	3.2	&	3.3	&	2.1	&	7.1	\\
70.5	&	8.7	&	7.6	&	1.6	&	4.7	&	1.7	&	1.3	&	4.0	&	1.5	&	2.1	&	3.8	&	1.5	&	3.1	&	3.9	&	1.8	&	6.8	\\
80.5	&	9.0	&	7.5	&	1.7	&	6.8	&	1.5	&	1.4	&	5.8	&	1.2	&	2.0	&	5.3	&	1.1	&	3.0	&	4.9	&	1.0	&	6.6	\\
90.5	&	8.0	&	7.2	&	1.9	&	4.1	&	1.4	&	1.3	&	3.4	&	1.3	&	2.2	&	3.0	&	1.4	&	3.3	&	3.0	&	1.8	&	7.0	\\
100.5	&	7.6	&	7.0	&	1.7	&	3.5	&	1.8	&	1.6	&	3.0	&	1.8	&	2.3	&	3.0	&	1.9	&	3.1	&	3.5	&	2.8	&	6.9	\\
110.5	&	7.5	&	6.7	&	1.7	&	3.7	&	1.5	&	1.6	&	3.1	&	1.4	&	2.5	&	2.9	&	1.6	&	3.6	&	3.2	&	2.3	&	7.5	\\
120.5	&	7.2	&	6.5	&	1.8	&	2.6	&	1.4	&	1.5	&	2.0	&	1.2	&	2.4	&	1.8	&	1.2	&	3.4	&	1.8	&	1.4	&	7.4	\\
130.5	&	7.5	&	6.9	&	2.1	&	2.6	&	1.8	&	1.8	&	2.0	&	1.5	&	2.7	&	1.7	&	1.4	&	3.7	&	1.6	&	1.4	&	7.7	\\
140.5	&	8.3	&	7.7	&	2.3	&	4.1	&	2.8	&	2.2	&	3.1	&	2.2	&	2.7	&	2.7	&	2.0	&	3.7	&	2.7	&	2.2	&	8.1	\\
150.5	&	9.4	&	9.0	&	2.0	&	5.3	&	3.0	&	2.1	&	4.1	&	2.4	&	2.8	&	3.6	&	2.5	&	4.0	&	3.5	&	2.7	&	8.2	\\
155.5	&	9.9	&	9.8	&	2.2	&	5.5	&	3.8	&	2.4	&	4.1	&	3.2	&	2.9	&	3.6	&	2.8	&	3.8	&	3.4	&	3.0	&	8.0	\\
160.5	&	10.6	&	10.5	&	1.7	&	6.1	&	4.8	&	2.5	&	4.8	&	4.0	&	2.7	&	4.2	&	3.9	&	3.6	&	4.2	&	4.1	&	7.3	\\
162.5	&	10.7	&	10.7	&	1.6	&	6.1	&	5.3	&	2.5	&	4.6	&	4.3	&	2.7	&	4.0	&	3.9	&	3.4	&	4.1	&	4.1	&	7.1	\\
164.5	&	10.7	&	10.7	&	1.5	&	6.2	&	5.6	&	2.6	&	4.7	&	4.6	&	2.8	&	4.1	&	4.1	&	3.4	&	4.5	&	4.5	&	7.3	\\
166.5	&	10.9	&	10.9	&	1.6	&	6.8	&	6.0	&	2.8	&	5.3	&	4.8	&	2.8	&	4.7	&	4.4	&	3.5	&	5.3	&	5.2	&	7.4	\\
168.5	&	11.2	&	11.2	&	1.7	&	7.1	&	5.9	&	2.3	&	5.2	&	4.8	&	2.9	&	4.4	&	4.2	&	3.6	&	4.6	&	4.5	&	7.4	\\
170.5	&	12.1	&	12.1	&	1.8	&	10.6	&	6.2	&	1.9	&	8.5	&	4.9	&	2.9	&	7.7	&	4.9	&	4.1	&	8.1	&	6.2	&	8.5	\\
171.5	&	12.4	&	12.4	&	1.8	&	10.5	&	7.0	&	2.3	&	8.0	&	5.7	&	3.0	&	7.1	&	5.4	&	4.1	&	7.5	&	7.0	&	8.5	\\
172.5	&	12.7	&	12.7	&	2.0	&	24.3	&	7.8	&	2.2	&	21.5	&	5.8	&	2.3	&	19.9	&	5.1	&	3.5	&	18.6	&	6.2	&	7.9	\\
173.5	&	12.6	&	12.6	&	1.9	&	14.5	&	9.0	&	2.4	&	10.8	&	6.6	&	2.5	&	9.1	&	5.5	&	3.3	&	8.2	&	6.0	&	7.6	\\
174.5	&	12.3	&	12.3	&	1.9	&	10.4	&	9.9	&	2.7	&	7.6	&	7.4	&	2.6	&	6.4	&	6.3	&	3.2	&	6.7	&	6.7	&	6.7	\\
175.5	&	12.0	&	12.0	&	1.6	&	9.3	&	9.2	&	2.5	&	6.6	&	6.6	&	2.5	&	5.3	&	5.3	&	3.0	&	4.9	&	4.9	&	6.4	\\
176.5	&	12.0	&	12.0	&	1.4	&	10.1	&	9.8	&	3.0	&	7.1	&	7.1	&	2.8	&	5.8	&	5.8	&	3.0	&	5.4	&	5.4	&	5.9	\\
177.5	&	12.4	&	12.4	&	1.3	&	12.5	&	14.7	&	2.3	&	9.7	&	9.3	&	3.8	&	8.3	&	8.2	&	4.1	&	9.0	&	9.0	&	7.7	\\
178.5	&	12.4	&	12.4	&	1.2	&	12.4	&	14.7	&	2.5	&	8.9	&	8.9	&	3.5	&	7.4	&	7.4	&	3.5	&	6.9	&	6.9	&	6.2	\\
179.5	&	12.7	&	12.7	&	0.9	&	12.4	&	12.4	&	2.9	&	8.9	&	8.9	&	2.7	&	7.0	&	7.0	&	2.6	&	6.0	&	6.0	&	4.9	\\

\\\hline

\end{tabular}
\end{center}
\caption
{
Haslam and WMAP: Original observational mean intensity $I_0 (\theta)$, final mean intensity after discarding the outliers $I(\theta)$, and standard deviation $\sigma(\theta)$, in units of $\rm{GeV\ cm^{-2}\ s^{-1}\ Hz^{-1}\ sr^{-1}}$.
}
\label{tabHaslam&WMAPIntensity}
\end{table*}

%__________________________________
\begin{table*}
\begin{center}
\begin{tabular}{cccccccccccccccc}

\\\hline
&\multicolumn{3}{c}{WMAP}
&\multicolumn{12}{c}{Fermi} \\
&\multicolumn{3}{c}{94 GHz} 
&\multicolumn{3}{c}{0.3-0.5 GeV} 
&\multicolumn{3}{c}{0.5-0.9 GeV} 
&\multicolumn{3}{c}{0.9-1.7 GeV} 
&\multicolumn{3}{c}{1.7-3.0 GeV}
\\\hline
\multicolumn{1}{c}{$\theta$}	
&\multicolumn{1}{c}{$I_0 (\theta)$} &\multicolumn{1}{c}{$I(\theta)$} &\multicolumn{1}{c}{$\sigma (\theta)$}
&\multicolumn{1}{c}{$I_0 (\theta)$} &\multicolumn{1}{c}{$I(\theta)$} &\multicolumn{1}{c}{$\sigma (\theta)$}	
&\multicolumn{1}{c}{$I_0 (\theta)$} &\multicolumn{1}{c}{$I(\theta)$} &\multicolumn{1}{c}{$\sigma (\theta)$}
&\multicolumn{1}{c}{$I_0 (\theta)$} &\multicolumn{1}{c}{$I(\theta)$} &\multicolumn{1}{c}{$\sigma (\theta)$}
&\multicolumn{1}{c}{$I_0 (\theta)$} &\multicolumn{1}{c}{$I(\theta)$} &\multicolumn{1}{c}{$\sigma (\theta)$}	\\
\multicolumn{1}{c}{$[^\circ]$}	
&\multicolumn{3}{c}{$\times 10^{-17}$}
&\multicolumn{3}{c}{$\times 10^{-30}$}
&\multicolumn{3}{c}{$\times 10^{-30}$}		
&\multicolumn{3}{c}{$\times 10^{-30}$}
&\multicolumn{3}{c}{$\times 10^{-30}$}
\\\hline

0.5	&	1097.8	&	1132.3	&	22.2	&	1500.2	&	1506.8	&	34.3	&	989.9	&	1016.7	&	18.7	&	713.6	&	725.8	&	9.1	&	367.7	&	371.1	&	4.6	\\
1.5	&	287.4	&	219.6	&	24.5	&	1295.6	&	1353.7	&	37.3	&	774.3	&	790.1	&	16.6	&	447.3	&	462.2	&	9.8	&	230.1	&	245.2	&	4.7	\\
2.5	&	140.3	&	66.9	&	21.4	&	1037.0	&	992.3	&	37.0	&	553.9	&	508.8	&	19.8	&	275.9	&	219.7	&	9.9	&	139.7	&	116.5	&	4.9	\\
3.5	&	114.6	&	52.3	&	21.2	&	826.8	&	689.0	&	34.2	&	418.4	&	304.9	&	16.7	&	211.3	&	129.4	&	8.8	&	106.7	&	66.6	&	4.0	\\
4.5	&	85.1	&	39.9	&	21.8	&	680.6	&	460.1	&	36.1	&	340.2	&	208.4	&	17.8	&	174.1	&	107.2	&	9.1	&	86.8	&	53.3	&	4.3	\\
5.5	&	69.9	&	25.3	&	19.2	&	585.4	&	358.1	&	34.3	&	293.0	&	170.4	&	16.8	&	146.5	&	82.5	&	8.3	&	71.8	&	36.7	&	4.4	\\
6.5	&	86.1	&	20.6	&	18.3	&	509.1	&	289.2	&	32.1	&	258.2	&	133.7	&	16.1	&	129.9	&	56.2	&	9.3	&	63.4	&	26.9	&	4.1	\\
7.5	&	66.6	&	20.8	&	17.8	&	470.1	&	213.9	&	38.6	&	243.3	&	101.3	&	19.3	&	126.2	&	47.8	&	7.7	&	61.7	&	23.8	&	3.9	\\
8.5	&	61.9	&	18.0	&	18.1	&	415.0	&	184.3	&	35.3	&	211.3	&	90.0	&	17.5	&	107.7	&	42.6	&	8.0	&	52.1	&	21.5	&	3.8	\\
9.5	&	43.2	&	15.7	&	18.6	&	367.1	&	165.0	&	33.0	&	181.9	&	80.7	&	16.7	&	89.5	&	38.4	&	8.1	&	42.9	&	19.5	&	3.6	\\
10.5	&	41.6	&	12.5	&	19.0	&	341.4	&	150.7	&	31.9	&	167.5	&	74.2	&	17.0	&	83.5	&	36.8	&	8.6	&	39.8	&	18.3	&	3.9	\\
12.5	&	33.5	&	9.1	&	17.5	&	285.3	&	132.9	&	30.0	&	137.5	&	61.5	&	14.8	&	67.8	&	30.3	&	7.7	&	32.2	&	14.8	&	3.5	\\
14.5	&	39.5	&	11.7	&	18.5	&	271.2	&	123.1	&	31.7	&	130.1	&	56.9	&	15.8	&	63.7	&	28.2	&	8.0	&	31.5	&	13.5	&	3.5	\\
16.5	&	36.6	&	14.3	&	18.5	&	272.5	&	117.7	&	35.1	&	135.2	&	55.0	&	16.8	&	70.1	&	27.2	&	8.3	&	34.5	&	13.2	&	3.9	\\
18.5	&	39.0	&	15.0	&	19.4	&	264.6	&	110.3	&	34.7	&	132.4	&	51.7	&	16.7	&	66.6	&	25.7	&	8.7	&	32.2	&	12.3	&	4.0	\\
20.5	&	27.8	&	13.2	&	20.5	&	241.1	&	106.6	&	33.4	&	118.3	&	49.9	&	16.6	&	57.9	&	23.7	&	8.1	&	28.0	&	11.5	&	4.0	\\
25.5	&	22.9	&	11.1	&	18.4	&	193.1	&	89.0	&	27.0	&	93.5	&	40.1	&	12.4	&	46.3	&	19.7	&	6.7	&	22.7	&	9.6	&	3.2	\\
30.5	&	20.9	&	7.0	&	17.4	&	160.3	&	76.2	&	21.4	&	78.9	&	36.0	&	11.1	&	39.1	&	17.7	&	6.1	&	18.6	&	8.3	&	2.7	\\
40.5	&	9.3	&	3.5	&	15.9	&	115.6	&	60.7	&	19.6	&	56.0	&	27.9	&	10.0	&	27.8	&	13.2	&	5.1	&	13.4	&	6.4	&	2.4	\\
50.5	&	9.4	&	4.4	&	15.2	&	87.4	&	50.2	&	16.8	&	41.5	&	22.6	&	8.3	&	19.8	&	10.6	&	4.5	&	9.7	&	5.2	&	2.2	\\
60.5	&	7.3	&	4.9	&	16.1	&	65.4	&	45.1	&	18.4	&	30.1	&	20.0	&	9.3	&	14.4	&	9.5	&	4.8	&	6.8	&	4.5	&	2.3	\\
70.5	&	7.3	&	4.0	&	15.1	&	63.3	&	43.4	&	16.7	&	29.2	&	19.1	&	8.4	&	13.6	&	8.7	&	4.3	&	6.5	&	4.1	&	2.0	\\
80.5	&	7.7	&	2.2	&	14.9	&	63.3	&	39.2	&	16.0	&	29.2	&	17.1	&	7.7	&	14.0	&	8.0	&	3.9	&	6.5	&	3.7	&	1.8	\\
90.5	&	6.4	&	3.9	&	15.6	&	59.4	&	37.4	&	15.7	&	27.4	&	16.3	&	8.2	&	13.0	&	7.8	&	4.6	&	6.0	&	3.6	&	2.1	\\
100.5	&	7.7	&	6.2	&	15.4	&	57.8	&	37.4	&	15.6	&	25.9	&	16.6	&	7.9	&	12.0	&	7.9	&	4.3	&	5.7	&	3.8	&	2.1	\\
110.5	&	6.9	&	5.1	&	16.6	&	54.1	&	36.8	&	15.0	&	25.2	&	15.9	&	7.6	&	11.9	&	7.4	&	4.0	&	5.6	&	3.5	&	2.0	\\
120.5	&	4.5	&	3.5	&	16.6	&	51.3	&	36.3	&	13.8	&	23.7	&	16.0	&	7.0	&	11.4	&	7.4	&	3.5	&	5.5	&	3.6	&	1.9	\\
130.5	&	4.1	&	3.7	&	17.6	&	57.2	&	40.2	&	15.6	&	26.1	&	17.5	&	7.7	&	12.3	&	8.2	&	3.8	&	5.8	&	3.8	&	1.8	\\
140.5	&	6.8	&	5.6	&	18.9	&	70.5	&	49.4	&	20.4	&	32.8	&	21.4	&	9.6	&	15.7	&	10.2	&	4.9	&	7.1	&	4.9	&	2.4	\\
150.5	&	8.6	&	7.0	&	19.0	&	83.2	&	60.0	&	26.1	&	39.3	&	27.2	&	12.5	&	19.1	&	13.4	&	6.6	&	8.7	&	6.1	&	2.9	\\
155.5	&	8.3	&	7.9	&	18.8	&	89.8	&	71.3	&	29.5	&	42.7	&	31.9	&	14.1	&	20.3	&	14.7	&	6.9	&	9.2	&	6.9	&	3.3	\\
160.5	&	10.3	&	10.1	&	16.9	&	100.4	&	84.8	&	31.2	&	46.4	&	39.5	&	15.4	&	21.6	&	18.6	&	7.7	&	9.8	&	8.5	&	3.4	\\
162.5	&	10.9	&	10.8	&	16.8	&	126.4	&	94.1	&	36.9	&	63.9	&	41.8	&	16.9	&	29.0	&	20.8	&	8.9	&	13.7	&	9.2	&	3.9	\\
164.5	&	12.3	&	12.3	&	16.8	&	156.4	&	102.5	&	38.7	&	92.6	&	47.2	&	19.3	&	62.2	&	22.1	&	9.2	&	33.5	&	9.8	&	3.9	\\
166.5	&	14.6	&	14.4	&	17.3	&	142.8	&	109.6	&	36.7	&	68.4	&	49.1	&	18.0	&	29.7	&	23.2	&	9.1	&	13.9	&	10.8	&	4.1	\\
168.5	&	13.5	&	13.3	&	17.3	&	141.5	&	121.4	&	34.7	&	65.7	&	51.8	&	14.7	&	30.2	&	25.2	&	7.7	&	13.7	&	11.4	&	3.7	\\
170.5	&	20.0	&	17.8	&	19.9	&	173.4	&	129.9	&	33.0	&	83.6	&	60.1	&	16.0	&	41.2	&	28.7	&	7.9	&	19.3	&	13.5	&	3.7	\\
171.5	&	19.7	&	19.7	&	19.5	&	194.1	&	138.3	&	31.3	&	96.7	&	62.7	&	14.3	&	47.6	&	30.7	&	7.5	&	22.6	&	14.1	&	3.6	\\
172.5	&	27.8	&	17.7	&	18.1	&	209.0	&	154.7	&	33.0	&	107.2	&	68.0	&	13.2	&	57.2	&	35.8	&	8.2	&	27.6	&	15.2	&	3.4	\\
173.5	&	20.9	&	19.0	&	17.5	&	212.6	&	164.2	&	29.3	&	106.9	&	79.6	&	16.3	&	55.3	&	39.7	&	7.8	&	26.1	&	17.3	&	3.9	\\
174.5	&	20.4	&	20.5	&	15.7	&	203.9	&	172.7	&	26.6	&	95.7	&	82.0	&	13.2	&	44.9	&	41.2	&	7.6	&	20.5	&	19.0	&	3.7	\\
175.5	&	16.4	&	16.4	&	14.7	&	196.7	&	179.6	&	25.3	&	90.4	&	85.5	&	12.7	&	42.5	&	42.3	&	7.6	&	19.4	&	19.4	&	3.6	\\
176.5	&	18.1	&	18.1	&	13.9	&	196.4	&	193.0	&	29.5	&	92.9	&	92.9	&	14.6	&	45.1	&	45.5	&	8.7	&	20.7	&	20.5	&	4.1	\\
177.5	&	26.8	&	26.6	&	18.5	&	203.1	&	203.1	&	26.9	&	98.7	&	98.7	&	14.9	&	49.5	&	53.3	&	9.2	&	23.0	&	23.1	&	4.5	\\
178.5	&	22.4	&	22.4	&	14.9	&	211.8	&	211.8	&	21.3	&	103.2	&	103.2	&	13.2	&	52.3	&	55.7	&	9.3	&	24.9	&	24.9	&	4.0	\\
179.5	&	21.0	&	21.0	&	13.2	&	217.6	&	217.6	&	10.8	&	105.3	&	105.3	&	7.0	&	54.5	&	54.5	&	6.4	&	26.2	&	26.2	&	2.1

\\\hline

\end{tabular}
\end{center}
\caption
{
WMAP and Fermi: Original observational mean intensity $I_0 (\theta)$, final mean intensity after discarding the outliers $I(\theta)$, and standard deviation $\sigma(\theta)$, in units of $\rm{GeV\ cm^{-2}\ s^{-1}\ Hz^{-1}\ sr^{-1}}$.
}
\label{tabWMAP&FermiIntensity}
\end{table*}

%__________________________________
\begin{table*}
\begin{center}
\begin{tabular}{ccccccccccccc}

\\\hline
&\multicolumn{12}{c}{Fermi} \\
&\multicolumn{3}{c}{3.0-5.3 GeV}
&\multicolumn{3}{c}{5.3-9.5 GeV}	
&\multicolumn{3}{c}{9.5-16.9 GeV}	
&\multicolumn{3}{c}{16.9-30.0 GeV}
\\\hline
\multicolumn{1}{c}{$\theta$}	
&\multicolumn{1}{c}{$I_0 (\theta)$} &\multicolumn{1}{c}{$I(\theta)$} &\multicolumn{1}{c}{$\sigma (\theta)$}
&\multicolumn{1}{c}{$I_0 (\theta)$} &\multicolumn{1}{c}{$I(\theta)$} &\multicolumn{1}{c}{$\sigma (\theta)$}	
&\multicolumn{1}{c}{$I_0 (\theta)$} &\multicolumn{1}{c}{$I(\theta)$} &\multicolumn{1}{c}{$\sigma (\theta)$}
&\multicolumn{1}{c}{$I_0 (\theta)$} &\multicolumn{1}{c}{$I(\theta)$} &\multicolumn{1}{c}{$\sigma (\theta)$}	\\
\multicolumn{1}{c}{$[^\circ]$}	
&\multicolumn{3}{c}{$\times 10^{-31}$} 		
&\multicolumn{3}{c}{$\times 10^{-31}$}
&\multicolumn{3}{c}{$\times 10^{-31}$}		
&\multicolumn{3}{c}{$\times 10^{-31}$}
\\\hline

0.5	&	1554.0	&	1574.6	&	19.2	&	482.4	&	488.9	&	7.5	&	188.8	&	195.6	&	3.1	&	74.4	&	75.7	&	1.8	\\
1.5	&	950.3	&	941.4	&	17.8	&	311.6	&	321.6	&	7.5	&	118.7	&	116.7	&	3.4	&	45.9	&	44.7	&	2.2	\\
2.5	&	577.1	&	468.4	&	20.9	&	201.6	&	163.2	&	7.3	&	71.7	&	60.3	&	3.7	&	29.3	&	23.3	&	1.9	\\
3.5	&	443.4	&	278.5	&	17.8	&	158.6	&	101.2	&	7.1	&	51.5	&	37.5	&	3.4	&	23.3	&	19.1	&	1.7	\\
4.5	&	360.0	&	220.5	&	18.4	&	129.0	&	82.1	&	7.9	&	42.7	&	27.8	&	3.6	&	19.3	&	16.0	&	2.0	\\
5.5	&	305.0	&	181.7	&	17.6	&	106.7	&	60.1	&	7.3	&	38.6	&	16.9	&	3.2	&	17.3	&	11.9	&	1.9	\\
6.5	&	274.4	&	149.0	&	15.7	&	93.0	&	53.1	&	6.1	&	36.6	&	15.6	&	3.4	&	15.6	&	9.4	&	2.1	\\
7.5	&	263.3	&	109.0	&	21.7	&	89.9	&	47.3	&	6.7	&	36.2	&	15.5	&	3.3	&	14.9	&	9.1	&	2.0	\\
8.5	&	221.0	&	94.7	&	17.7	&	78.7	&	41.4	&	8.0	&	30.5	&	13.9	&	3.4	&	13.4	&	7.2	&	2.2	\\
9.5	&	182.7	&	85.8	&	15.3	&	67.2	&	36.6	&	8.2	&	24.7	&	11.9	&	3.1	&	11.3	&	6.1	&	1.9	\\
10.5	&	169.4	&	78.0	&	17.2	&	63.0	&	33.2	&	7.8	&	22.4	&	11.2	&	3.2	&	10.4	&	5.7	&	1.8	\\
12.5	&	140.5	&	66.4	&	16.2	&	52.6	&	27.1	&	6.7	&	20.3	&	10.0	&	3.2	&	8.8	&	4.8	&	1.8	\\
14.5	&	132.9	&	65.9	&	16.5	&	49.9	&	25.6	&	6.6	&	19.5	&	9.5	&	3.1	&	9.1	&	5.4	&	1.9	\\
16.5	&	146.3	&	61.7	&	16.4	&	52.4	&	24.0	&	7.1	&	18.9	&	8.6	&	2.5	&	8.8	&	4.6	&	1.7	\\
18.5	&	136.3	&	53.9	&	16.5	&	50.5	&	22.6	&	6.6	&	18.0	&	8.5	&	2.8	&	7.5	&	3.9	&	1.7	\\
20.5	&	116.8	&	51.4	&	17.0	&	44.6	&	20.4	&	5.9	&	17.0	&	8.5	&	3.0	&	7.9	&	4.0	&	1.7	\\
25.5	&	94.5	&	42.6	&	15.2	&	35.2	&	17.1	&	5.3	&	12.9	&	6.0	&	2.5	&	6.0	&	2.8	&	1.5	\\
30.5	&	79.0	&	36.8	&	10.6	&	29.7	&	14.5	&	5.0	&	10.9	&	6.3	&	2.2	&	5.2	&	3.0	&	1.4	\\
40.5	&	57.8	&	29.6	&	10.6	&	21.2	&	11.2	&	4.7	&	8.1	&	4.8	&	2.2	&	3.6	&	2.2	&	1.1	\\
50.5	&	39.8	&	21.8	&	8.7	&	15.7	&	9.5	&	4.3	&	5.9	&	4.1	&	2.0	&	2.7	&	1.9	&	1.2	\\
60.5	&	28.3	&	19.9	&	9.5	&	11.6	&	8.2	&	3.6	&	4.5	&	3.3	&	1.7	&	2.1	&	1.6	&	1.1	\\
70.5	&	27.0	&	18.1	&	9.0	&	11.1	&	7.8	&	4.1	&	4.6	&	3.3	&	1.9	&	2.1	&	1.5	&	1.1	\\
80.5	&	27.7	&	16.5	&	7.5	&	11.3	&	7.2	&	3.8	&	4.5	&	3.0	&	1.7	&	2.2	&	1.7	&	1.2	\\
90.5	&	25.7	&	16.0	&	8.6	&	9.7	&	6.9	&	3.5	&	3.8	&	2.9	&	1.8	&	1.8	&	1.5	&	1.0	\\
100.5	&	23.6	&	16.2	&	8.6	&	9.2	&	6.7	&	4.0	&	3.6	&	3.0	&	1.9	&	1.7	&	1.4	&	1.1	\\
110.5	&	23.8	&	15.7	&	8.4	&	9.3	&	6.9	&	3.7	&	3.9	&	3.0	&	1.9	&	2.1	&	1.7	&	1.2	\\
120.5	&	22.6	&	15.4	&	7.9	&	8.8	&	6.5	&	3.4	&	3.5	&	2.7	&	1.7	&	1.7	&	1.5	&	1.0	\\
130.5	&	24.3	&	17.8	&	9.2	&	9.9	&	7.8	&	4.3	&	4.0	&	3.0	&	1.9	&	1.9	&	1.8	&	1.2	\\
140.5	&	30.0	&	21.3	&	11.2	&	10.9	&	8.6	&	4.5	&	4.4	&	3.7	&	2.1	&	2.0	&	1.7	&	1.2	\\
150.5	&	35.2	&	26.1	&	12.5	&	12.9	&	10.7	&	5.3	&	4.6	&	3.8	&	2.1	&	1.9	&	1.7	&	1.2	\\
155.5	&	38.6	&	29.0	&	14.1	&	13.9	&	12.0	&	5.8	&	4.7	&	4.4	&	2.2	&	2.3	&	2.3	&	1.3	\\
160.5	&	39.4	&	35.3	&	15.1	&	13.6	&	12.5	&	4.7	&	5.4	&	5.3	&	2.4	&	2.7	&	2.6	&	1.5	\\
162.5	&	54.9	&	39.4	&	18.1	&	17.6	&	14.2	&	6.0	&	6.2	&	5.8	&	2.4	&	2.7	&	2.4	&	1.4	\\
164.5	&	134.4	&	42.9	&	17.0	&	39.3	&	16.1	&	5.8	&	8.8	&	5.6	&	2.7	&	2.9	&	2.4	&	1.5	\\
166.5	&	55.7	&	46.1	&	18.1	&	18.6	&	15.5	&	5.1	&	6.9	&	6.3	&	2.3	&	2.9	&	2.8	&	1.6	\\
168.5	&	54.8	&	46.9	&	16.9	&	19.9	&	17.6	&	5.6	&	7.0	&	6.6	&	2.5	&	3.3	&	3.0	&	1.6	\\
170.5	&	82.7	&	53.7	&	14.5	&	30.4	&	21.6	&	6.2	&	11.2	&	6.8	&	2.7	&	5.3	&	3.2	&	1.4	\\
171.5	&	94.7	&	54.1	&	14.7	&	35.3	&	22.6	&	6.2	&	13.0	&	6.7	&	2.7	&	5.9	&	3.3	&	1.5	\\
172.5	&	111.4	&	58.4	&	17.4	&	41.5	&	22.9	&	6.1	&	16.2	&	7.4	&	2.9	&	7.7	&	3.6	&	1.6	\\
173.5	&	103.5	&	65.2	&	15.4	&	37.0	&	22.5	&	5.9	&	15.0	&	8.7	&	2.8	&	7.2	&	3.7	&	1.8	\\
174.5	&	79.0	&	69.2	&	11.6	&	27.6	&	24.0	&	6.5	&	10.2	&	9.4	&	2.1	&	4.6	&	3.9	&	1.8	\\
175.5	&	75.4	&	74.7	&	14.8	&	26.1	&	24.8	&	5.6	&	9.2	&	9.0	&	2.2	&	3.9	&	3.9	&	1.3	\\
176.5	&	81.8	&	81.8	&	17.2	&	28.4	&	28.3	&	6.1	&	9.9	&	9.2	&	2.4	&	4.3	&	4.3	&	1.5	\\
177.5	&	93.7	&	98.6	&	18.4	&	32.7	&	31.9	&	7.2	&	11.2	&	11.0	&	2.4	&	4.8	&	3.9	&	1.4	\\
178.5	&	105.6	&	109.0	&	22.0	&	32.5	&	32.4	&	7.1	&	10.3	&	10.3	&	1.5	&	4.6	&	4.2	&	1.7	\\
179.5	&	116.3	&	116.3	&	14.9	&	29.1	&	29.1	&	3.6	&	8.6	&	8.6	&	0.5	&	3.8	&	3.8	&	0.8	

\\\hline

\end{tabular}
\end{center}
\caption
{
Fermi: Original observational mean intensity $I_0 (\theta)$, final mean intensity after discarding the outliers $I(\theta)$, and standard deviation $\sigma(\theta)$, in units of $\rm{GeV\ cm^{-2}\ s^{-1}\ Hz^{-1}\ sr^{-1}}$.
}
\label{tabFermiIntensity1}
\end{table*}

%__________________________________
\begin{table*}
\begin{center}
\begin{tabular}{ccccccccc}

\\\hline
&\multicolumn{8}{c}{Fermi} \\
&\multicolumn{3}{c}{30.0-53.3 GeV}
&\multicolumn{3}{c}{53.3-94.9 GeV}	
&\multicolumn{1}{c}{94.9-168.7 GeV}	
&\multicolumn{1}{c}{168.7-300.0 GeV}
\\\hline
\multicolumn{1}{c}{$\theta$}	
&\multicolumn{1}{c}{$I_0 (\theta)$} &\multicolumn{1}{c}{$I(\theta)$} &\multicolumn{1}{c}{$\sigma (\theta)$}
&\multicolumn{1}{c}{$I_0 (\theta)$} &\multicolumn{1}{c}{$I(\theta)$} &\multicolumn{1}{c}{$\sigma (\theta)$}	
&\multicolumn{1}{c}{$I_0 (\theta)$} &\multicolumn{1}{c}{$I_0(\theta)$} 	\\
\multicolumn{1}{c}{$[^\circ]$}	
&\multicolumn{3}{c}{$\times 10^{-32}$}						
&\multicolumn{3}{c}{$\times 10^{-32}$}	
&\multicolumn{1}{c}{$\times 10^{-32}$}						
&\multicolumn{1}{c}{$\times 10^{-32}$}
\\\hline

0.5	&	253.5	&	263.1	&	13.8	&	130.5	&	134.0	&	8.7	&	75.5	&	56.9	\\
1.5	&	189.2	&	183.6	&	14.5	&	96.3	&	101.4	&	9.4	&	50.7	&	31.4	\\
2.5	&	144.4	&	113.8	&	12.9	&	63.8	&	50.9	&	8.6	&	32.1	&	13.4	\\
3.5	&	115.9	&	76.8	&	12.3	&	44.4	&	37.2	&	9.4	&	22.3	&	5.4	\\
4.5	&	97.8	&	58.0	&	13.2	&	37.4	&	33.1	&	8.4	&	17.5	&	4.8	\\
5.5	&	86.2	&	47.6	&	11.4	&	34.1	&	27.0	&	7.5	&	12.4	&	5.6	\\
6.5	&	71.5	&	39.8	&	9.9	&	32.1	&	15.3	&	9.3	&	9.6	&	5.7	\\
7.5	&	71.0	&	32.1	&	12.8	&	32.5	&	17.3	&	9.1	&	10.0	&	7.5	\\
8.5	&	64.3	&	29.7	&	12.5	&	30.1	&	18.1	&	9.3	&	9.5	&	6.6	\\
9.5	&	55.4	&	27.6	&	11.1	&	28.2	&	14.3	&	8.6	&	9.6	&	5.2	\\
10.5	&	55.1	&	28.4	&	11.9	&	29.8	&	14.0	&	9.0	&	11.9	&	4.7	\\
12.5	&	48.5	&	25.3	&	11.8	&	27.8	&	12.6	&	8.4	&	10.7	&	4.6	\\
14.5	&	44.0	&	20.9	&	11.4	&	20.1	&	10.1	&	6.9	&	8.8	&	1.9	\\
16.5	&	43.7	&	21.0	&	10.2	&	17.2	&	7.3	&	6.0	&	8.2	&	4.2	\\
18.5	&	39.7	&	21.9	&	11.1	&	17.9	&	8.0	&	7.0	&	9.1	&	3.7	\\
20.5	&	39.2	&	22.7	&	11.4	&	13.4	&	7.8	&	6.1	&	9.5	&	2.5	\\
25.5	&	28.9	&	16.5	&	9.8	&	14.4	&	9.1	&	6.6	&	7.1	&	3.5	\\
30.5	&	21.8	&	13.7	&	9.4	&	10.9	&	6.3	&	6.0	&	6.3	&	2.0	\\
40.5	&	19.1	&	13.4	&	8.6	&	8.5	&	6.6	&	6.2	&	4.2	&	2.0	\\
50.5	&	11.1	&	8.0	&	6.7	&	6.8	&	4.4	&	5.0	&	4.6	&	1.1	\\
60.5	&	9.4	&	7.7	&	6.9	&	6.2	&	5.0	&	5.5	&	2.4	&	0.8	\\
70.5	&	11.4	&	8.9	&	8.2	&	5.9	&	4.8	&	5.1	&	3.6	&	2.3	\\
80.5	&	11.6	&	8.6	&	7.2	&	5.9	&	4.1	&	5.1	&	2.8	&	1.1	\\
90.5	&	8.9	&	7.9	&	7.7	&	5.5	&	4.8	&	5.5	&	2.4	&	1.5	\\
100.5	&	8.3	&	7.5	&	6.5	&	5.2	&	4.8	&	5.4	&	2.7	&	0.9	\\
110.5	&	9.4	&	8.3	&	7.9	&	5.1	&	4.0	&	5.2	&	2.7	&	1.1	\\
120.5	&	8.3	&	7.7	&	7.0	&	5.6	&	5.3	&	5.4	&	2.6	&	1.2	\\
130.5	&	11.5	&	10.6	&	8.5	&	5.0	&	4.5	&	5.4	&	2.4	&	1.4	\\
140.5	&	9.6	&	8.7	&	8.0	&	6.0	&	4.9	&	5.4	&	3.5	&	1.5	\\
150.5	&	10.9	&	9.3	&	7.9	&	6.4	&	5.2	&	5.4	&	2.8	&	1.1	\\
155.5	&	9.6	&	8.4	&	7.5	&	5.3	&	5.1	&	5.4	&	3.3	&	1.2	\\
160.5	&	11.3	&	10.6	&	8.3	&	5.7	&	5.2	&	5.7	&	3.1	&	2.7	\\
162.5	&	9.8	&	9.0	&	7.6	&	6.0	&	5.0	&	5.9	&	3.4	&	2.6	\\
164.5	&	12.4	&	11.6	&	9.1	&	6.9	&	6.3	&	6.0	&	3.8	&	1.5	\\
166.5	&	12.8	&	11.6	&	7.5	&	5.6	&	5.5	&	5.5	&	3.4	&	1.7	\\
168.5	&	11.6	&	11.5	&	8.7	&	6.6	&	5.9	&	5.8	&	3.5	&	1.8	\\
170.5	&	20.9	&	13.3	&	9.5	&	12.1	&	5.9	&	5.2	&	5.3	&	2.6	\\
171.5	&	23.4	&	12.9	&	9.4	&	13.7	&	6.3	&	5.1	&	6.9	&	3.4	\\
172.5	&	27.9	&	12.0	&	9.1	&	18.2	&	8.0	&	6.3	&	11.0	&	4.7	\\
173.5	&	26.4	&	14.0	&	8.5	&	16.7	&	7.9	&	6.8	&	9.6	&	4.9	\\
174.5	&	19.3	&	17.7	&	9.7	&	9.4	&	8.0	&	6.7	&	5.5	&	3.3	\\
175.5	&	17.6	&	16.8	&	9.6	&	6.3	&	6.3	&	5.5	&	6.5	&	1.3	\\
176.5	&	16.3	&	16.3	&	9.5	&	5.6	&	5.6	&	5.3	&	7.3	&	0.9	\\
177.5	&	17.8	&	17.8	&	9.1	&	6.4	&	6.4	&	5.8	&	4.6	&	3.5	\\
178.5	&	25.0	&	25.0	&	10.3	&	6.9	&	6.9	&	6.0	&	1.3	&	5.2	\\
179.5	&	29.7	&	29.7	&	9.3	&	6.3	&	6.3	&	4.0	&	0.2	&	3.4

\\\hline

\end{tabular}
\end{center}
\caption
{
Fermi: Original observational mean intensity $I_0 (\theta)$, final mean intensity after discarding the outliers $I(\theta)$, and standard deviation $\sigma(\theta)$, in units of $\rm{GeV\ cm^{-2}\ s^{-1}\ Hz^{-1}\ sr^{-1}}$.
}
\label{tabFermiIntensity2}
\end{table*}

Intensity maps $I(l,b)$ in Galactic coordinates, masked residual maps $I(l,b)-I(\theta)$, and spherically-averaged intensities for all radio, microwave, and gamma-ray frequencies are shown in Figures~\ref{figHaslam&WMAPMap}, \ref{figFermiMap1}, and~\ref{figFermiMap2}.
Numeric values of the original mean intensity $I_0(\theta)$, the mean intensity $I(\theta)$ after discarding the outliers, and the standard deviation $\sigma(\theta)$ are quoted in
Tables~\ref{tabHaslam&WMAPIntensity}, \ref{tabWMAP&FermiIntensity}, \ref{tabFermiIntensity1}, and \ref{tabFermiIntensity2}.

%__________________________________

\section{Cylindrical boundary conditions}
\label{secCylindrical}

%__________________________________
\begin{figure}
\centering \includegraphics[width=8cm]{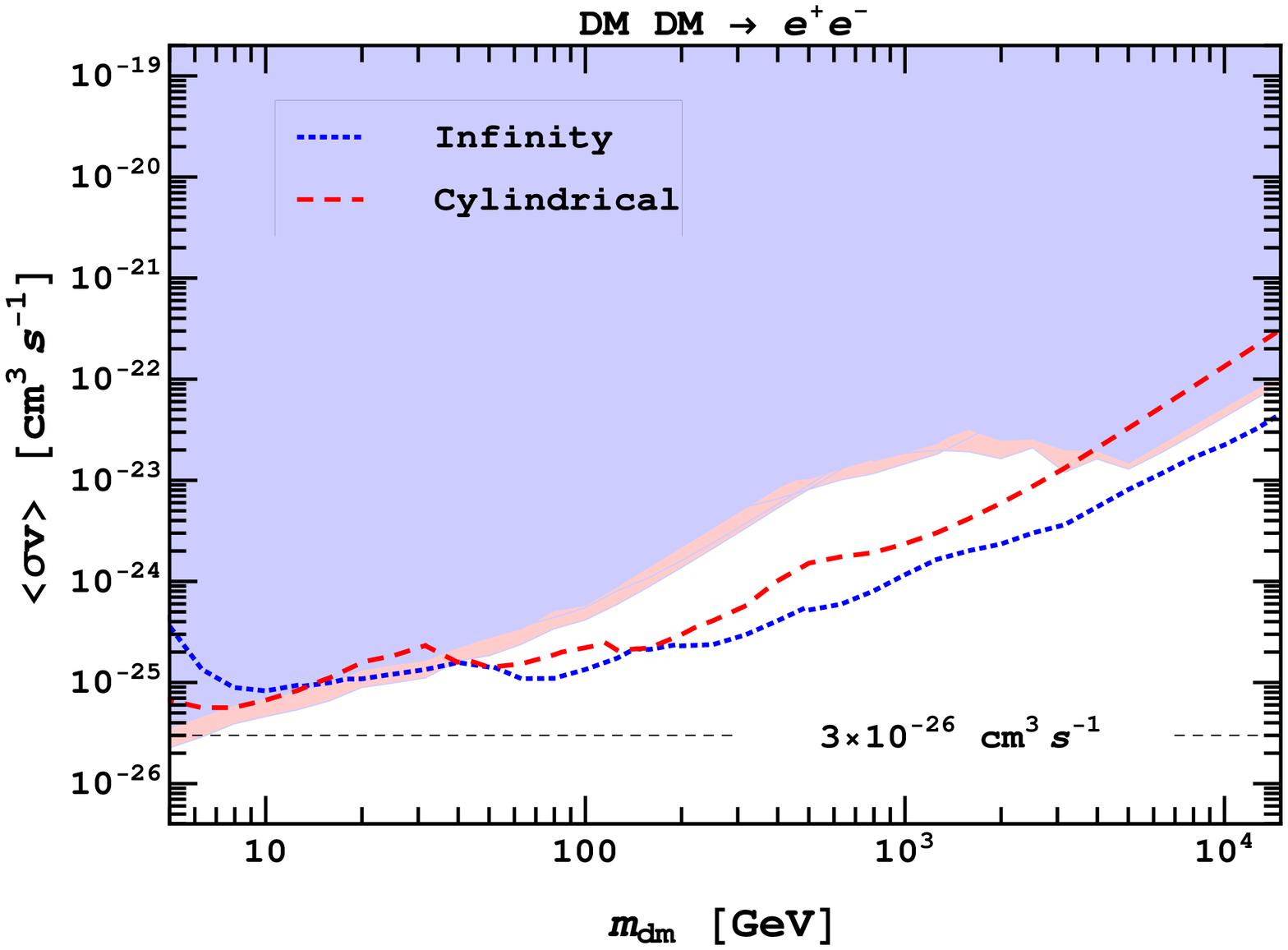}
\caption
{
Effect of the boundary conditions on the upper limits derived from inverse Compton scattering (lines) and the local electron and positron spectra (shaded regions) for dark matter annihilating directly into electron-positron pairs.
Red colours correspond to cylindrical boundary conditions \Referee{with MED model and vertical height $L=4$ kpc}, whereas blue is used to represent our adopted boundary conditions at infinity.
}
\label{figCylindrical}
\end{figure}
%__________________________________

The cosmic-ray propagation in our Galaxy is often modelled by imposing null boundary conditions on a cylinder of finite height and radius \citep[see e.g.][]{Maurin+01}.
Moreover, the adopted normalisation and spectral index of the diffusion coefficient are based on analyses implementing a cylindrical diffusion zone, and therefore our use of boundary conditions at infinity is not entirely self-consistent.

The effect of the adopted boundary conditions is illustrated in Figure~\ref{figCylindrical} for the case of dark matter annihilation into electron-positron pairs.
Results for null boundary conditions at infinity (i.e. the same as those shown on the top-left panel of Figure~\ref{figChannels_sigma}) are compared with the upper limits derived from the ICS emission and the local electron and positron spectra estimated by the routines in {\sc pppc4dmid} \citep{Cirelli+11}, based on cylindrical boundary conditions 
\Referee{with MED model and vertical height $L=4$ kpc.}

% XXX...
The most important differences occur at the high-mass end, where the ICS constraints are weakened by almost an order of magnitude.
For dark matter masses below 1~TeV, the choice of boundary conditions modifies the results by about a factor of two.

\newpage~~\newpage~~
\newpage~~\newpage~~
\newpage~~\newpage~~
\newpage~~\newpage~~
\newpage~~\newpage~~
\newpage~~\newpage~~

\end{document}